\documentclass[
  aps,
  prd,
  twocolumn,
  nofootinbib,
  superscriptaddress
]{revtex4-1} \usepackage{graphicx}
\usepackage{rotating}
\usepackage[T1]{fontenc}

\usepackage{graphicx}	
\usepackage{amsmath}	
\usepackage{amssymb}	
\usepackage{acronym}
\usepackage{comment}    
\usepackage{ulem}       
\usepackage{natbib}
\usepackage{hyperref}

\def\i{{\rm i}}

\def\Msun{{\rm M_{\odot}}}
\def\GMc2{{\rm G M_{\odot} c^{-2}}}

\def\O{\mathcal{O}}

\def\Mo{{\rm M_{\odot}}}
\def\Msun{\Mo}
\def\kt2{\kappa^\text{T}_2}

\def\R14{R_\text{1.4}^\text{TOV}}

\def\eg{{\it e.g.}}

\usepackage{pifont} 

\newcommand{\thc}{\texttt{THC}~}

\newcommand{\athena}{\texttt{Athena++}~}
\newcommand{\gra}{\texttt{GR-Athena++}~}
\newcommand{\knec}[1]{\texttt{KNEC{#1}}~}

\newcommand{\snec}{\texttt{SNEC}~}
\newcommand{\skynet}{\texttt{SkyNet}~}
\newcommand{\winnet}{\texttt{WinNet}~}

\newcommand{\be}{\begin{equation}}
\newcommand{\ee}{\end{equation}}
\newcommand{\bea}{\begin{eqnarray}}
\newcommand{\eea}{\end{eqnarray}}
\newcommand{\bel}{\begin{align}}
\newcommand{\eel}{\end{align}}

\usepackage{color}
\definecolor{cyan}{rgb}{0,0.9,0.9}
\definecolor{orange}{rgb}{0.9,0.5,0}
\definecolor{magenta}{rgb}{1,0,1}
\definecolor{purple}{rgb}{0.8,0.4,0.8}
\definecolor{gray}{rgb}{0.8242,0.8242,0.8242}

\definecolor{teal}{rgb}{0,0.55,0.55}

\acrodef{BNS}[BNS]{binary neutron star}
\acrodef{NR}[NR]{Numerical relativity}
\acrodef{GR}[GR]{general relativity}
\acrodef{EoS}[EOS]{equation of state}
\acrodef{GW}[GW]{gravitational wave}
\acrodef{EM}[EM]{electromagnetic}
\acrodef{MHD}[MHD]{magnetohydrodynamics}
\acrodef{NS}[NS]{neutron star}
\acrodef{BH}[BH]{black hole}
\acrodef{LK}[LK]{leakage}
\acrodef{GRB}[SGRB]{short-gamma-ray burst}
\acrodef{GRLES}[GRLES]{general relativistic Large-Eddy-Simulations}

\begin{document}

\title{3D Binary Neutron Star Merger Ejecta Evolution up to Seconds Timescale:\\
  Dynamics, Element Distribution, and Light Curves}

\author{Lu\'is Felipe Longo Micchi}
\affiliation{Theoretisch-Physikalisches Institut,
Friedrich-Schiller-Universit{\"a}t Jena,
Fr{\"o}belstieg 1, 07743 Jena, Germany}

\author{Fabio Magistrelli}
\affiliation{Theoretisch-Physikalisches Institut,
Friedrich-Schiller-Universit{\"a}t Jena,
Fr{\"o}belstieg 1, 07743 Jena, Germany}

\author{Maximilian Jacobi}
\affiliation{Theoretisch-Physikalisches Institut,
Friedrich-Schiller-Universit{\"a}t Jena,
Fr{\"o}belstieg 1, 07743 Jena, Germany}

\author{Sebastiano Bernuzzi}
\affiliation{Theoretisch-Physikalisches Institut,
Friedrich-Schiller-Universit{\"a}t Jena,
Fr{\"o}belstieg 1, 07743 Jena, Germany}

\date{\today}

\begin{abstract}
We present long-term, three-dimensional simulations of ejecta from four binary neutron star mergers to second-long time-scales. Numerical-relativity data serve as boundary conditions for a general-relativistic hydrodynamics evolution incorporating an equation of state valid outside nuclear statistical equilibrium and an effective nuclear-heating prescription based on reaction-network calculations. We investigate the ejecta's dynamical and geometrical properties, the impact of nuclear heating, the formation and spatial distribution of elements, and compute multi-angle kilonova light curves.
Nuclear heating significantly affects ejecta dynamics, delaying homologous expansion beyond second time-scales and reshapes the spatial distribution of heavy nuclei. This effect is largest for asymmetric binaries with long-lived remnants; extending the evolution from ${\sim}150$~ms to ${\sim}1$~s widens the angular polar region containing 90\% of the heavy-element mass (\eg, $Z=56$, $Z=79$) from $|\theta|\lesssim15^{o}$ to $|\theta|\lesssim30^{o}$. Our nucleosynthesis results confirm that the $^{56}$Ni$\rightarrow^{56}$Co$\rightarrow^{56}$Fe decay chain dominates the heating at $\sim$100 days, with cobalt decay producing gamma-ray lines at 846.77 and 1238.288~keV.
Comparing kilonova ray-by-ray light curves obtained from multi-angle 3D profiles and averaged 2D profiles, we find the latter approach broadly robust, although 2D light curves should be treated as upper limits. Increasing dimensionality generally lowers the bolometric luminosity, with binary asymmetry strengthening the viewing-angle dependence. For observers aligned with a lanthanide curtain's densest region, 3D emission can match its 2D counterpart in brightness. We conclude that increased dimensionality alone is unlikely to reconcile current theoretical models with AT2017gfo observations.
\end{abstract}

\maketitle

\section{Introduction}
\label{sec:intro}

During the coalescence process, binary neutron star (BNS) mergers emit gravitational waves (GW) and eject hot and neutron-rich material to their surroundings \cite{Radice:2020ddv}.
As the ejecta expand, the rapid neutron capture process ($r$-process) generates heavy elements and powers an electromagnetic (EM) transient known as a kilonova (KN) \cite{Li:1998bw,Lattimer:1974slx,Metzger:2010sy}.
BNS studies have received ever growing attention since the detection of GW170817 \cite{Abbott:2018hgk,Abbott:2018wiz}, the first BNS event detected by the LIGO-Virgo collaboration associated with the KN AT2017gfo \cite{GBM:2017lvd,Coulter:2017wya,Arcavi:2017a,Cowperthwaite:2017dyu,Diaz:2017uch,Drout:2017ijr,Evans:2017mmy,Kasliwal:2017ngb,Pian:2017gtc,Smartt:2017fuw,Troja:2017nqp,Utsumi:2017cti,Valenti:2017ngx,Villar:2017wcc}.

Many works have focused on the modeling of the post-merger phase of these events and investigated the different ejecta components that emerge from BNS events. Ejecta are launched at various stages of the merger and remnant dynamics by different mechanisms and can be roughly classified as dynamical, spiral, neutrino-driven and viscous-driven ejecta \cite{Bernuzzi:2020tgt}. Dynamical ejecta originate from the bounce shock \cite{Radice:2018pdn} and tidal arms during the GW-driven merger and post-merger phase, e.g. \cite{Rosswog:1998hy,Rosswog:2013,Perego:2019adq}. For this reason they are composed of neutron-rich material mostly around the orbital plane of the binary. Spiral-wave winds are also generated during the post-merger early phase on time-scales of $20-100$~ms from spiral-density waves transporting angular momentum outwards \cite{Nedora:2019jhl}. Their dynamics and composition can be further influenced by neutrino irradiation~\cite{Radice:2023xxn} and magnetic fields \cite{Combi:2022nhg}. Neutrino-driven ejecta originate from the energy deposition due to neutrino absorption at the edge of the remnant disk and above the remnant~\cite{Dessart:2008zd,Perego:2014fma,Martin:2015hxa}. Large neutrino fluxes both unbind fluid elements and force neutrino capture by neutrons; neutrino-driven ejecta are therefore characterized by a different geometry than the dynamical ejecta and the spiral-wave winds, and are composed of proton-rich material. Lastly, viscous ejecta are produced at longer time-scales (${\sim}0.1-1.0$~s) from the remnant disk and are driven by a combination of neutrino absorption and magnetic field effects. They are believed to constitute the most massive ejecta component, although ab-initio numerical-relativity studies are limited due to the long time-scales and complex physics \cite{Kiuchi:2022nin}.

The different ejecta components undergo distinct nucleosynthesis processes. Dynamical ejecta, spiral-wave-wind and part of the viscous equatorial early ejecta produce heavy (A$\gtrsim 100$) unstable neutron-rich isotopes \cite{Metzger:2010sy,Goriely:2011vg,Korobkin:2012uy} from $r$-process nucleosynthesis. Proton-rich winds tend instead to form symmetric isotopes such as $^{56}$Ni \cite{Perego:2014fma,Martin:2015hxa,Jacobi:2025eak}. Viscous ejecta are expected to undergo both weak and strong $r$-processes~\cite{Beloborodov:2002af,Siegel:2017nub,Sprouse:2023cdm}. Most of the first isotopes to be formed are unstable and decay on varying time-scales. Their decays power the electromagnetic (EM) emission of the KN. Due to their different compositions, different ejecta components have different opacities, and therefore leave different imprints on the KN light curves. Material with $Y_e\lesssim0.25$ is known to synthesize lanthanides and to shift the peak of the EM emission towards the near-infrared (NIR) band \cite{Kasen:2014toa,Tanaka:2017qxj}. For this reason the dynamical and spiral ejecta are often related to the ``red'' component of the KN. On the other hand, $Y_e\gtrsim0.25$ ejecta are associated with emission peaking in the UV/optical spectrum and is often referred to as the KN's ``blue'' component \cite{Perego:2014fma,Kasen:2014toa}. However, both observations and theoretical studies indicate a more complex picture than the ``red/blue'' scheme \cite{Perego:2017wtu,Nedora:2020hxc}.

One of the great challenges in modeling KN and their corresponding nucleosynthesis is the different scales over which different physical effects work.
Given the high computational cost, numerical relativity (NR) simulations of BNS mergers typically cover a few hundreds milliseconds of post-merger dynamics.
Nevertheless, KN emission evolves and is typically in the band of current detectors on time-scales of days to months.
The nucleosynthesis involves different nuclear reactions acting on $\sim10^{-3}-1$~second time-scales, up tothe decay of actinides on the days to years time-scale.
Also contributing to the difficulty of simulating long time-scales is the large number of degrees of freedom in nuclear network codes, where $\sim{6-7}\times10^{3}$ different nuclei and their reactions are usually implemented, \eg \cite{Lippuner:2017tyn, Reichert:2023xqy}, and the related computational cost of in-situ nuclear network~\cite{Magistrelli:2024zmk,Magistrelli:2025xja}.

In order to build the bridge between these different time-scales, past works evolved BNS mergers to $\O({100})$~ms and used the resulting data as input for further modeling  \cite{Radice:2016dwd, Radice:2018ozg, Nedora:2020hxc, Camilletti:2022jms,Neuweiler:2022eum, Magistrelli:2024zmk,Neuweiler:2025klw,Magistrelli:2025xja}.
Many of these works have used this hand-off system to investigate the nucleosynthesis by feeding the NR data to nuclear network codes such as \skynet \cite{Lippuner:2017tyn} and \winnet \cite{Reichert:2023xqy} in post-processing.
One such early work extended to 100 years the initial $t\sim{20}$~ms post-merger evolution of a BNS merger, performed with a Newtonian smoothed particle hydrodynamics (SPH) code \cite{Rosswog:2013kqa}. The evolution included the effect of nuclear heating by means of fits, made available in \cite{Korobkin:2012uy}, from nuclear network trajectories. More recently, \cite{Kawaguchi:2024hdk} also performed a hydrodynamical evolution of the ejecta of a BH-NS merger system for $t\sim10^{4}$ seconds.
Other works employed NR data as input to radiative transport codes \cite{Morozova:2015bla, Bulla:2019muo} to compute their expected light curves.
Even though much progress has been achieved into modeling KN light curves from first principles, there is still a gap between the ab-initio simulated light curves and AT2017gfo data \cite{Collins:2022ocl,Groenewegen:2025ezj,Breschi:2021tbm}.
This hints to a missing component (due to the finite duration of NR simulations), 3D effects and/or missing physics in current models.
Nevertheless, analytical multi-parameter fitting KN models do achieve better agreement with observations. Such models usually divide the ejecta powering the KN into two or three components according to their composition and emission properties \cite{Cowperthwaite:2017dyu,Villar:2017wcc,Kasen:2017sxr,Metzger:2019zeh,Nicholl:2021abc}.

More recent efforts, such as \cite{Neuweiler:2022eum}, have extended the hydrodynamical evolution of BNS dynamical ejecta out to $\sim$1 second, tracking its geometric evolution before mapping the resulting profiles onto a Monte Carlo radiative-transfer calculation that computes KN light curves from the ejecta's density, composition, and velocity structure together with time-dependent nuclear heating rates (HR) and opacities. The authors use this pipeline to test the robustness of the resulting light curve against the choice of profile extraction time. \cite{Just:2023wtj} instead mapped azimuthally-averaged SPH profiles onto a special-relativistic hydrodynamics code \cite{Just:2015fda,Obergaulinger:Thesis:2008} with an energy-dependent two-moment neutrino treatment, extending the post-merger evolution for a range of binary mass ratios. As in the previous work, nuclear heating was not coupled directly to the hydrodynamics but instead evaluated by post-processing the resulting profiles. \cite{Kawaguchi:2024hdk} took a complementary approach, evolving an NS-BH merger over second-long time-scales with a hydrodynamic code assuming a non-rotating BH background, coupling the evolution directly to an analytic nuclear heating formula rather than applying it in post-processing.

In this paper, we consider evolutions of NR ejecta up to second time-scales. We investigate the 3D ejecta dynamics under the influence of nuclear heating, the element formation in the ejecta and their distribution for four binaries. We also compare synthetic light curves generated from data at different stages of the ejecta dynamics.

Our setup to perform general-relativistic three-dimensional ejecta evolutions and their post-processing is described in Sec.~\ref{sec:method}. It allows us to input ejecta data from NR as a boundary condition, and minimizes geometry and data manipulation. It incorporates an out of nuclear-statistical-equilibrium (NSE) equation of state (EOS) at low densities and nuclear burning using fitted (effective) heating rates.
Results on ejecta kinematics, geometry and composition are presented in Sec.~\ref{sec:res:dyn}.
Element formation in the ejecta is discussed in Sec.~\ref{sec:res:elem}, which presents nucleosynthesis calculations and the element distribution in the ejecta.
KN light curves are discussed in Sec.~\ref{sec:res:lc}.
In Sec.~\ref{sec:con}, we summarize our findings and discuss future steps of investigation.

\section{Methods}
\label{sec:method}

We perform long-term hydrodynamical ejecta evolutions using the \athena{} code \cite{Stone:2020} and ab-initio ejecta data from the NR simulations of Refs.~\cite{Radice:2023zlw,Bernuzzi:2024mfx,Gutierrez:2024pch}. The NR data are injected as a boundary condition at the bottom of the \athena spherical-shell grid and evolved up to $\mathcal{O}(1~\mathrm{s})$. High-density matter with typical densities $10^{3}\lesssim \rho/  ({\rm g/cm}^{3})  \lesssim 2 \times 10^{15} $ and temperatures $0.05\lesssim T/{\rm MeV} \lesssim 100$
is described with the same micro-physical EOS of the NR data, which assumes NSE conditions.
In order to describe matter evolving towards lower densities and temperatures, the EOS is extended using Timmes' EOS \cite{Timmes:2000a}.
The latter includes contributions from radiation, completely ionized nuclei, and degenerate and relativistic electrons and positrons.
In these thermodynamical states we account for nuclear burning using the tabulated heating rates of \cite{Magistrelli:2025xja}, which constitute an extension of the fits of \cite{Wu:2021ibi}.
In order to obtain detailed predictions of the nucleosynthesis of our models, their ejecta are post-processed with the \winnet nuclear reaction network code \cite{Reichert:2023xqy}.
KN light curves are computed by mapping the homologously expanding 3D \athena profiles onto the ray-by-ray radiation-hydrodynamics \snec/\knec{} code \cite{Morozova:2015bla,Wu:2021ibi,Magistrelli:2024zmk,Magistrelli:2025xja}.
Details of our setup are given in what follows.

\subsection{\athena setup}
\label{sec:method:athena}

\athena~hydrodynamical simulations are performed assuming a Schwarzschild background metric with mass parameters compatible with the NR binary merger simulations.
We adopt a spherical-shell 3D grid without symmetries and use radial, latitude and longitude resolutions, respectively, of $N_{r}=256$, $N_{\theta}=16$, $N_{\phi}=32$. 
Throughout this paper we assume $\theta=0$ to be aligned with the equatorial plane.
The radial extent of the grid is $300 \leq r/M_{\odot} \leq 200000$, whose upper limit is approximately one light-second. 
The radial grid is geometrically spaced with $\Delta r$ increasing outwards with a geometric ratio of $q_{dr}=1.035$. 
Hydrodynamics is solved using the HLLE flux and piece-wise linear method (PLM) reconstruction.

The time integration is performed with a second-order accurate van Leer predictor-corrector scheme subject to a Courant–Friedrichs–Lewy (CFL) condition.
The built-in time stepping condition in \athena for the general relativistic case ($dt \propto 1/c$, where $c$ is the speed of light) is too stringent as our simulations do not contain highly-relativistic motions for most of the run-time. Therefore, we adopt a hybrid method to speed up the computations. For times smaller than $5M_{\odot}$, we use $dt \propto 1/c$ in order to capture the injection of the boundary data into the grid (initially populated by floor values of pressure and density, see below). Afterwards, we use a CFL condition based on the acoustic wave characteristics, as prescribed in \cite{Banyuls:1997zz}.

We have implemented in \athena the use of tabulated nuclear EOS, an extension to Timmes/Helmholtz EOS and nuclear burning using fitted heating rates, which are described next.

\subsection{Binary models and Data injection}
\label{sec:method:nr}

\begin{table*}[t]
\centering
\caption{Parameters of the four BNS merger models used in this work. All masses are gravitational masses. $t_{\rm collapse}$ denotes the post-merger time at which the remnant collapses to a black hole. $t_{\rm start}$ denotes the start time of our simulation relative to merger. Different start times are chosen so as to initialize the simulation close to the moment when the ejecta cross the \thc extraction radius. All our simulations end after $\sim$1 second of evolution on \athena. For all models, the transition EOS defined by Eqs.~\eqref{eq:wdefinition}--\eqref{eq:wndefinition} is applied between $T_{\rm trans}^{\rm start}=0.7$~MeV and $T_{\rm trans}^{\rm end}=0.05$~MeV, and between $n_{\rm trans}^{\rm start}=10^{-10}$~fm$^{-3}$ and $n_{\rm trans}^{\rm end}=10^{-12}$~fm$^{-3}$.}
\label{tab:bnsmodels}
\begin{tabular}{|c| c| c| c| c| c| c|}
\hline\hline
Model & EOS & $M_{1}[M_{\odot}]$ & $M_{2}[M_{\odot}]$ & Remnant & $t_{\rm collapse}$~(ms) & $t_{\rm start}$~(ms) \\
\hline\hline
BLh\_q1.43 & BLh  & 1.635 & 1.146 & BH & 114  & 16.9 \\
\hline
SFHo\_q1.0 & SFHo & 1.35  & 1.35  & BH & 8.3  & 0.05 \\
\hline
DD2\_q1.0  & DD2  & 1.35  & 1.35  & NS & --   & 0.05 \\
\hline
DD2\_q1.67 & DD2  & 1.80  & 1.08  & NS & --   & 0.05 \\
\hline\hline
\end{tabular}
\end{table*}

We use four BNS merger simulations from \cite{Radice:2023zlw,Bernuzzi:2024mfx,Gutierrez:2024pch}; some properties of the simulated binaries are summarized in Tab.~\ref{tab:bnsmodels}.
The neutron stars are described by three micro-physical EOS:  BLh \cite{Bombaci:2018ksa,Logoteta:2020yxf}, SFHo \cite{Steiner:2010fz},  and DD2 \cite{Typel:2009sy,Hempel:2009mc}.
We consider two equal-mass and two unequal-mass binaries with mass ratios $q\simeq{1.43}$ and $q\simeq{1.67}$.
The binaries were initially evolved for approximately 100~ms after merger with the (3+1)D GRHD \thc{} code which incorporates a grey M1 neutrino scheme with complete neutrino-matter coupling and a realistic subgrid model for magneto-hydrodynamics-driven turbulent viscosity \cite{Radice:2012cu,Radice:2013hxh,Radice:2013xpa,Radice:2015nva,Radice:2016dwd,Radice:2017zta,Radice:2020ids,Radice:2021jtw}.
The two DD2 binary mergers result in a massive remnant neutron star, while the SFHo and BLh mergers produce a short- and a long-lived remnant that collapse to a black hole at $t\sim{8}$~ms and $t\sim{114}$~ms post-merger, respectively.

Beyond the standard GRHD primitive variables, among which the fluid three-velocity $v^{i}$ with its associated Lorentz factor $W$, \athena additionally evolves a set of passive scalars: the electron fraction $Y_{e}$, the average nuclear mass $\bar{A}$, the entropy $s$, the inverse timescale $\tau^{-1}$, the binding energy $\epsilon_B$, and a flag $F$ tracking the origin of each fluid element, defined as 

\begin{equation}
F = \begin{cases} 0, & \text{initial atmosphere},\\
1, & \text{injected \texttt{THC} data},\\
2, & \text{late-time outflow boundary condition}.\end{cases}
\label{eq:Flag}
\end{equation}

From each one of these simulations, the properties of the ejecta crossing a coordinate radius of 300$M_{\odot}$ are recorded as a function of time on a 2D spherical surface of resolution $(N_{\theta},N_{\phi})=(51,92)$. These ejecta data are used as boundary data for \athena simulations.
\thc ejecta at a given radius are used to populate the inner radial boundary of the \athena spherical grid.
The boundary data are linearly interpolated as a function of the angular coordinates of the spherical surface and time. This way, we construct a time dependent boundary condition.
When the injected data reaches its end, the inner boundary condition becomes a simple outflow condition, imposing that the ghost zones be copies of the adjacent physical cells.
This boundary condition can still lead to spurious injection even after the end of the NR \thc data. Fluid elements injected from this boundary condition are marked with $F=2$, as in Eq.~\eqref{eq:Flag}.
At $t=0$, our physical grid is fully populated with floor values for pressure ($p=5.55\times10^{13}$ dyn/cm$^{2}$), density ($\rho=2.47\times 10^{-6}$ g/cm$^{3}$) and temperature ($T=1\times10^{-7}$~MeV).
For the passive scalars, we start our atmosphere with $Wv^{i}=0$, $Y_{e}=0.5$, $\bar{A}=2.0$, $s= 10^{-45} k_{B}$/baryon, and $F=0$.

We note that during the boundary injection we do not require that fluid elements have strictly positive velocities. Therefore, we allow for all dynamical effects as, for example, fallback. In Appendix~\ref{app:mflux} we show a direct comparison of the mass flux with \thc data at early times. The fluxes closely reproduce those from \thc{}.

\subsection{Transition EOS and Nuclear Heating Rates}
\label{sec:method:HR}

Aiming for a second-long evolution of the ejecta, material reaches densities and temperatures that are not supported by the usual tabulated nuclear EOS.
Our EOS tables of choice for example display a minimum density of $\sim10^{2}-10^{3}$g/cm$^{3}$ and a minimum temperature of 0.01-0.1~MeV, orders of magnitude above the values expected at the late times of our simulations.
For this reason we implement a smooth transition scheme between a tabulated nuclear EOS of choice and the tabulated EOS provided by \cite{Timmes:2000a}.
This latter EOS has lower limits in density and temperature of $\sim 10^{-12}$~g/cm$^{3}$ and ${\sim}0.086$~eV, respectively.
To reach these low densities and temperatures, \cite{Timmes:2000a} calculated the EOS using the Helmholtz free energy and a thermodynamically consistent interpolation scheme, therefore it is often referred to as Helmholtz EOS.
As this EOS includes contributions from degenerate and relativistic electrons and positrons, this constitutes an improvement with respect to the EOS used in \cite{Kawaguchi:2024hdk}.

In the transition regime we define a transition parameter $w$ to smoothly connect both of the EOS tables. It is constructed as

\begin{align} 
w &=
\begin{cases}
1, & \ln n>\ln n_{\mathrm{trans}}^{\mathrm{start}} 
\ \text{and}\ 
T>T_{\mathrm{trans}}^{\mathrm{start}}, \\[2pt]
1, & \ln n>\ln n_{\mathrm{Helm}}^{\mathrm{max}}, \\[2pt]
0, & \ln n<\ln n_{\mathrm{trans}}^{\mathrm{end}} 
\ \text{or}\ 
T<T_{\mathrm{trans}}^{\mathrm{end}}, \\[2pt]
w_T\,w_n, & \text{otherwise},
\end{cases} 
\label{eq:wdefinition}\\[6pt]
w_T &= f\!\left(T;\,T_{\mathrm{trans}}^{\mathrm{end}},\,T_{\mathrm{trans}}^{\mathrm{start}}\right),
\label{eq:wTdefinition}\\
w_n &= f\!\left(\ln n;\,\ln n_{\mathrm{trans}}^{\mathrm{end}},\,\ln n_{\mathrm{trans}}^{\mathrm{start}}\right),
\label{eq:wndefinition}
\end{align}

with the shared clamped linear ramp
\begin{equation}
f(x;x_a,x_b) \equiv \dfrac{x-x_a}{x_b-x_a}.
\label{eq:framp}
\end{equation}

In the previous equations, $n$ refers to the number density and $T$ is the temperature of the fluid.
While $n_{\mathrm{trans}}^{\mathrm{start}}$ and $n_{\mathrm{trans}}^{\mathrm{end}}$ are the boundaries of the transition region with respect to $n$, $T_{\mathrm{trans}}^{\mathrm{start}}$ and $T_{\mathrm{trans}}^{\mathrm{end}}$ represent the boundaries with respect to the temperature.

As discussed before, the definition of the parameter $w$ is made to control the transition between two EOS tables of different validity regimes.
When the NSE assumptions are used, the variables used by the nuclear EOS solver are ($\rho,T,Y_{e}$).
Helmholtz EOS calls take as input ($\rho,T,Y_{e},\bar{A}$), justifying the inclusion of $\bar{A}$ as an extra passive scalar.
The EOS solvers are called independently to solve for a thermodynamic quantity $q = w \times q_{\mathrm{nuclear}} + (1-w) \times q_{\mathrm{Helm}}$, where $q$ is any EOS related quantity to be reconstructed from the primitive variables. 
If $w=0(1)$, only the Helmholtz (nuclear) EOS solver is used.

Accessing the nuclear heating from the fits requires keeping $\bar{A}$ and $Y_{e}$ constant for a given fluid element at their value at ejection, rather than letting them evolve self-consistently alongside nuclear burning as the fluid element is advected.
Storing both an evolved and fixed value for each quantity throughout the simulation would substantially increase the required memory usage of the code.
We instead assess the impact of this simplification a posteriori.
We verify in post-processing, for the BLh\_q1.43 model's tracer trajectories, that 90\%(50\%) of the ejecta mass has a maximum relative error on the pressure smaller than $\sim12\%(8\%)$.
This shows that the contributions of varying $\bar{A}$ and $Y_{e}$ are subdominant in the analysis.

\begin{figure}
\centering
\includegraphics[width=1.0\linewidth]{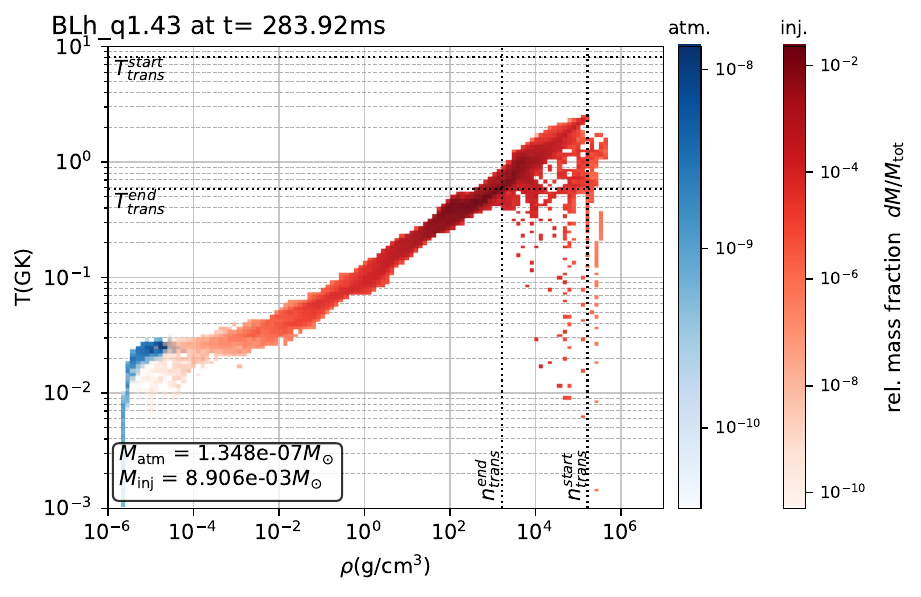}
\caption{Scatter plot of temperature of fluid elements as a function
  of density for BLh\_q1.43 model at
  a time where the boundary injection is still taking place. Dashed
  red lines mark the parameters for the EOS transition
  regime. Atmosphere points are marked with a different color.
  Our data demonstrate a smooth transition between the two regimes of
  the hybrid EOS.
  The points of high density and low temperature are most likely related to failures of the primitive solver. However, we verify that they constitute a negligible number of points in our simulations.
  The reported injected mass still includes masses that will fallback to the remnant.
  }
\label{fig:Trho}
\end{figure}

The parameters that guide the transition are the number density and temperature, as seen in Eqs.~\eqref{eq:wdefinition}--\eqref{eq:wndefinition}.
The values of these parameters were chosen so as to achieve a smooth transition in the $T-\rho$ plane, and can be found in Tab.~\ref{tab:bnsmodels}.
An example of this transition is illustrated in Fig.~\ref{fig:Trho}.
The latter shows an early snapshot of the $T(\rho)$ profile for the BLh\_q1.43 model.

The heating rates are included in the hydro evolution by adding an extra source term to the advection equation of the specific binding energy ($\epsilon_{B}$).
However, in our code this is handle in an operator splitting fashion, where on top of the stantard advection equation
\begin{equation}
	\partial_{t}(\sqrt{-g} \rho W \epsilon_{B}) + \partial_{i} [\sqrt{-g} \rho W \epsilon_{B} v^{i} ] = 0
\label{eq:BindEnergyEvol}
\end{equation}

we add further evolve the binding energy as in 

\begin{equation}
	\epsilon_{B}(t+dt) = \epsilon_{B}(t) - (1-w) \alpha \rho\dot{\epsilon} dt.
\label{eq:BindEnergyEvol2}
\end{equation}

The $w$ function corresponds to the same one used in the transition EOS and defined in Eq.~\eqref{eq:wdefinition}.
The transition function $w$ multiplies the heating rates, as the nuclear EOS assumes NSE and therefore the changes in composition and the associated heating are already accounted for.
In practice the source term in Eq.~\eqref{eq:BindEnergyEvol2} is added in the Chemistry class \cite{Gong:2023abc}.
Eq.~\eqref{eq:BindEnergyEvol2} is integrated using a forward Euler scheme with a time step matching that of the hydrodynamical evolution, rather than the usual substepping approach.
This choice improves computational efficiency while maintaining accuracy, as the heating rate, $\dot{\epsilon}$, varies on timescales much longer than the hydrodynamical time step.

The initial value of $\epsilon_{B}$ for a given fluid element is obtained by

\begin{equation}
	\epsilon_{B} = \epsilon_{\rm NSE} - \epsilon_{\rm Helm}
\end{equation}
evaluated while populating the time-dependent boundary condition responsible for the data injection.
Finally, the nuclear burning is taken into account by adding the evolved specific binding energy to the specific internal energy calls in our EOS implementation.
As for the other thermodynamical quantities, the binding energy is also smoothly added to the specific internal energy $\epsilon$ during the transition regime.
This is achieved by adding the binding energy only to the Helmholtz EOS' specific energy calls,
\begin{equation}
	\bar{\epsilon}_{\rm Helm} = \epsilon_{\rm Helm} + \epsilon_{B}
\end{equation}
which is later combined to the NSE value as

\begin{equation}
	\epsilon = (1-w) \times \bar{\epsilon}_{\rm Helm} + w \times \epsilon_{\rm NSE}
\end{equation}

In this scheme no source terms are required in the energy evolution equation.
In recent works, new terms arise in the conservation of energy equation \cite{Just:2025hyy}, but these terms are associated with neutrinos, which are not accounted for in our simulations.

The right-hand side of Eq.~\eqref{eq:BindEnergyEvol} is in principle composed of a sum of cooling/heating terms corresponding to the individual implemented reactions.
We follow the prescription of \cite{Magistrelli:2025xja,Wu:2021ibi}, where $\dot{\epsilon}$ for a given fluid element is fitted as a time-dependent piece-wise function in the Lagrangian frame.
This fit takes into account the energy released by nuclear reactions but ignores the effect of neutrino cooling.
The parameters of this fit are obtained as functions of three micro-physical parameters at the time of the mass ejection: $Y_{e}$, $s$, and the expansion time-scale $\tau$, implying $\dot{\epsilon}=\dot{\epsilon}(t | Y_{e,0},s_0,\tau)$. To access these values, we inject 6 extra passive scalars ($Y_{e,0},s_{0},\tau^{-1}_{0},\bar{A}_{0},F_{0},\epsilon_{B}$)  to be purely advected on top of the injected primitive variables.
The injection flag $F$ is useful for diagnostics and for a smooth activation of HR.
The HR are activated only if $0.9<F<1.1$. In our post-processing ejecta analysis we usually constrain ourselves to $|F-1|<0.01$.
The less stringent constraint in the activation of the HR is chosen to minimize the effects of steep gradients on the evolution.
Meanwhile, the more stringent constraint on the post-processing is required to avoid contamination from unphysical atmospheric fluid elements.
The quantity $\tau$ is the expansion time-scale. Its inverse is advected instead, which allows for a cleaner flooring policy corresponding to $\tau=\infty$ in the atmosphere.
The expansion time-scale is first computed as
\begin{equation}
	\tau =  \dfrac{e}{\pi} \dfrac{R_{0}}{v^{r}_{0}}
\end{equation}
where $R_{0}$ is the injection radius, and $v^{r}_{0}$ is the radial velocity at injection time.

While the Helmholtz EOS is described in \cite{Timmes:2000a}, the primitive solver for the nuclear EOS was ported from \gra \cite{Cook:2023bag}, and it is based on \cite{Kastaun:2020uxr}.
The evolution equations implemented in \athena use different conservative variables than \gra (cf. \cite{Stone:2020} with \cite{Cook:2023bag} for details); denoting $D_{\rm GR},S_{i,{\rm GR}},E_{{\rm GR}}$ as the conserved (Eulerian) density, momentum and energy density used in the \gra code, the translation between them and those employed in \athena's primitive solver is given by:
\begin{equation}
  \frac{D_{\rm GR}}{\gamma} = \alpha D,
  \qquad \frac{S_{i,\rm GR}}{\gamma} = \alpha S_{i},
\end{equation}
and
\begin{equation}\label{eq:energytranslation}
  \frac{E_{\rm GR}}{\gamma} = \alpha^{2} g^{0i} S_{i} - E - \alpha D,
\end{equation}
where $\gamma$ is the determinant of the 3-metric and $\alpha$ the lapse function.
Note that $E = T^{0}_{0}$ is conserved for stationary space-times, while $E_{\rm GR} = T^{\mu \nu} n_{\mu}n_{\nu}$ in Eq.~\eqref{eq:energytranslation} is the energy associated with the Eulerian observers commonly used in NR.

\subsection{Tracers and Nucleosynthesis}
\label{sec:method:nucleo}

After our relativistic hydrodynamic evolution simulations reach 1 second, we stop the evolution and use the 3D \athena data to construct tracers.
We post-process our data by recovering fluid-element trajectories with a backward time integration method.
Having 3D data uniformly sampled with $\delta t \approx 20 M_{\odot}$, we choose a linear integrator with a maximum time step of $200M_{\odot}$ to allow a faster integration at later times, when homologous expansion is expected. All tracers are integrated from  $\lesssim$1 s until the injection radius ($R_{\rm inj.}=300M_{\odot}$) is reached, or the condition $|F-1|<0.01$ is no longer satisfied, where $F$ is defined in Eq.~\eqref{eq:Flag}.

We spread the tracers over the original \athena simulation domain using $N_{\theta}=24$ and $N_{\phi}=48$ points, while the radial coordinates are selected on an iso-volumetric grid of 50 points.
All positions are later randomized within the size of the grid cell where the tracers were initially seeded in order to minimize possible grid effects on our analysis.
Of the 57600 initially seeded tracers, at least 39000 satisfy the injection flag condition for each simulation and are considered representative for our nucleosynthesis calculations.
We note that the number of tracers was selected to resolve the total ejecta mass in our simulations. In Tab.~\ref{tab:EjectaMasses}, one can find the ejecta mass recovered by our tracers.

The nucleosynthesis data is obtained by means of the open-source\footnote{\url{https://github.com/nuc-astro/WinNet}} \winnet code \cite{Reichert:2023xqy}.
This is a single-zone nuclear reaction network. As such, each fluid element is treated as moving only radially, excluding mixing between neighboring cells.
We employ a network of 6757 isotopes up to $^{337}$Rg together with the REACLIB reaction rate database \cite{Cyburt:2010a}, supplemented by tabulated theoretical $\beta^{\pm}$-decay and $e^{\pm}$-capture rates \cite{Fuller:1985,Oda:1994,Langanke:2001,PruetFuller:2003,Suzuki:2016} and by fission reactions, for which we include $\beta$-delayed \cite{Mumpower:2022}, neutron-induced \cite{Panov:2010} and spontaneous \cite{Khuyagbaatar:2020} fission rates, all evaluated for the fission barriers of \cite{Moller:2015}, together with the fission fragment distribution of \cite{Panov:2001}.
We further include parameterised $\alpha$-decays \cite{DongRen:2005}, high-temperature partition functions \cite{Rauscher:2003}, Coulomb screening in the prescription of \cite{Kravchuk:2014}, and the tabulated electron-positron chemical potential of \cite{TimmesArnett:1999}.
In our set-up, we assume no neutrino absorption (in agreement with the free-streaming assumption), and a free density expansion with an associated adiabatic evolution of the temperature, as proposed in \cite{Korobkin:2012uy}.

\subsection{\knec{} Light Curves}
\label{sec:method:LC}

After one second of evolution, our 3D profiles are mapped to radial profiles by separating individual angular sections of given fixed pair ($\theta,\phi$) that match our \athena grid.
This procedure obtains 512 individual radially-organized sections containing the necessary thermodynamic quantities to feed the ray-by-ray Lagrangian radiation-hydrodynamics code \knec{} \cite{Wu:2021ibi,Magistrelli:2024zmk,Magistrelli:2025xja,Morozova:2015bla}.
Each of the sections is treated as an individual ray in \knec.
\knec\ independently evolves each angular section by including the power released by nuclear reactions and using a selected thermalization scheme and opacity model.
We evolve the fluid with the Helmholtz EOS \cite{Timmes:2000a} at high temperatures and switch to the Paczynski EOS \cite{Paczynski:1986px} at lower temperatures as described in \cite{Magistrelli:2024zmk}.

Consistently with the relativistic hydrodynamics evolution of the ejecta in \athena{}, we calculate the nuclear power from the analytical fits of \cite{Magistrelli:2025xja,Wu:2021ibi}, based on a grid of tabulated thermodynamic trajectories analyzed with the nuclear network (NN) \skynet \cite{Lippuner:2017tyn}.
The NN includes 7836 isotopes up to $^{337}$Cn and uses the JINA REACLIB \cite{Cyburt:2010a} and the
same setup as in \cite{Lippuner:2015gwa, Perego:2020evn}.
The power emitted in electrons, positrons and $\alpha$ particles is thermalized assuming an analytical prescription for the released energy and thermalization factor $f_\text{th}$ as in \cite{Kasen:2018drm}.
A constant $f_\text{th} = 0.5$ is assumed for the remaining nuclear power, similar to the prescription of \cite{Magistrelli:2025xja,Wu:2021ibi}.
As in \cite{Magistrelli:2025xja, Wu:2021ibi}, we assume a constant opacity for the ejecta only set by their initial electron fraction.
See \cite{Magistrelli:2024zmk, Magistrelli:2025xja} for the effects of including an in-situ NN and employing more detailed thermalization and opacity models.

Each of the profiles is filtered in order to exclude fluid elements that have negative velocity, that have $|F-1|>0.1$, or that are bound according to the Bernoulli criterion.
We use a resolution of 100 points in the radial direction.
All the radial sections are evolved individually for 30 days.
As a final post-processing stage, all 1D profiles are combined with angle-of-view dependent geometric weights $p_{k}$ \cite{Martin:2015hxa} to calculate the light curves.
These weights represent the projection of the solid angle subtended by each profile onto the line of sight of a distant observer.
The flux factors $p_{k}$ are defined as
\begin{equation}
	p_{k} = \iint_{\bf{ \hat{n}_{k}} \cdot \bf{\hat{q}}} \bf{\hat{q}} \cdot d\bf{\vec{\Omega}},
\end{equation}
where the $k$ index refers to the different bins over a unit sphere centered at the origin.
Each of the bins is constructed to match the angular resolution of the \athena grid.
The vectors $\bf{\hat{n}_{k}}$ and $\bf{\hat{q}}$ are the normal to the bin's surface and the unit vector pointing to the observer, respectively, and $d\bf{\vec{\Omega}}$ is the usual vector solid-angle element.

\section{Results}
\label{sec:res}

\begin{table*}[t]
\centering
	\caption{In this table we show the ejecta and fallback mass for each one of our models. We also show the amount of $^{56}$Ni produced in each of them.
The ejecta masses are obtained by integrating the unbound mass (by the Bernoulli criterion) over our grid after 1~second of evolution. The $^{56}$Ni masses are obtained in post-processing via the \winnet code (see \cite{Reichert:2023xqy} for more details).
The values of ejecta mass in parentheses refer to the ejecta mass recovered by our tracers.
The last 3 columns are the 2D equivalent quantities obtained in \cite{Jacobi:2025eak}.
While our current models systematically recover lower ejecta and $^{56}$Ni masses, their ratio remains in the same window of a few percent.
The only exception to this is the SFHo\_q1.0 model.}
\label{tab:EjectaMasses}
\begin{tabular}{|c | c| c| c| c| c| c|c|}
\hline\hline
Model
& $M_{\mathrm{ej}}[M_{\odot}]$
& $M_{\mathrm{fall}}[M_{\odot}]$
& $M_{\mathrm{ej}}^{^{56} \mathrm{Ni}}[M_{\odot}]$
& $M_{\mathrm{ej}}^{^{56} \mathrm{Ni}}/M_{\mathrm{ej}}[\%]$
& $M_{\mathrm{ej, 2D}}[M_{\odot}]$
& $M_{\mathrm{ej,2D}}^{^{56}\mathrm{Ni}}[M_{\odot}]$
& $M_{\mathrm{ej,2D}}^{^{56} \mathrm{Ni}}/M_{\mathrm{ej, 2D}}[\%]$   \\
\hline\hline
	BLh\_q1.43 & 8.08(7.91)$\times 10^{-3}$ & 1.57$\times 10^{-3}$ & $ 4.34\times 10^{-4}$ & 5.5 & 1.08$\times 10^{-2}$ & 8.4$\times 10^{-4}$ & 7.8\\
\hline
	SFHo\_q1.0 & 4.47(4.20)$\times 10^{-3}$ & 1.15$\times 10^{-3}$ & $ 1.03\times 10^{-6}$ & 0.03 & 6.25$\times 10^{-3}$ & 1.0$\times 10^{-5}$ & 0.2\\
\hline
	DD2\_q1.0  & 4.54(4.34)$\times 10^{-3}$ & 7.73$\times 10^{-4}$ & $ 1.59\times 10^{-4}$ & 3.7 & 5.80$\times 10^{-3}$ & 4.1$\times 10^{-4}$ & 7.1\\
\hline
	DD2\_q1.67 & 1.37(1.35)$\times 10^{-2}$ & 4.02$\times 10^{-3}$ & $ 3.08\times 10^{-4}$ & 2.3 & 2.53$\times 10^{-2}$ & 4.3$\times 10^{-4}$ & 1.7\\ [1ex]
\hline\hline
\end{tabular}
\end{table*}

\subsection{Ejecta dynamics}
\label{sec:res:dyn}

\begin{figure}[t]
    \centering
    \includegraphics[width=\linewidth]{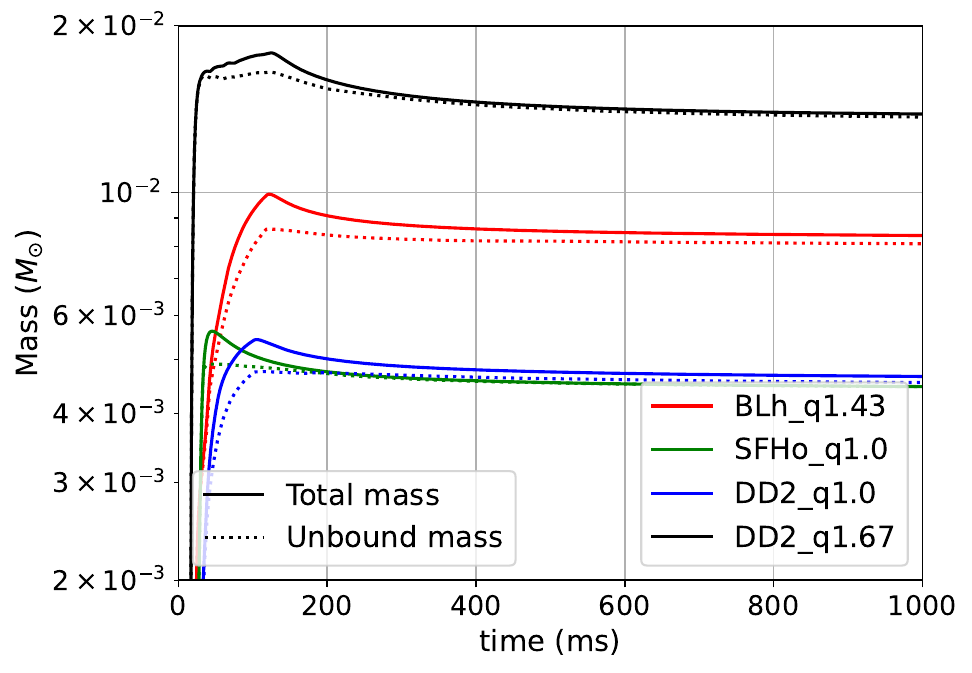}
	\caption{Total and unbound masses in our simulations as a function of time.
	We found a considerable amount of fallback material in the DD2\_q1.67 case ($\sim{4}\times 10^{-3}M_{\odot}$).
	The fallback material is the reason for the decreasing mass at $t\sim100$~ms.
	While this may be unphysical for most of our models due to their unfinished ejection process within the duration of the \thc simulations, the $\sim{1} \times 10^{-3} M_{\odot}$ fall back in the SFHo\_q1.0 model is understood as physical, as its remnant is short-lived and collapses into a BH at $t\sim{8}$~ms after merger, so that no significant disk and no later ejecta are expected.}
    \label{fig:EjectaMassTime}
\end{figure}

All our models are evolved beyond ${\sim}1$~s post-merger, a time-scale 10 times larger than the duration of the original \thc simulations.
The ejecta baryon mass for each model is reported in Tab.~\ref{tab:EjectaMasses} and its evolution is shown in Fig.~\ref{fig:EjectaMassTime}.
The figure reports both the total mass and the mass flagged as unbound during the simulation according to the Bernoulli criterion.
We find DD2\_q1.67 has the most massive ejecta of order ${\sim1.4}\times10^{-2}\Msun$, while SFHo\_q1.0 has the smallest mass ${\sim}4.5\times10^{-3}\Msun$. 
There is a clear hierarchy in our models: stiffer EOS and unequal-mass mergers produce long-lived remnants and generate more massive ejecta, while softer EOS and equal-mass binaries produce short-lived remnants and less massive ejecta.
For $q=1$, the ejecta masses obtained with the SFHo (soft) and DD2 (stiff) equations of state are $4.47\times10^{-3}\,M_{\odot}$ and $4.54\times10^{-3}\,M_{\odot}$, respectively, showing no significant difference between the two models.
While softer equations of state typically produce larger ejecta masses for equal-mass binaries, the longer-lived remnant in the DD2 case enhances the mass ejection, resulting in an ejecta mass comparable to that of the SFHo model.
Note that the mass ejection from long-lived remnants would still continue past the NR simulated time~\cite{Radice:2023zlw,Bernuzzi:2024mfx}.

We also observe material fallback, especially at early times, see Tab.~\ref{tab:EjectaMasses}.
DD2\_q1.67 has about ${\sim 4} \times 10^{-3}M_{\odot}$ fallback material, which may be in part overestimated, since the ejecta would be supported by material pushed out by the remnant at times beyond the duration of the \thc simulations.
SFHo\_q1.0 has ${\sim} 1 \times 10^{-3} M_{\odot}$ of fallback material, which is likely a good estimate at these time-scales since the remnant has collapsed and the remnant disk has baryon mass ${\sim}0.011\Mo$ \cite{Gutierrez:2024pch}.
Given that the disk is understood to be able to eject up to $\sim30\%$ of its mass as viscous ejecta, the expected viscous ejecta mass is of the same order as the fallback material found in our analysis.
Consequently, the fallback material may suppress the viscous ejection, or at least significantly reduce its velocity.

\begin{figure*}[t]
    \centering
    \includegraphics[width=0.49\linewidth]{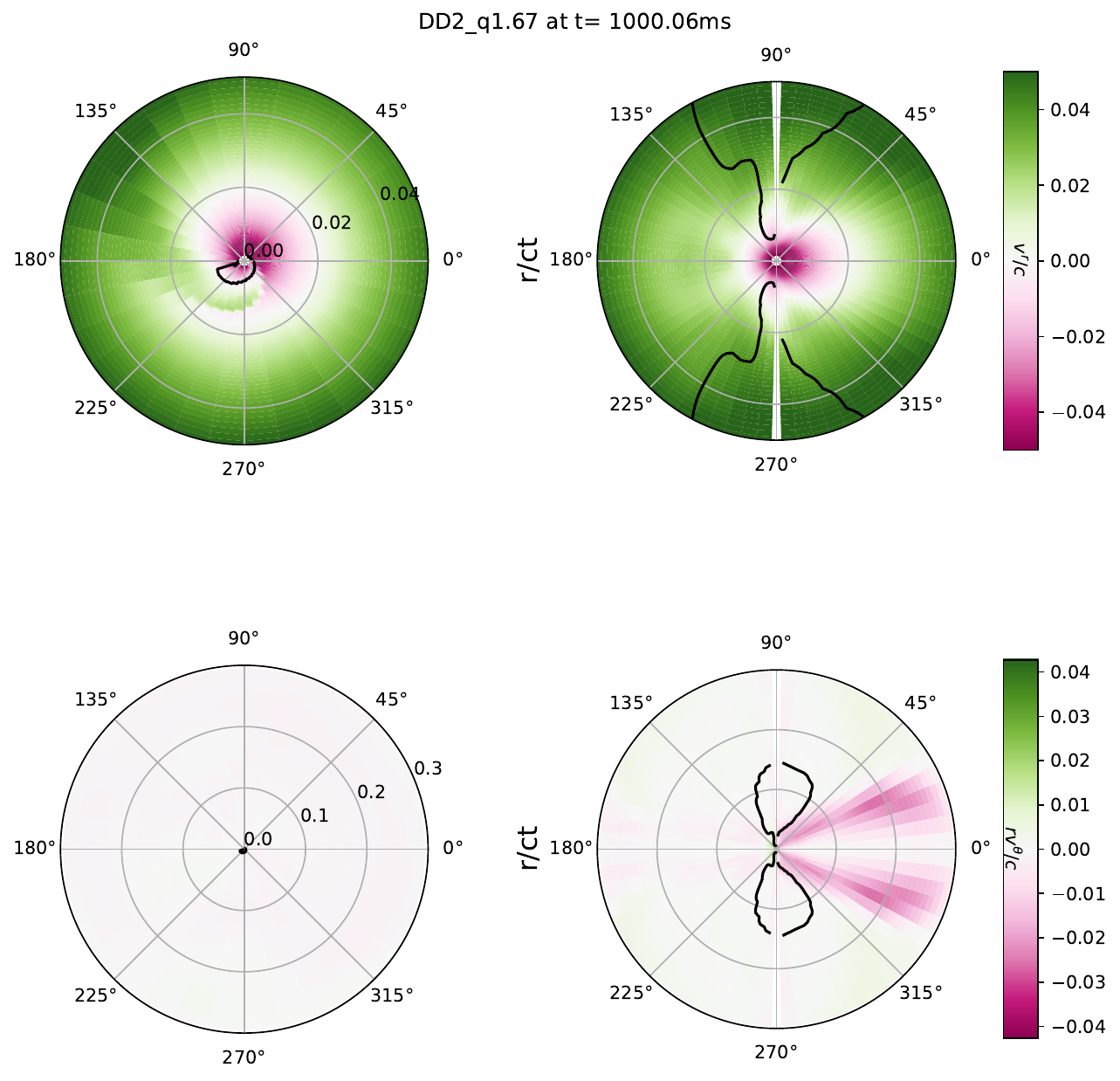}
    \includegraphics[width=0.49\linewidth]{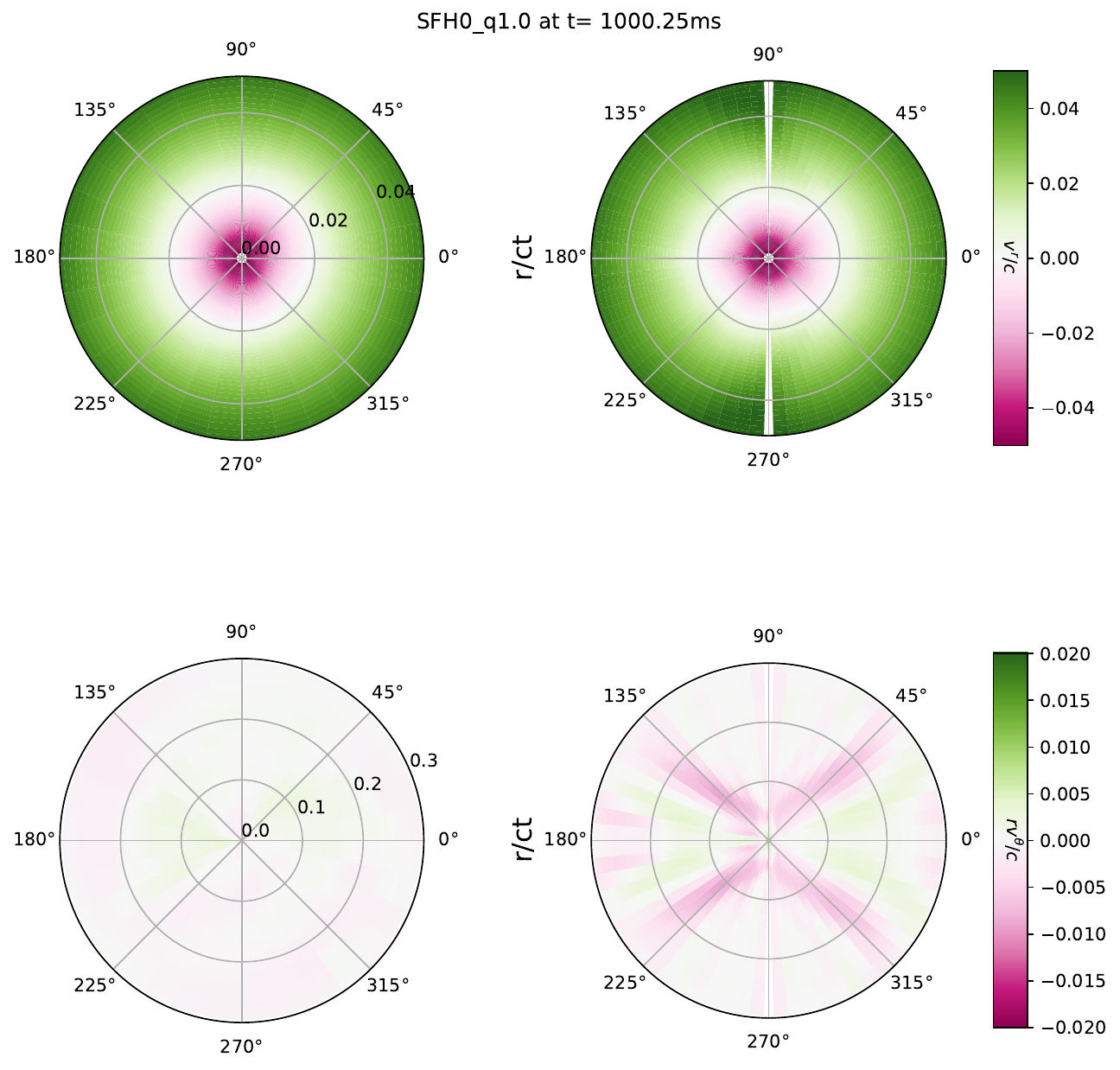}
	\caption{Components of the three-velocity of the ejecta for the DD2\_q1.67 and SFHo\_q1.0 model. While the left columns of each panel represent the equatorial plane, the right columns show the perpendicular plane at $\phi=0^{\circ}\cup 180^{\circ}$. In our definition of velocity, the fluid is moving towards the poles when $v^{\theta} <0$. Therefore it is clear from this snapshot that the non-radial motion present is mainly due to an expansion of the tidal arm. The black line shows the contour of our injection flag, separating the ejecta (outer) material from the (inner) spurious material injected after the NR data end. The hour-glass shaped surface is due to the neutrino-driven wind pushing the ejecta away from the poles. We can also see that $v^{r}$ goes to zero close to the equatorial plane, implying that the fallback occurs in this region.}
	\label{fig:Vfdd2a}
\end{figure*}

During the evolution, the ejecta typically attains higher velocities above the poles of the remnant.
This behavior is driven by the late-time neutrino-driven wind, which is strongest in the polar regions.
As a result, the polar ejecta is accelerated to larger radii than its equatorial counterpart.
Consequently, the surface of last injection (the boundary between $F=1$ and $F=2$, see Equation~\ref{eq:Flag}) expands more rapidly along the polar direction.
Figure~\ref{fig:Vfdd2a} shows the different components of the 3-velocity in the orbital and polar planes for the DD2\_q1.67 and SFHo\_q1.0 binaries.
The faster polar expansion is clearly visible in the DD2\_q1.67 model through the hourglass-shaped contour in the right panel.
In contrast, the SFHo\_q1.0 model does not develop such a morphology because its neutrino-driven wind is weaker and shorter-lived, owing to black hole formation at $t\sim8\,\mathrm{ms}$.
The hourglass morphology coincides with the region where $v^{r}\rightarrow0$, effectively confining fallback to the equatorial plane.
By contrast, in the SFHo\_q1.0 model, $v^{r}\rightarrow0$ isotropically as $r\rightarrow R_{\rm inj.}$, permitting a more symmetric fallback, as shown in the left panel of Fig.~\ref{fig:Vfdd2a}.

\begin{figure}[t]
    \centering
    \includegraphics[width=\linewidth]{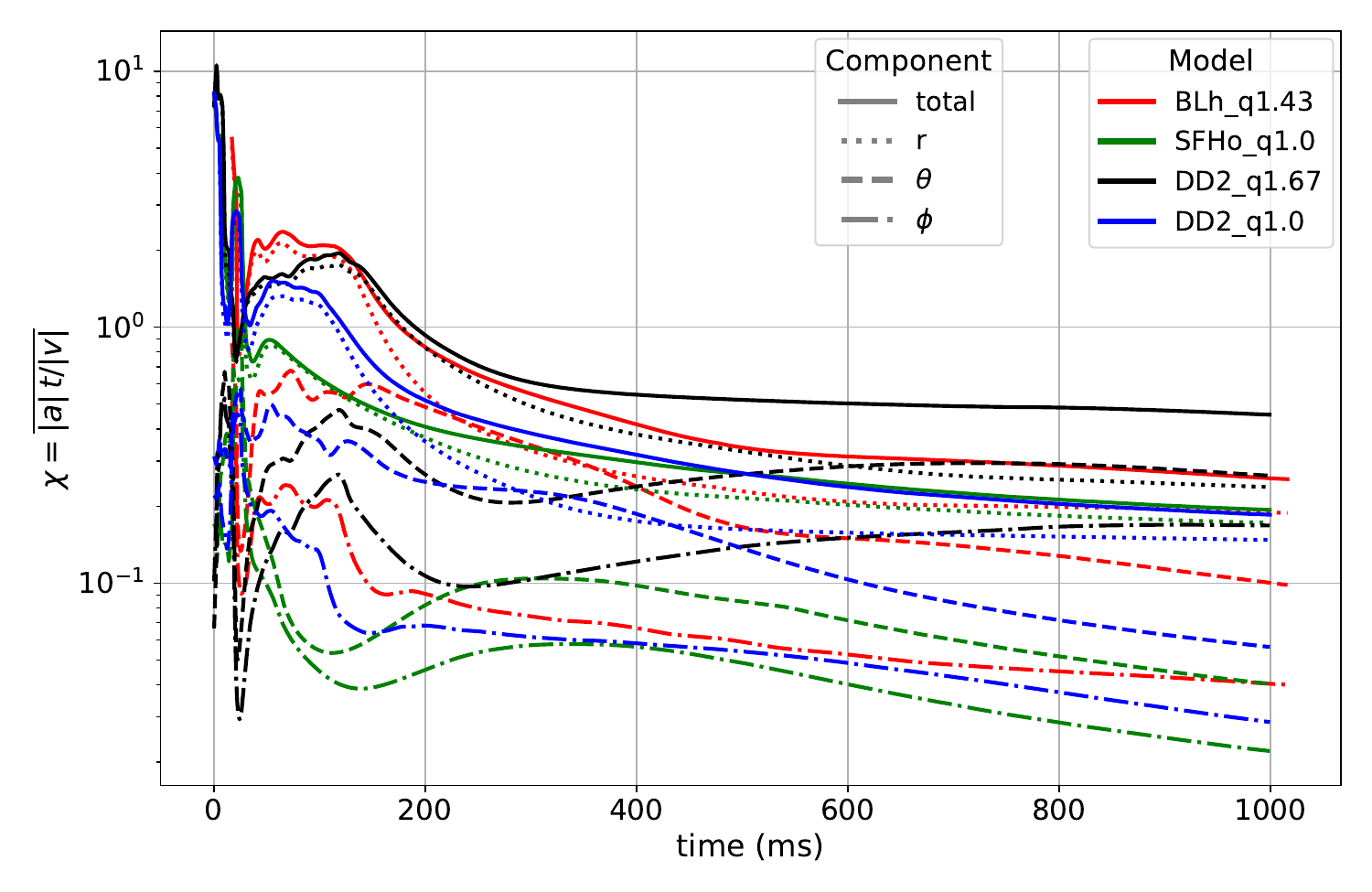}
	\caption{Time evolution of the homology parameter $\chi$ for all four merger models. Line color indicates the model and linestyle indicates the $\chi$ component. Note that for DD2\_q1.67, the non-radial contributions to $\chi$ remain comparable to the radial one even at $t\sim1$~s. This underlines the fact that the homologous expansion approximation may not yet hold at the second time-scale, especially for highly asymmetric binaries.}
    \label{fig:Homology}
\end{figure}

We investigate the ejecta expansion by means of a Newtonian homology parameter \cite{Rosswog:2013kqa}
\begin{equation}
	\chi = \overline{\left(\dfrac{|a|t}{|v|}\right)},
\end{equation}
where the overbar denotes the mass average, and $a$ and $v$ are the acceleration and the velocity in the Lagrangian reference frame.
As usual for our analysis, we only consider fluid elements where $|F-1.0|<0.01$.
The homology parameter is zero for perfectly ballistically expanding ejecta.
We also separate the contribution of the individual directions to understand the impact of non-radial motion on the ejecta by considering

\begin{equation}
	\chi^{i} = \overline{\left(\dfrac{|a^{i}| t }{|v|}\right)}
\end{equation}
where the accelatation in the Lagrangian grame in each direction is computed as
\begin{equation}
	a^{i}  = \partial_t v^{i} + v^{j} \partial_{j} v^{i} + \Gamma^{i}_{jk}v^{j}v^{k}  ,
\end{equation}
and $\Gamma^{i}_{jk}$ is the usual Christoffel symbol for spherical coordinates.

The evolution of the homology parameter is shown in Fig.~\ref{fig:Homology}.
While individual contributions are shown in discontinous linestyles, solid lines represent the total value of $\chi$.
The homology parameter decreases rapidly during the first 500~ms and reaches values of $\chi\sim0.1$. None of the ejecta considered reaches homologous expansion in the second time-scales, however we do not observe the systematic increase in $\chi$ reported in \cite{Rosswog:2013kqa}.
This difference may be attributed to the different nuclear heating models and the different durations of the merger simulations. Ref.~\cite{Rosswog:2013kqa} deals only with the first $\sim$20 ms of ejected material. Ref.~\cite{Neuweiler:2022eum} showed no systematic decrease or increase of $\chi$, but they did not couple nuclear heating to the hydrodynamics and only followed its evolution up to $t\sim0.1$~s.
The slow decay of $\chi$ for DD2\_q1.67 is a direct consequence of the higher heating rates in this particular model.

\begin{figure*}[t]
  \centering
  \includegraphics[width=0.9\linewidth]{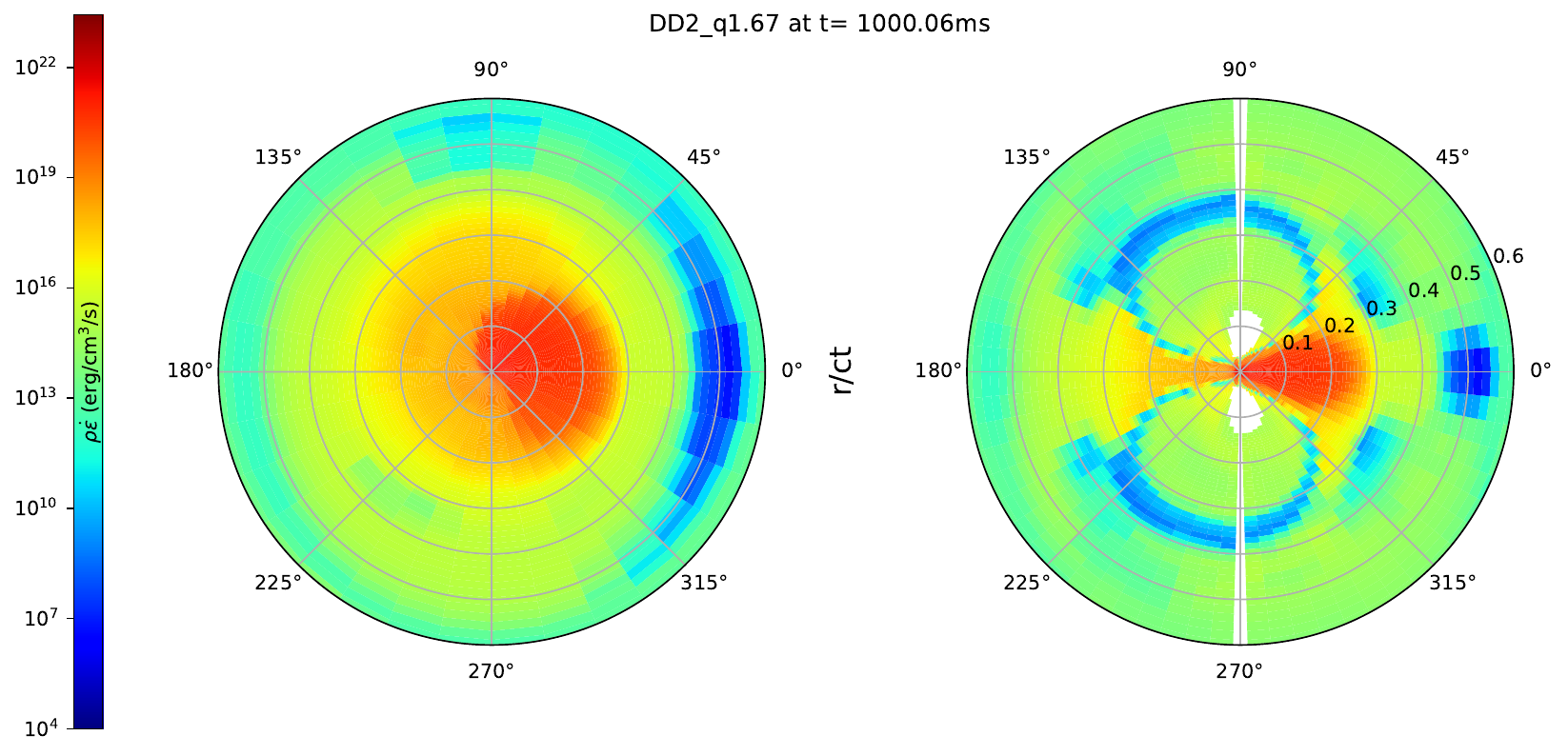}
  \caption{Nuclear heating rates obtained through the fit formulas of \cite{Magistrelli:2025xja, Wu:2021ibi} for DD2\_q1.67 at $t\sim1$s after merger. While the left column represents the equatorial plane, the right column shows the perpendicular plane at $\phi=0^{\circ}\cup180^{\circ}$. We note that at this time DD2\_q1.67 has the largest heating, followed by BLh\_q1.43. This heating is located at the tidal arm and is found to be correlated with the persistence of non-radial motions at late times ($t\sim1$~s) of the hydrodynamical evolution, see Fig.~\ref{fig:Vfdd2a} and Fig.~\ref{fig:Energies}.  }
  \label{fig:Heating}
\end{figure*}

In all the models, the largest contribution to $\chi$ is due to the radial motion, as demonstrated by Fig.~\ref{fig:Homology}.
However, the most asymmetric model DD2\_q1.67 has angular accelerations (specifically $a^{\theta}$) of the same order of magnitude as $a^{r}$, with the non-radial contribution maintained up to the second time-scale. This feature is due to the nuclear heating in the tidal arms.
Figure~\ref{fig:Heating} shows snapshots of the HR $\dot{\epsilon}$ in the orbital and perpendicular ($\phi=0^{\circ}\cup180^{\circ}$) planes.
The tidal arm of DD2\_q1.67 has the highest HR among our models.

\begin{figure}
   \centering
	\includegraphics[width=1.1\linewidth]{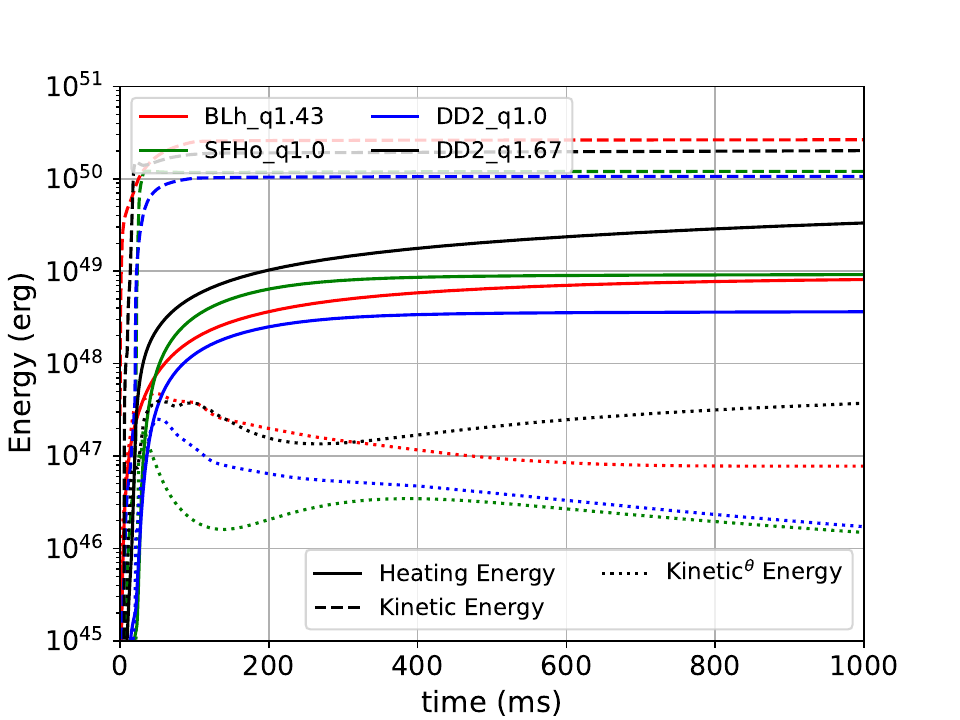}
	\caption{Kinetic and cumulative heating energies as function of time. One can see that DD2\_q1.67 has the largest contribution from heating among all our models, with the heating energy affecting the energy of the motion along the $\theta$ direction. This reinforces the understanding that the heating affects this model the most and complements the results shown in Fig.~\ref{fig:Vfdd2a}, Fig.~\ref{fig:Homology}, and Fig.~\ref{fig:Heating}. }
   \label{fig:Energies}
\end{figure}

We compare three diagnostic energies to further investigate the impact of this region of high heating. The total heating energy and total relativistic kinetic energy are defined as
\begin{equation}
  E_{\rm Heat} = \int \dot{\epsilon} \rho dV dt,
  \qquad
  E_{\rm Kinetic} = \int (W-1) \rho dV,
\end{equation}
where $W$ is the Lorentz factor and the integrals are performed only over volume elements satisfying our ejecta flag condition. Given the non-radial motion found at 1~s for DD2\_q1.67, we are particularly interested in knowing if this motion is due to the heating of the tidal arm. Therefore we define, in a Newtonian approximation, the energy associated with the $\theta$ motion as:
\begin{equation}
  E^{\theta}_{\rm Kinetic} = \int \dfrac{1}{2} (rv^{\theta})^{2} \rho dV,
\end{equation}
where $v^{\theta}$ is the velocity along the $\theta$ direction, and the integral is yet again restricted by our ejecta flag. Although this last definition is constructed from a Newtonian approximation, for most of our data we have $r \gg 2M$ and $W\lesssim 1.1$, justifying our approximation.

Figure \ref{fig:Energies} shows the results of our analysis.
The total kinetic energy of all four models plateaus shortly after the injection of \thc data is over ($t\sim150$~ms).
Meanwhile, the cumulative heating energy of all models but DD2\_q1.67 stops growing past $5-9\times10^{48}$~ergs.
Consequently, only DD2\_q1.67 shows a growing kinetic energy associated with the $\theta$ motion past $t\sim 400$~ms.
The growth of $E^{\theta}_{\rm Kinetic}$ for DD2\_q1.67 is guided by nuclear heating from $t\sim 250$~ms ($\sim 150$~ms after the end of injection) until the end of our simulations.
We understand that the motion in the $\theta$ direction is more affected than in $\phi$ due to the steeper gradient of $\dot{\epsilon}$ along the polar coordinate when compared to the azimuthal one. This result illustrates the impact of nuclear heating rates on the hydrodynamics of BNS ejecta.

\begin{figure*}[t]
    \centering
    \includegraphics[width=0.49\linewidth]{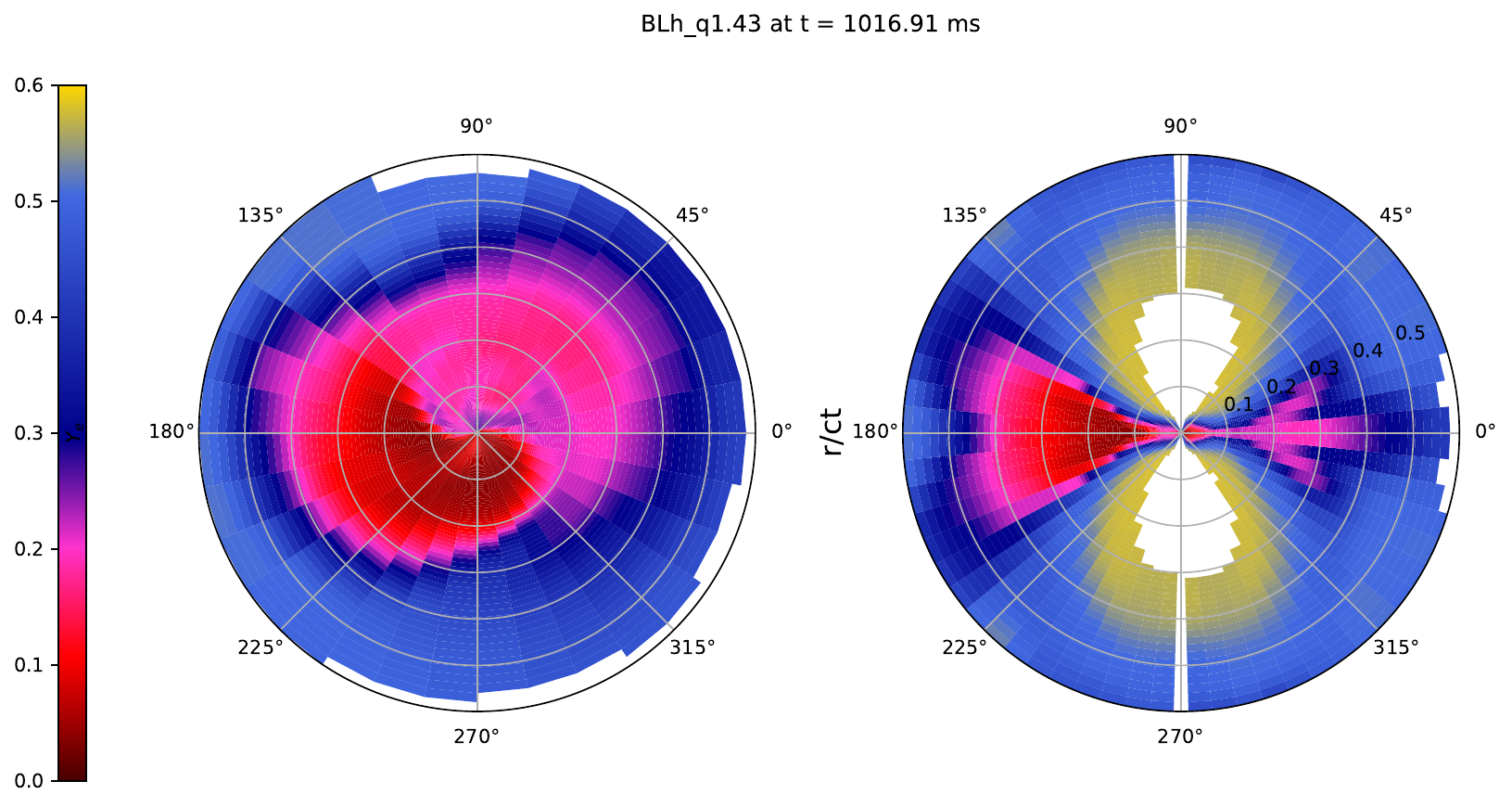}
    \includegraphics[width=0.49\linewidth]{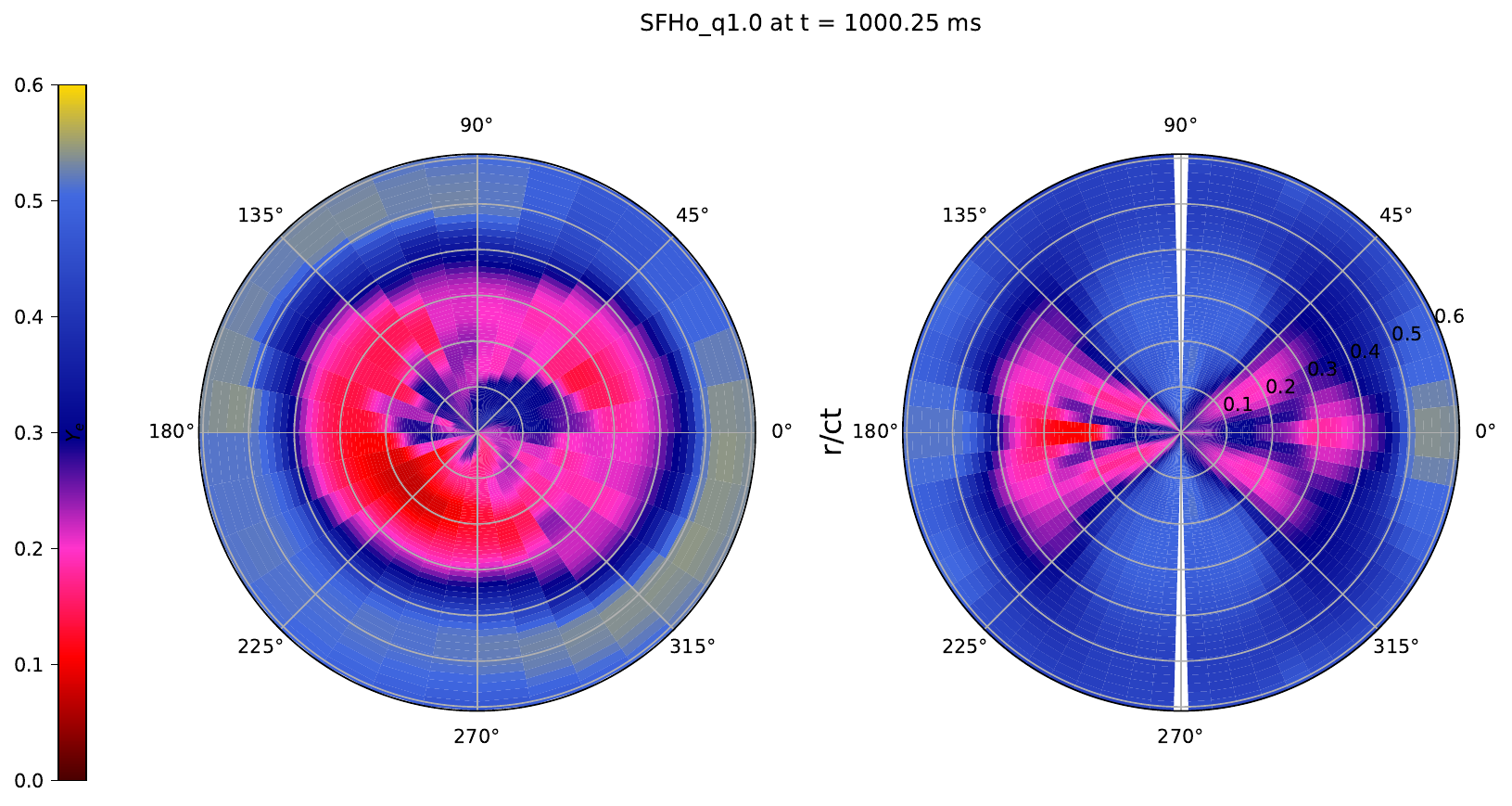}
    \includegraphics[width=0.49\linewidth]{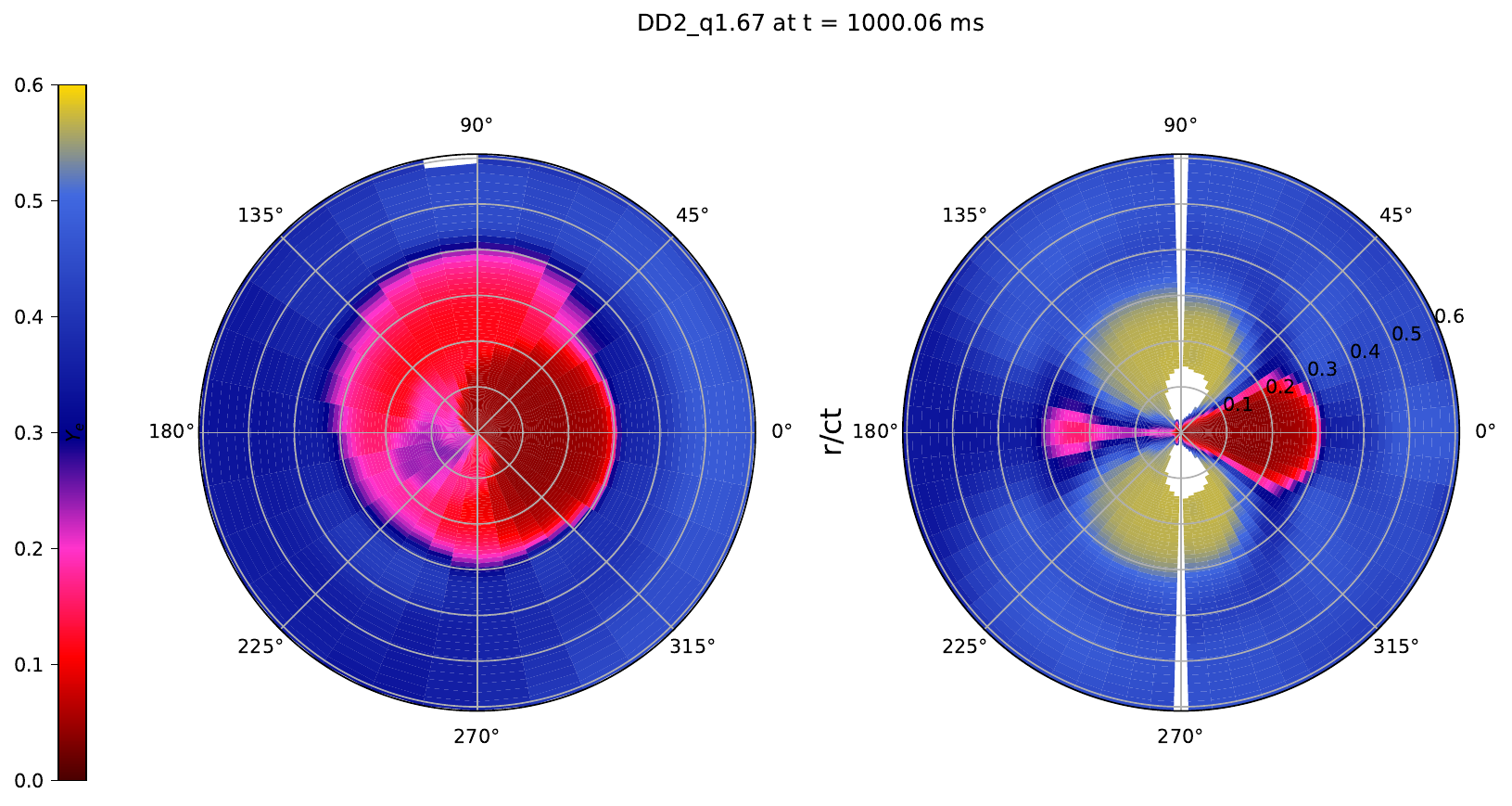}
    \includegraphics[width=0.49\linewidth]{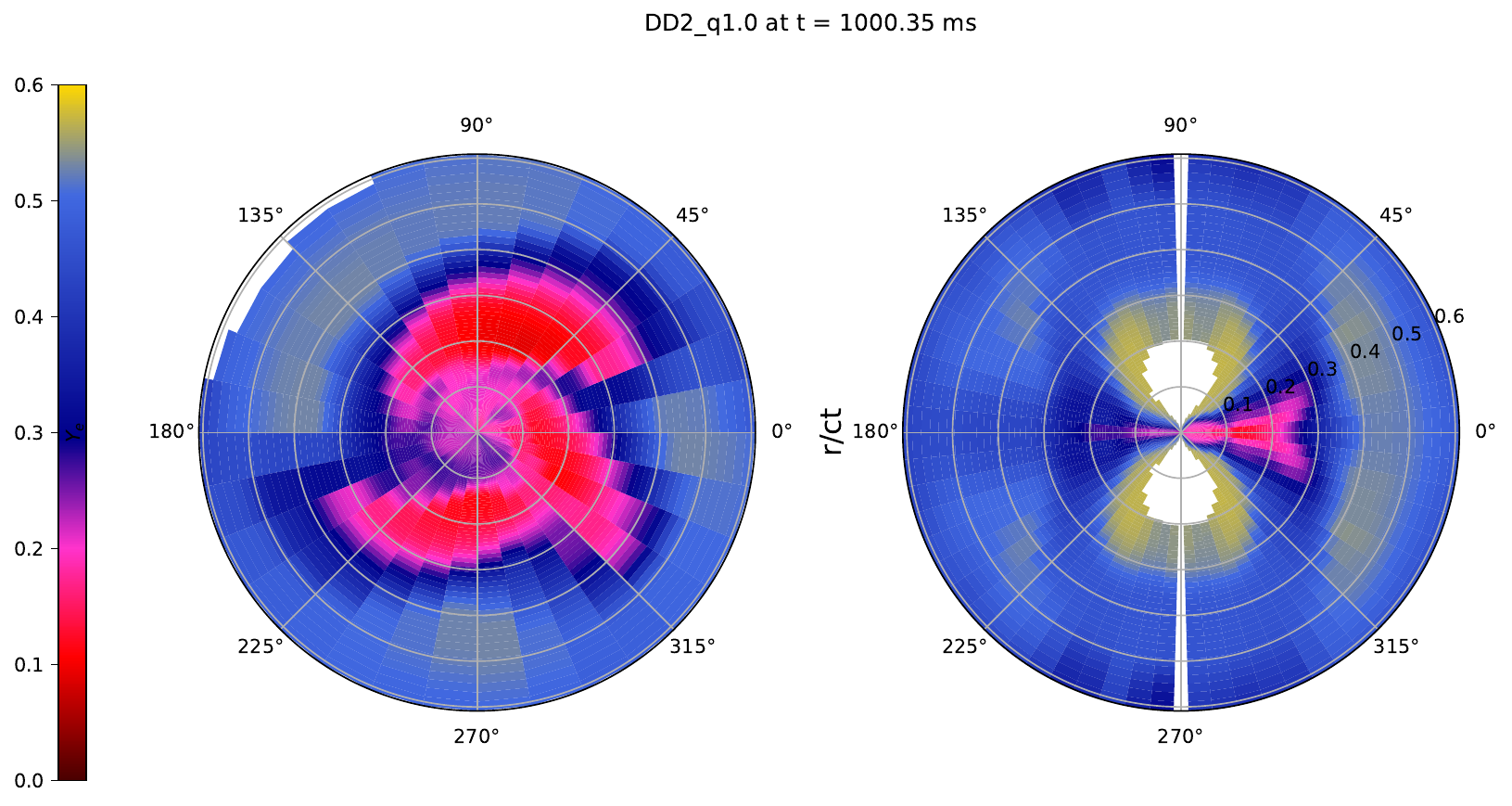}
	\caption{ $Y_{e}$ profile of our models close to $t\sim1$s. While the left inset of each panel represents the equatorial plane, the right ones show the perpendicular plane at $\phi=0^{\circ}\cup180^{\circ}$. While the DD2\_q1.0 (SFHo\_q1.0) displays a symmetric double-arm (disk-like) structure, DD2\_q1.67 (BLh\_q1.43) shows a one-arm structure with lower electron fraction at $\sim 270^{\circ}$ ($0^{\circ}$). We also notice that SFHo\_q1.0 has the lowest electron fraction at the poles among our models due to the lack of a neutrino-driven wind from the BH remnant.}
    \label{fig:Ye}
\end{figure*}

In our set-up, $Y_{e}$ is initially computed with a full M1 neutrino transport scheme within \thc \cite{Radice:2012cu,Radice:2013hxh,Radice:2015nva}.
From the injection time in \athena, it is treated as a passive scalar and simply advected by the fluid's flow.
The resulting electron fraction profiles at $t\sim1$~s are shown in Fig.~\ref{fig:Ye}.
Although the neutrinos can be assumed to be free-streaming for $r\gtrsim 300M_{\odot}$, processes such as $\beta$-decay and $e^{\mp}$ capture can drive the $Y_{e}$ to $\sim0.4$ in the second time-scale.
For this reason our final electron-fraction profiles, seen in Fig.~\ref{fig:Ye}, still display apparently very neutron-rich material in the tidal arms at the 1~s time-scale, which is not representative of the true composition at this time.
With these caveats in mind, we show the advected electron fraction as a way to track the evolution of the tidal arm.
In a later section we correlate our nucleosynthesis results (Sec.~\ref{sec:res:elem}) with the kinematics of our models.
While SFHo\_q1.0 and DD2\_q1.0 are mainly symmetric around the axis of rotation, their tidal arms are contained at $\theta\lesssim45^{\circ}$ and $\theta\lesssim30^{\circ}$ respectively.
With a more asymmetric profile, BLh\_q1.43 (DD2\_q1.67) has the bulk of its (originally) neutron-rich material at $135^{\circ} \lesssim \phi \lesssim 315^{\circ} ( -90^{\circ}\lesssim \phi \lesssim 90^{\circ})$ with an opening angle of $\theta\lesssim 40^{\circ}$.

\begin{figure*}
        \centering
        \includegraphics[width=0.8\linewidth]{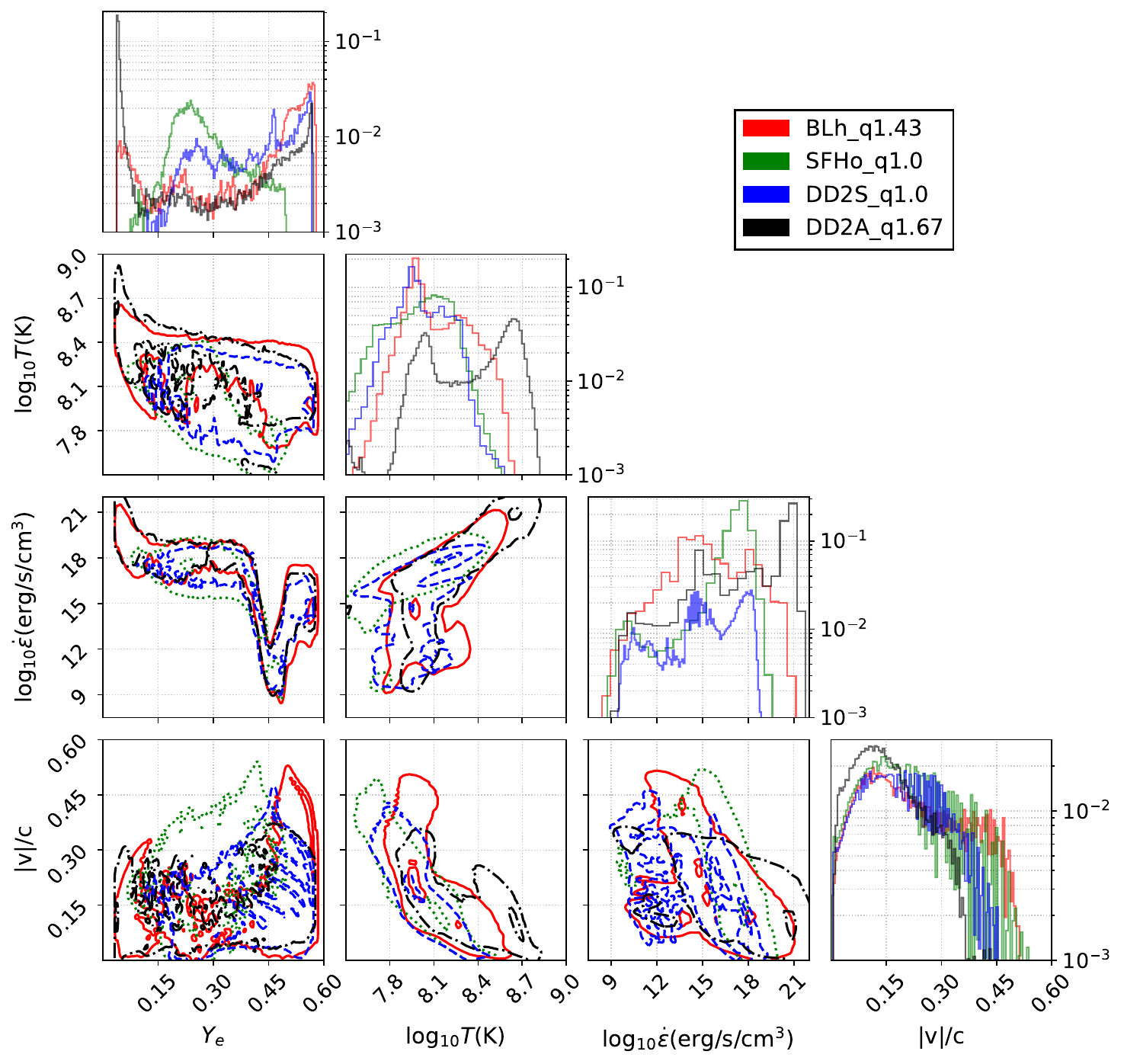}
        \caption{In this corner plot we show the distribution and correlation of several quantities of interest for our models by ${\sim} 1$~s. Here we see that DD2\_q1.67 retains the highest amount of heating among our models, correlating with low-velocity, low-$Y_{e}$, and high temperature fluid elements. }
	\label{fig:Corner}
\end{figure*}

In Fig.~\ref{fig:Corner} we compare the different thermodynamical profiles of the four binaries after ${\sim}1$~s of evolution in the \athena code.
The corner plot highlights the correlations among different quantities.
Material with $Y_{e}\lesssim 0.15$ displays higher temperatures for BLh\_q1.43 and DD2\_q1.67 models, a direct imprint of the higher heating rates found in these fluid elements.
This reinforces our understanding that the tidal arm of DD2\_q1.67 is still expanding by the end of our simulations due to thermal heating.

This figure also shows us that the symmetric binary models (SFHo\_q1.0 and DD2\_q1.0) have the least amount of heating, which we attribute to their $Y_e$ distribution peaking at $\sim0.2$, noticeably higher than the peak found for the asymmetric models, which display a peak at $Y_e\sim 0.1$.
While a $Y_e\sim0.2$ peak is still low enough to enable $r$-process nucleosynthesis, it favors the production of lighter, less neutron-rich species compared to the more neutron-rich conditions of the asymmetric models.
This results in a less efficient $r$-process and correspondingly lower radioactive heating, consistent with the lower temperatures found for symmetric mass-ratio models.

\subsection{Element Production and Distribution}
\label{sec:res:elem}

\begin{figure*}[t]
  \includegraphics[width=1.0\linewidth]{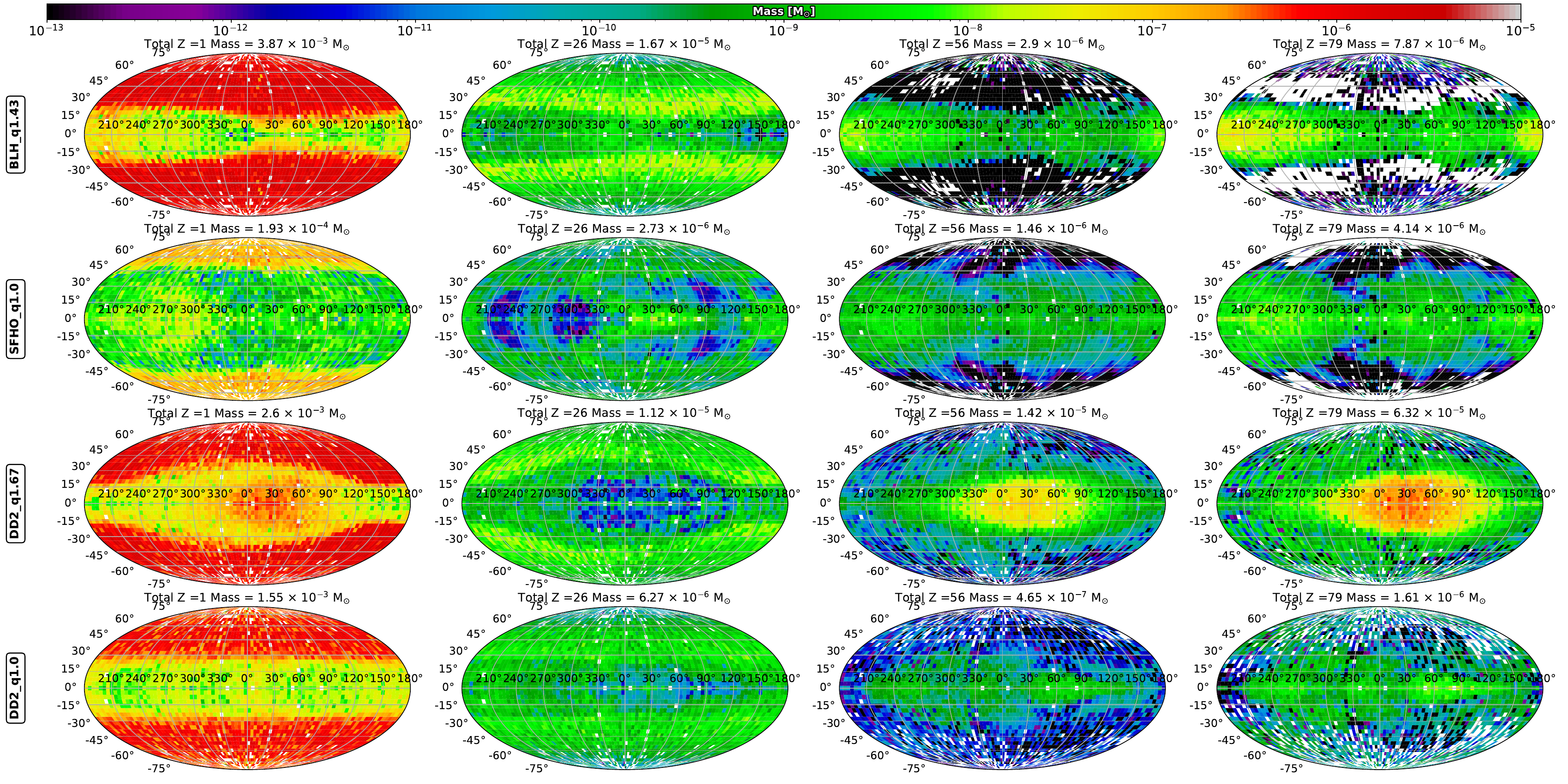}
  \caption{Sky location of elements as a function of atomic number for our models. We note the large asymmetry in the DD2\_q1.67 model.
	   The presence of the lanthanide curtain only for $-80^{\circ} \lesssim \phi \lesssim 120^{\circ}$ leads us to anticipate a bluer and shorter light curve for observers diametrically opposed to this region.
	   This effect is also anticipated for BLh\_q1.43 but to a lesser extent as the asymmetry is less prominent.
	   }
  \label{fig:SkyMaps}
\end{figure*}

\begin{figure*}
    \centering
    \includegraphics[width=.49\linewidth]{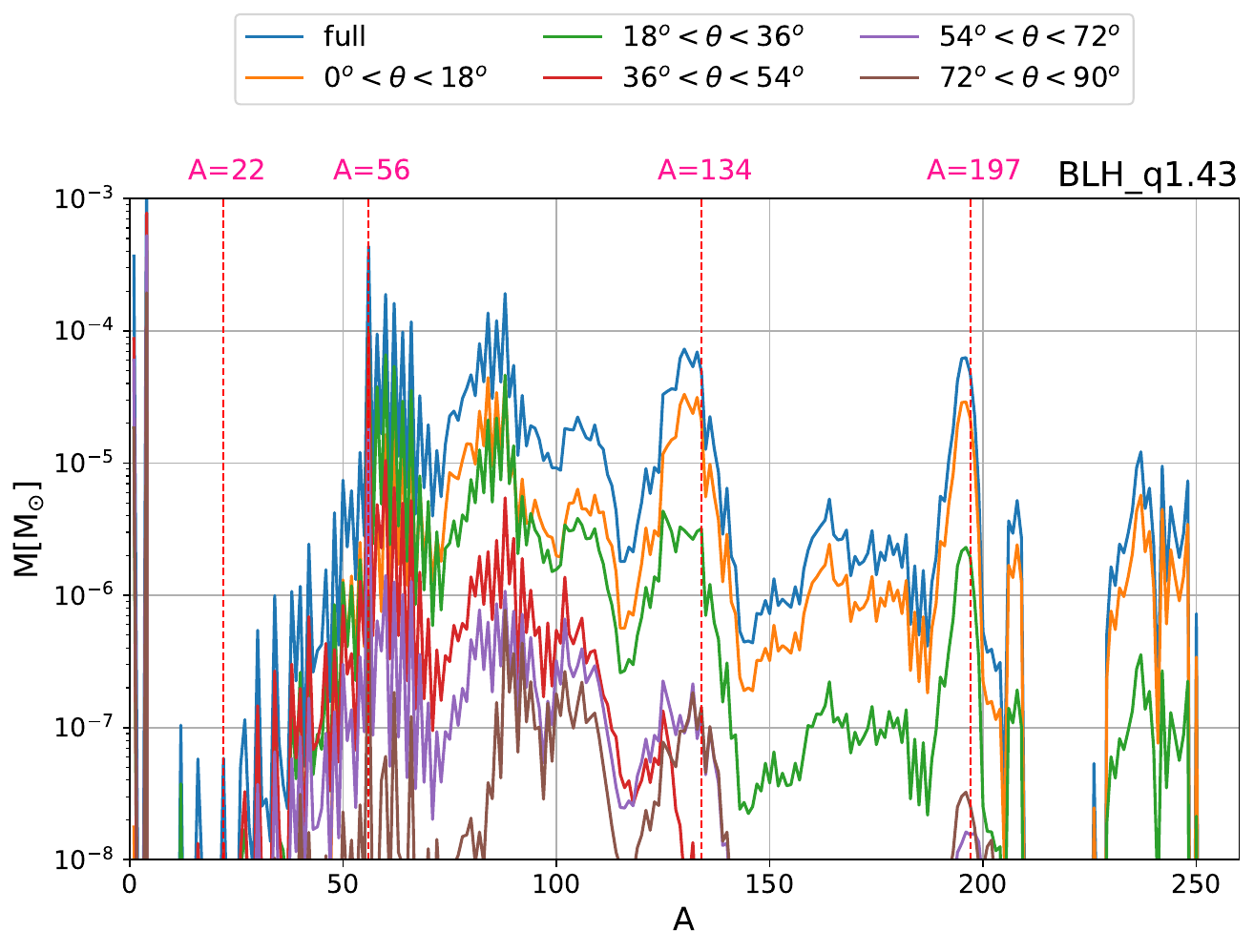}
    \includegraphics[width=.49\linewidth]{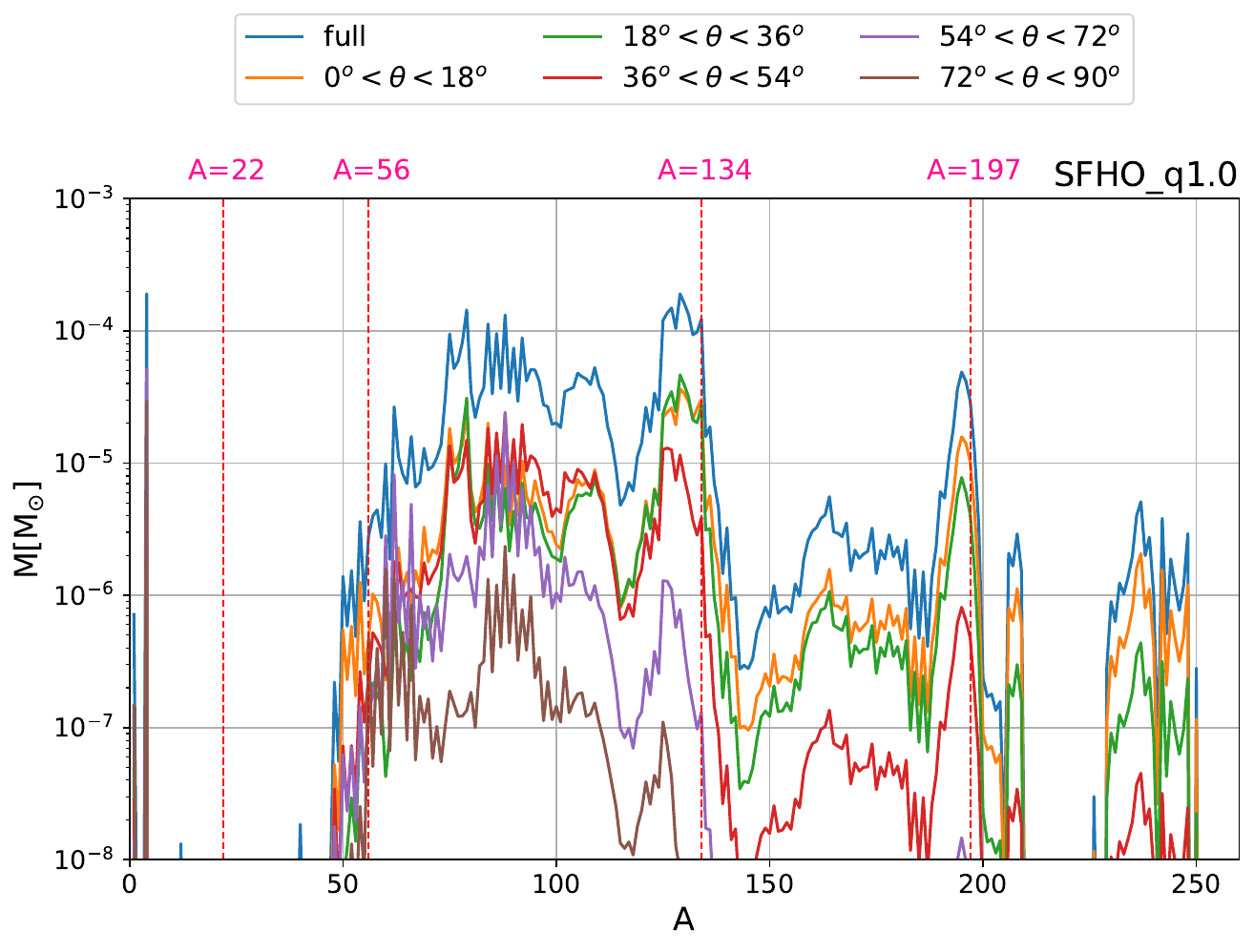}
    \includegraphics[width=.49\linewidth]{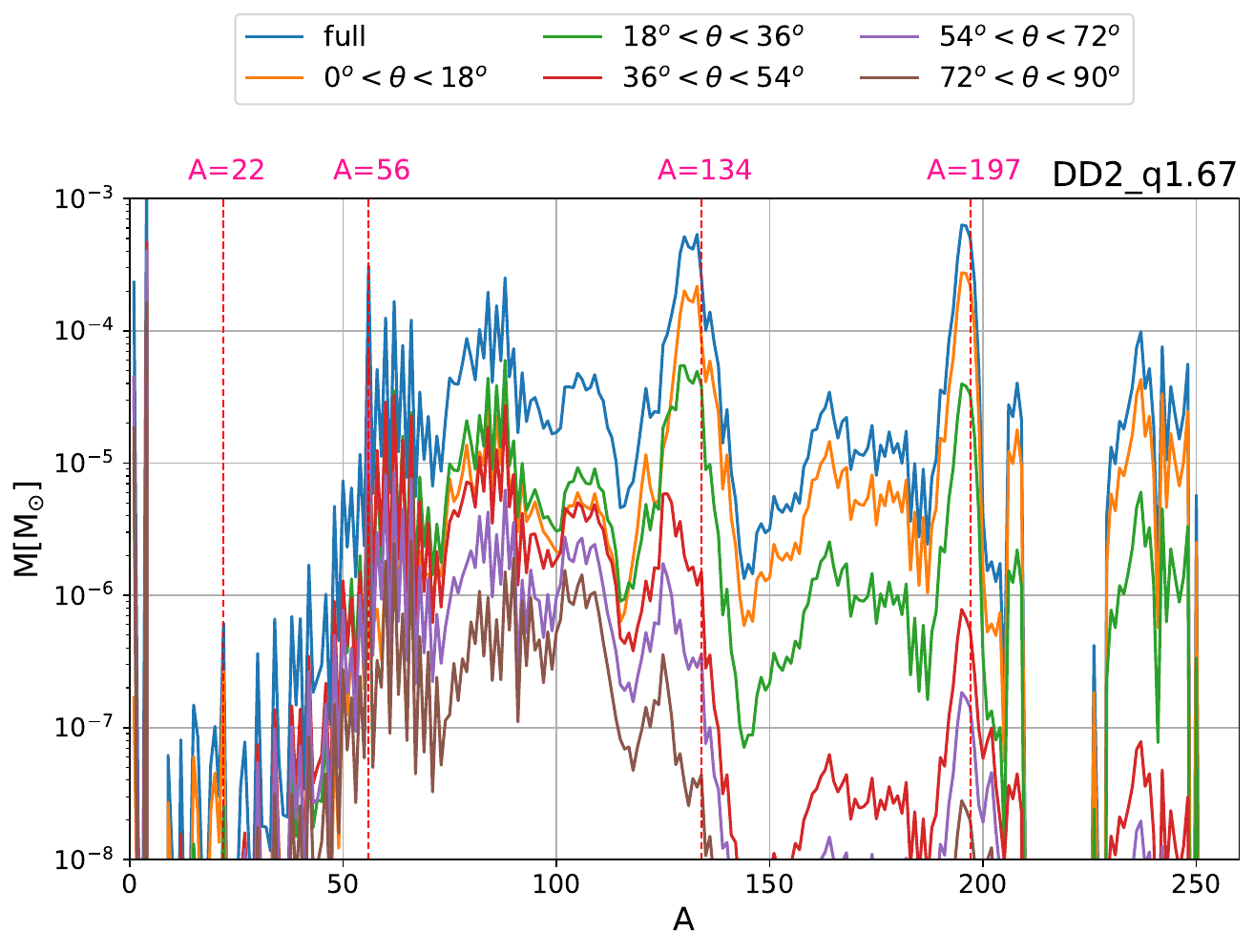}
    \includegraphics[width=.49\linewidth]{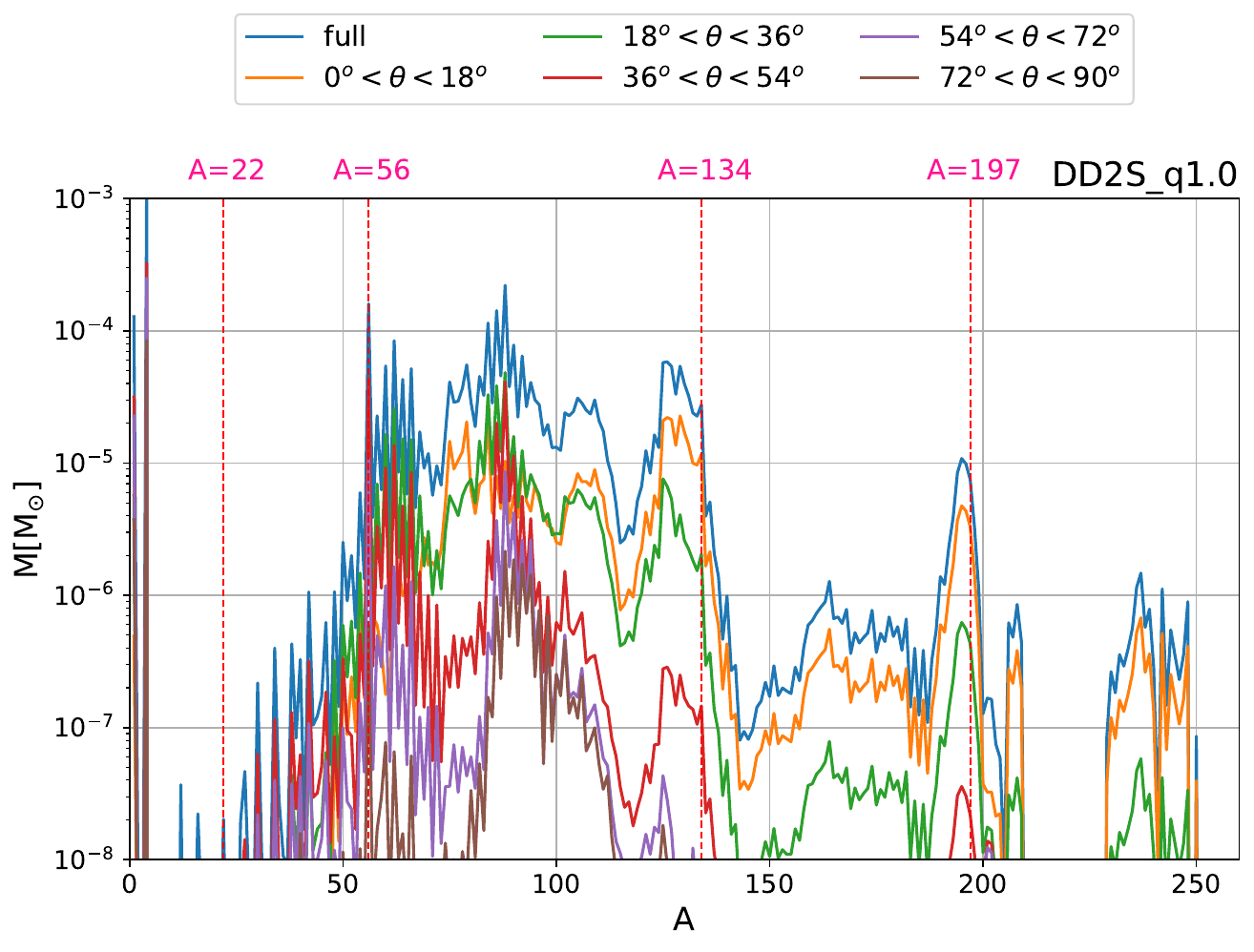}
    \caption{Element masses as a function of sections of varying $\theta$ ranges for our models. As expected, all models exhibit a larger amount of heavy $r$-process elements in the equatorial plane. For all models, the bulk of heavy $r$-process elements are found at $\theta<36^{\circ}$.}
    \label{fig:AbsAngAlltheta}
\end{figure*}

Tracers evolved with \winnet are sorted according to their final position and a sky-map is constructed with their composition.
Figure~\ref{fig:SkyMaps} shows a selection of sky-maps for our models, where we select $^{1}$H,  $^{56}$Fe,  $^{137}$Ba, and  $^{197}$Au as representative nuclei.
Regarding the distribution of $^{1}$H, all models present higher masses at the polar regions, with a larger production for asymmetric binaries.
These binaries also display a broader spread of $^{1}$H, as BLh\_q1.43 and DD2\_q1.67 display high concentrations of hydrogen for $\theta$ as low as $15^{\circ}$.
The equal-mass binaries DD2\_q1.0 (SFHo\_q1.0) display a reduced region of high $^{1}$H concentration, $\theta\gtrsim30^{\circ}$ ($50^{\circ}$).

Following the production of $Z=56$ and $Z=79$ in Figure~\ref{fig:SkyMaps}, we see a very strong dependence on the mass ratio of the original binary.
While the $q=1.0$ models show a lanthanide curtain that weakly depends on $\phi$, BLh\_q1.43 (DD2\_q1.67) shows a depletion of such elements on the $-30^{\circ}\lesssim \phi \lesssim 150^{\circ} $ ($120^{\circ}\lesssim \phi \lesssim 300^{\circ} $).
These depletion regions are correlated to the distribution of the tidal ejecta.
More specifically, the parts of the tidal arms with $Y_e \lesssim 0.2$ ($\gtrsim 0.2$) correlate with the regions of higher (lower) lanthanide production.
While SFHo\_q1.0 and DD2\_q1.0 have low-$Y_{e}$ tidal ejecta (roughly) symmetrically distributed on the equatorial plane, $q>1.0$ binaries show a one-sided tidal arm structure.
In the extremal case of DD2\_q1.67, we find a lanthanide bullet instead of a curtain, where by bullet we mean a very collimated region with high lanthanide production diametrically opposed to a region depleted of such elements.

As seen in Fig.~\ref{fig:SkyMaps}, $^{56}$Fe, as a decay product of $^{56}$Ni, is expected to form in $Y_{e}\sim 0.5$ regions normally associated with the neutrino-driven wind.
This element tends to form away from the poles for BLh\_q1.43 and DD2\_q1.0, where the $\theta>60^{\circ}$ polar caps contain systematically less mass than the $30^{\circ}\lesssim \theta \lesssim 60^{\circ}$ region.
We understand this as a reflection of the high entropy content of the poles, which suppresses Ni production.
This effect is weaker in DD2\_q1.67, because tidal-arm heating has a stronger influence on the non-radial motion of the ejecta, pushing fluid elements towards the poles, see Fig.~\ref{fig:Vfdd2a} and Fig.~\ref{fig:Energies}.
SFHo\_q1.0 shows negligible iron compared to the other models, due to its weaker neutrino-driven wind.

The final abundances obtained with \winnet for each tracer are combined to compose the mass distribution according to nuclear mass $A$.
Fig.~\ref{fig:AbsAngAlltheta} shows the final element mass patterns as a function of the polar angle.
As anticipated from the location of the neutron-rich material in Fig.~\ref{fig:Ye}, the bulk of the heavy element production takes place around the equatorial plane, $\theta\lesssim 36^{\circ}$ for all models.
For example, in DD2\_q1.67 the $A=196$ and $A=134$ elements reach masses of $\sim 3\times10^{-4}M_{\odot}$ in equatorial regions ($\theta \lesssim 36^{\circ}$), see Fig.~\ref{fig:AbsAngAlltheta}, meanwhile their masses stay below $ \sim10^{-6} M_{\odot}$ for $\theta\gtrsim 36^{\circ}$.
Meanwhile all models display negligible masses of these elements above $\theta \sim 54^{\circ}$.
Regarding the production of $A\gtrsim70$ elements, the final total element patterns are more similar within the pairs SFHo\_q1.0/DD2\_q1.0 and BLh\_q1.43/DD2\_q1.67.
While BLh\_q1.43 and DD2\_q1.67 show very similar total masses of second and third $r$-process peaks, the total mass of the third peak for symmetric binaries is $\sim 3$ times smaller than for the second peak.

Below $A\lesssim70$, the element pattern of SFHo\_q1.0 shows a strong attenuation compared to other models, a clear impact of its weaker neutrino wind.
In particular, the total $^{56}$Fe mass produced by this model is of order $\sim3 \times 10^{-6} M_{\odot}$, two orders of magnitude smaller than for all other models.
This is also the model with the lowest overall masses of hydrogen and helium, reaching only $\sim10^{-6}M_{\odot}$ and $\sim10^{-4}M_{\odot}$, respectively.
Meanwhile all the other models have a total mass of hydrogen (helium) of $\gtrsim 10^{-4}(10^{-3}) M_{\odot}$.
These mass values for $^{1}$H and $^{4}$He are again a direct reflection of the amount of free neutrons and $\alpha$ particles made available by the rapid neutrino-driven wind.

\begin{figure*}
    \centering
    \includegraphics[width=.49\linewidth]{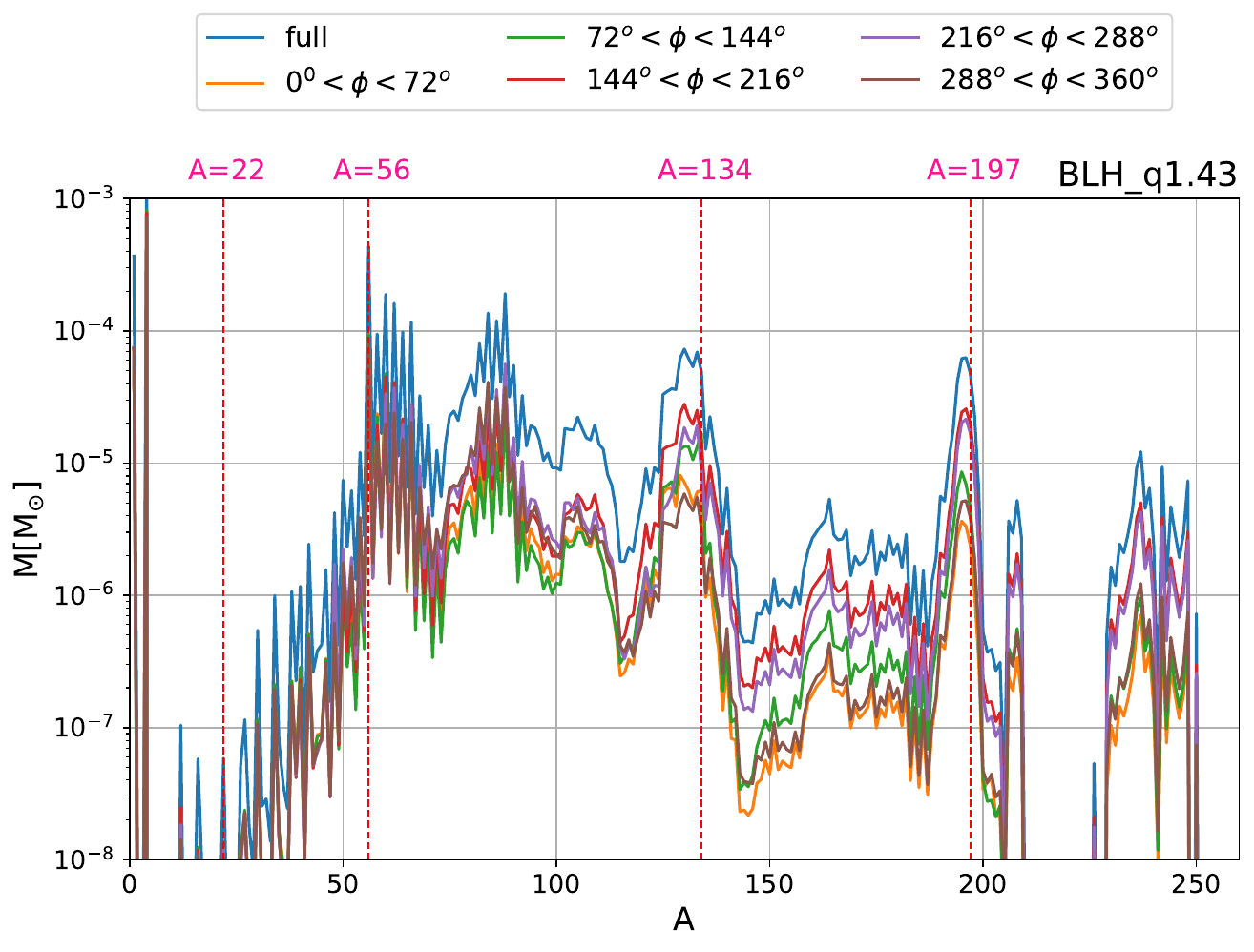}
    \includegraphics[width=.49\linewidth]{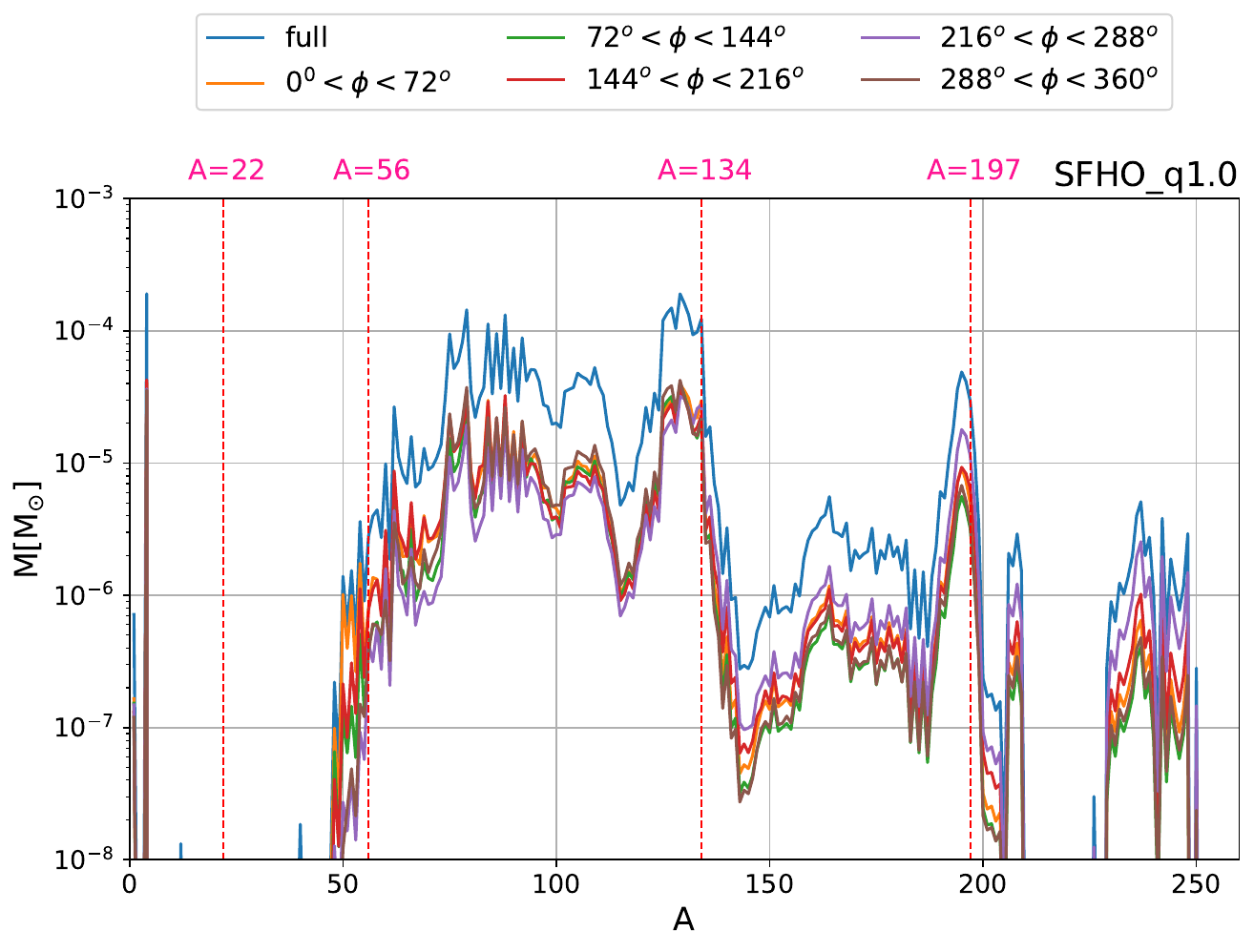}
    \includegraphics[width=.49\linewidth]{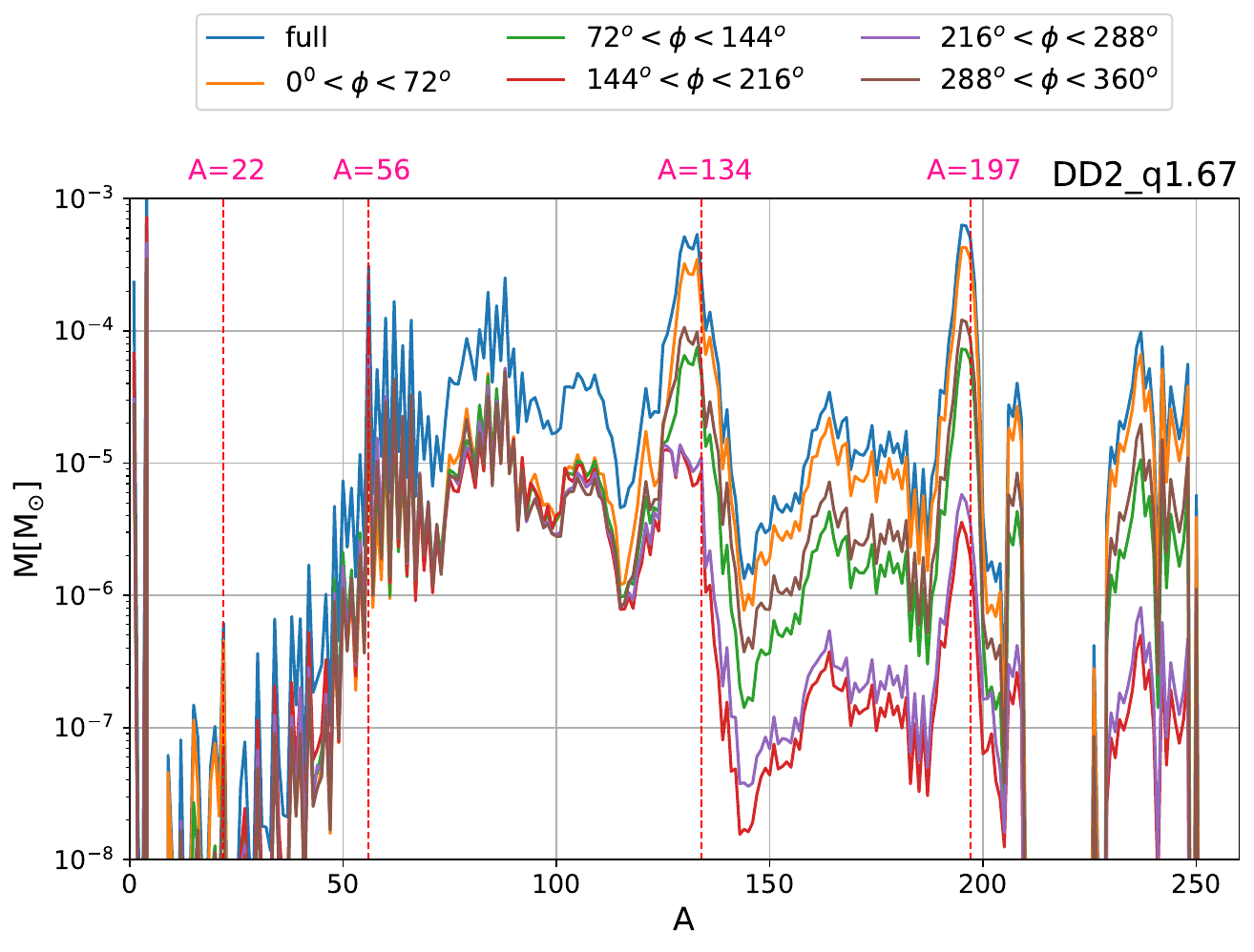}
    \includegraphics[width=.49\linewidth]{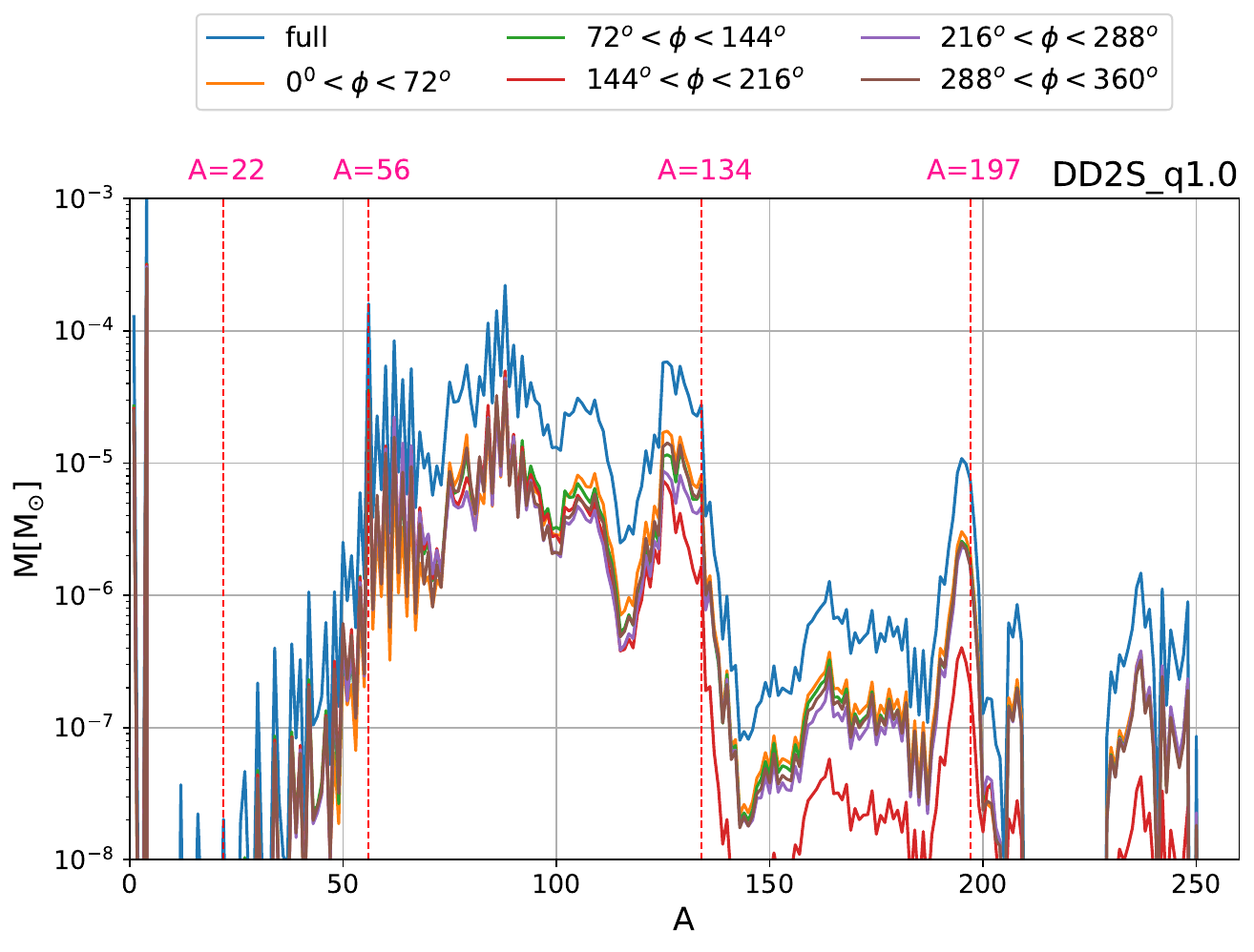}
    \caption{Element masses as a function of sections of varying $\phi$ ranges for the four models. BLh\_q1.43 and DD2\_q1.67 show the highest degree of asymmetry, and the highest amount of $A>100$ for $\phi\sim250^{\circ}$ and $\phi\sim40^{\circ}$, respectively. The element yields for SFHo\_q1.0 (DD2\_q1.0) reflect their symmetric mass ratio, but one can still see a slightly higher production of heavy $r$-process elements for $\phi\sim250^{\circ}(\sim40^{\circ})$. These locations agree with the lower electron fraction part of the tidal arm of each model, see Fig.~\ref{fig:Ye}. }
    \label{fig:AbsAngAllphi}
\end{figure*}

In Fig.~\ref{fig:AbsAngAllphi} we also divide the tracers according to their final $\phi$ position and investigate the asymmetry with respect to the rotation axis. Models of symmetric mass ratio present nearly indistinguishable profiles as a function of $\phi$. The strongest influence of the azimuthal angles on the abundance patterns of SFHo\_q1.0 and DD2\_q1.0 is seen on elements heavier than $A=134$, never surpassing a factor $\sim$2 in difference. BLh\_q1.43 and DD2\_q1.67 also vary the most for $A>134$ across different azimuthal regions. Contrary to the symmetric models, we find a variation by a factor of $\sim{5-10}$ in heavy element abundances when comparing diametrically opposed profiles in the DD2\_q1.67 and BLh\_q1.43 models.

\begin{figure}[t]
    \centering
    \includegraphics[width=\linewidth]{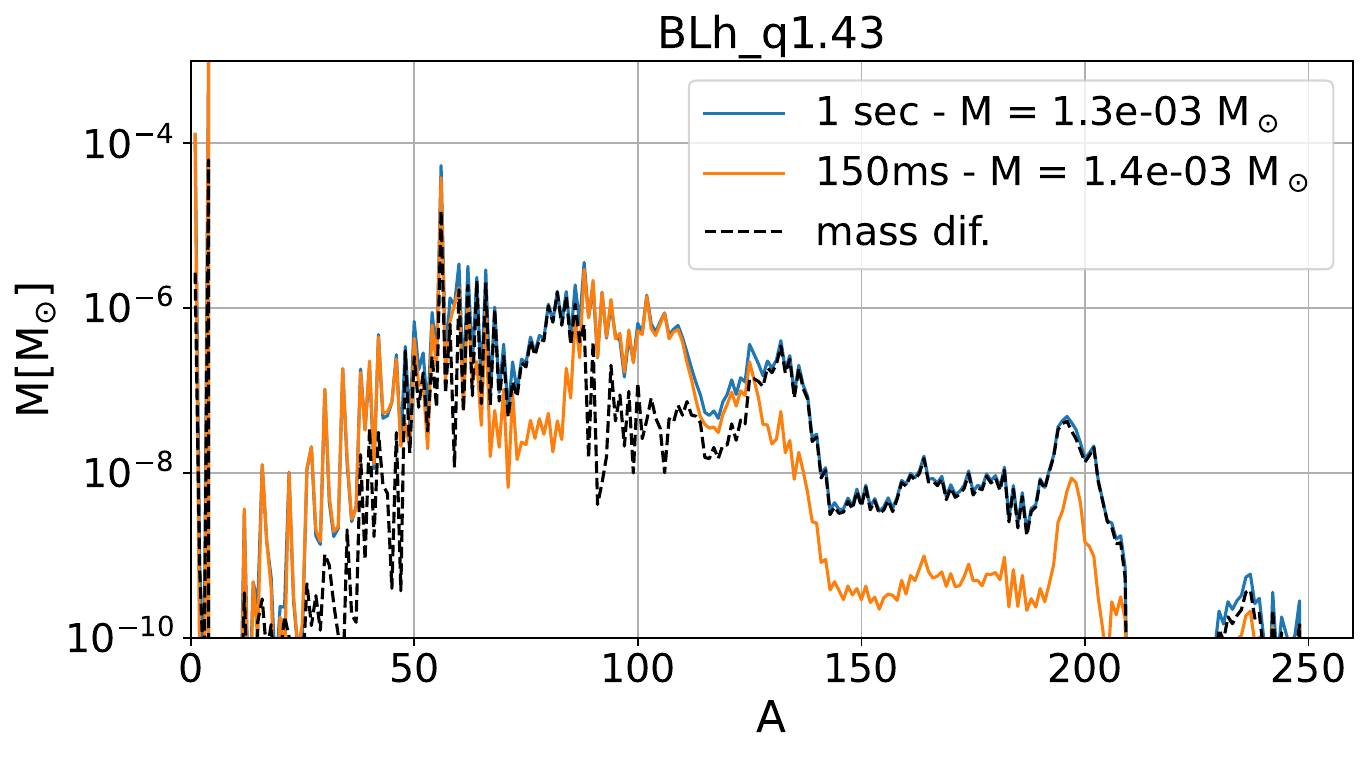}
    \includegraphics[width=\linewidth]{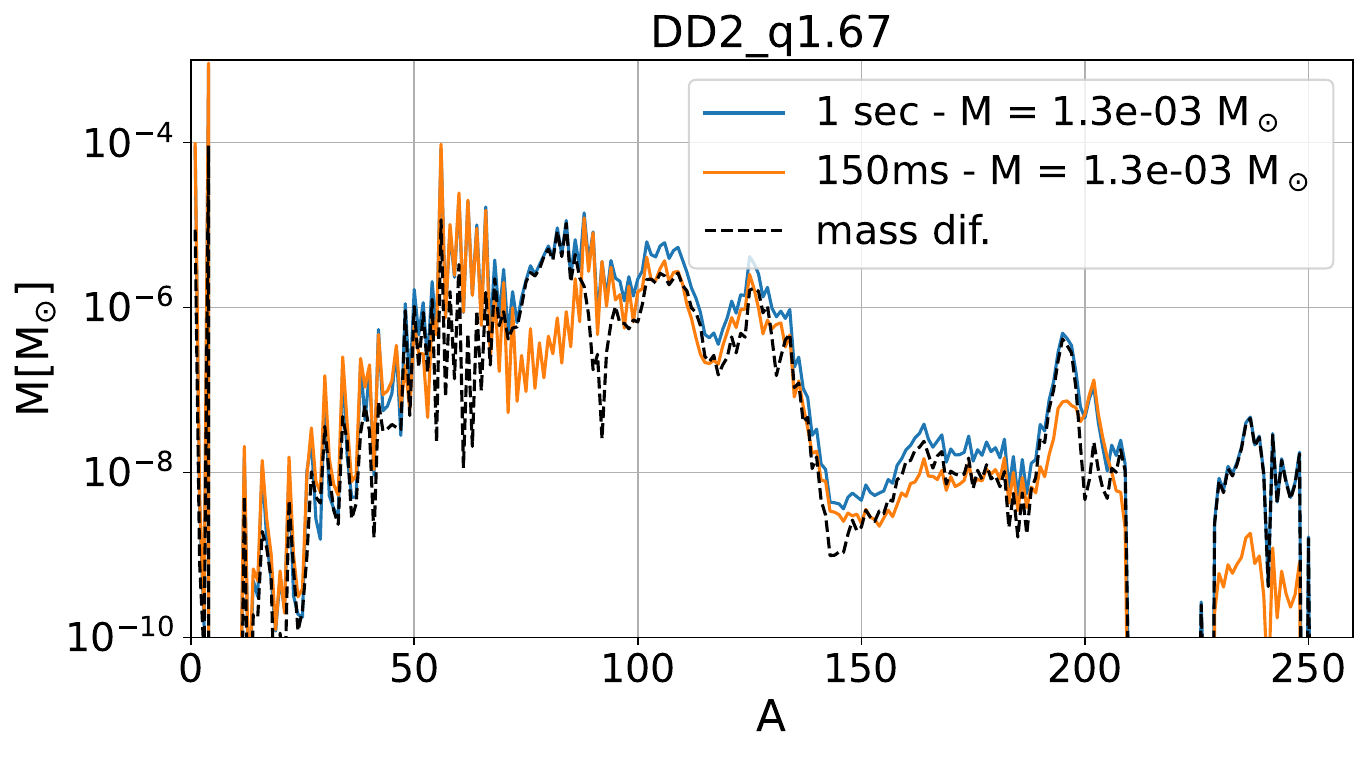}
	\caption{In this figure we show the impact of an early assumption of homology on the polar ($\theta>45^{\circ}$) element distribution. Blue (orange) lines represent the element formation computed when using the profiles obtained with 150~ms (1~s) of hydrodynamical evolution with \athena. Black dashed lines represent the absolute difference between the two data sets. While the element distribution is not affected for symmetric binaries (not shown here), BLh\_q1.43 and DD2\_q1.67 show signs of expansion of the tidal arm. Our results indicate that the lanthanide curtain is still expanding at the analyzed times, especially in asymmetric binaries. }
    \label{fig:WinNetTimes}
\end{figure}

It is a common practice in the literature to perform hydrodynamic simulations of the ejecta up to hundreds of milliseconds before using their data as input for nuclear networks both in 3D \cite{Curtis:2021guz,Ricigliano:2024lwf}, and 2D ray-by-ray \cite{Jacobi:2025eak}.
A notable exception is \cite{Kawaguchi:2024hdk}, where they continuously inject data on the grid until $t\sim{1}~$s.
Our current set-up allows us to assess first-hand the impact of such assumptions.
As seen in Fig.~\ref{fig:Homology}, the parameter $\chi$ is (conservatively) non-negligible up to  $t=150$~ms after the start of the injection in the inner boundary.
Taking this time as a benchmark for comparison, we truncate our tracer trajectories and reuse them as input to the nuclear network code.
The comparison between these two data sets is performed in Fig.~\ref{fig:WinNetTimes}.
In this figure we show only the polar contribution ($\theta>45^{\circ}$) for BLh\_q1.43 and DD2\_q1.67, where we observe the largest impact.
For the tested  models (BLh\_q1.43 and DD2\_q1.67) the angle-integrated nucleosynthesis remains effectively unchanged, with individual element mass fractions changing by $\mathcal{O}(30\%)$ or less.
Nevertheless, our analysis shows that the \emph{distribution} of $r$-process elements, in particular from the third $r$-process peak (180$\lesssim A \lesssim$ 200) and heavy isotopes (225$\lesssim A \lesssim $250), can be significantly affected by the early termination of the hydrodynamical evolution.
Systematic differences were found for third $r$-process peak elements in BLh\_q1.43, where the polar masses of elements such as gold increase by a factor of $\sim5$ ($\sim9\times{10}^{-9}M_{\odot}$ to $\sim5\times10^{-8}M_{\odot}$).
Very heavy unstable nuclei ($A\gtrsim 225$) make their way towards the poles in DD2\_q1.67 as our simulation evolves. While these individual elements display $\sim 10^{-9}M_{\odot}$ at the poles at 150~ms, at 1~s these masses increase towards $\sim{10}^{-7}M_{\odot}$.
These two effects are understood to be related to the expansion of the tidal arm where higher heating rates take place.
Nevertheless, the masses of the first $r$-process peak ($A \sim80$) elements at the polar region are affected by a factor of $\sim{10}$ for both shown models.

\begin{figure}[t]
    \centering
    \includegraphics[width=\linewidth]{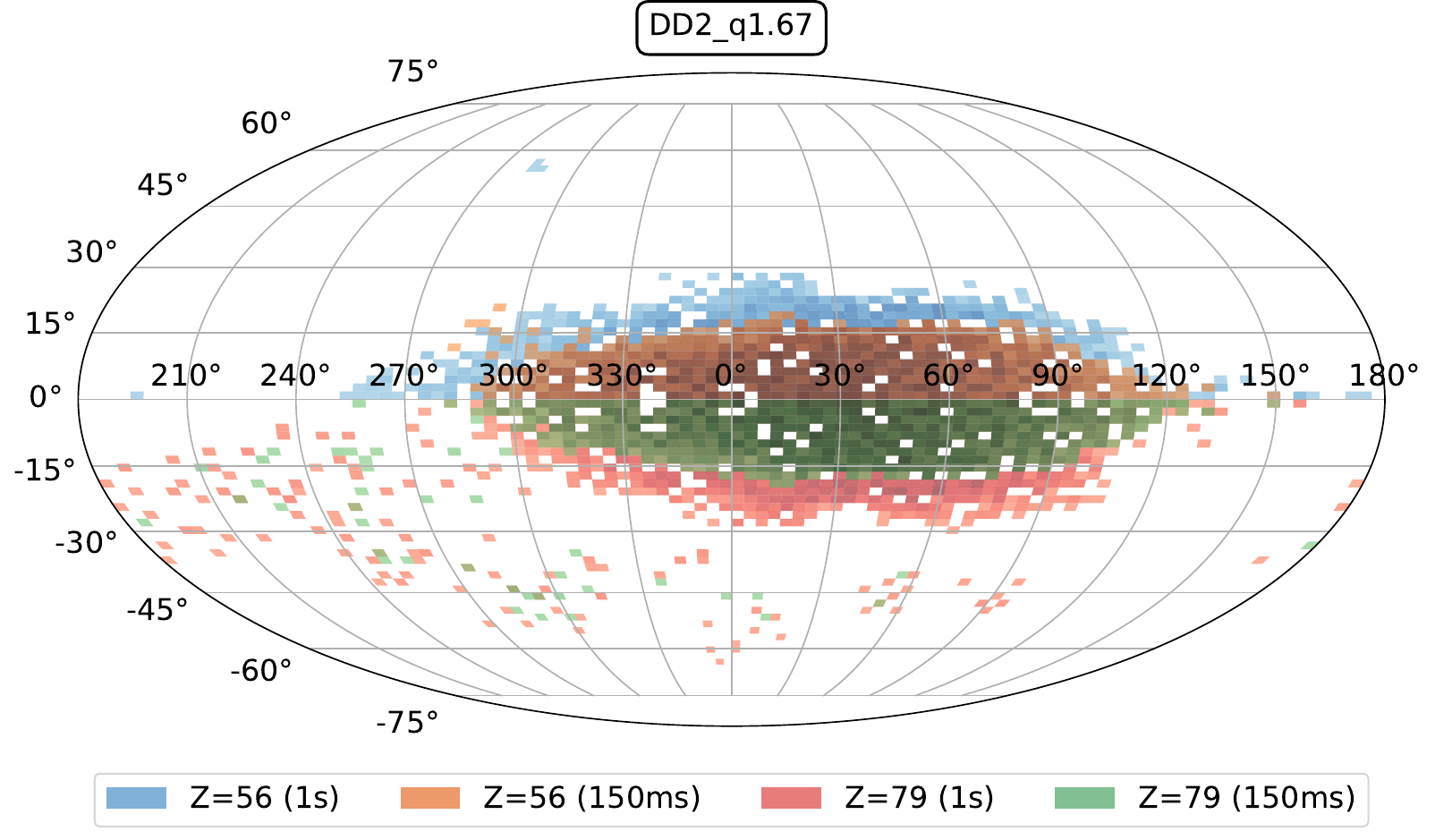}
	\caption{ Here we display the sky regions containing 90\% of the mass of two selected elements for DD2\_q1.67, when homology is assumed from 150~ms and from 1~s. We show this model as it presents the largest differences in the polar yields (Fig.~\ref{fig:WinNetTimes}). For the 1~s data-set, the location of the tidal arm (here correlated with the presence of $Z=56$ and $Z=79$) is more spread in the $\theta$ direction. While relatively high concentrations ($\sim{10}^{-7}M_{\odot}$ per bin) of these heavy elements are seen at $\theta \lesssim 30^{\circ}$ for a later assumption of homology, the earlier use of this approximation results in a lanthanide curtain constrained to $\theta  \lesssim 15^{\circ}$. These effects are explained by non-radial velocities even at 1~s for DD2\_q1.67. Free protons seem to be more robust with respect to the assumed homology time, and are therefore not shown here for brevity. }
        \label{fig:Redistribution}
\end{figure}

\begin{figure}[t]
    \centering
    \includegraphics[width=\linewidth]{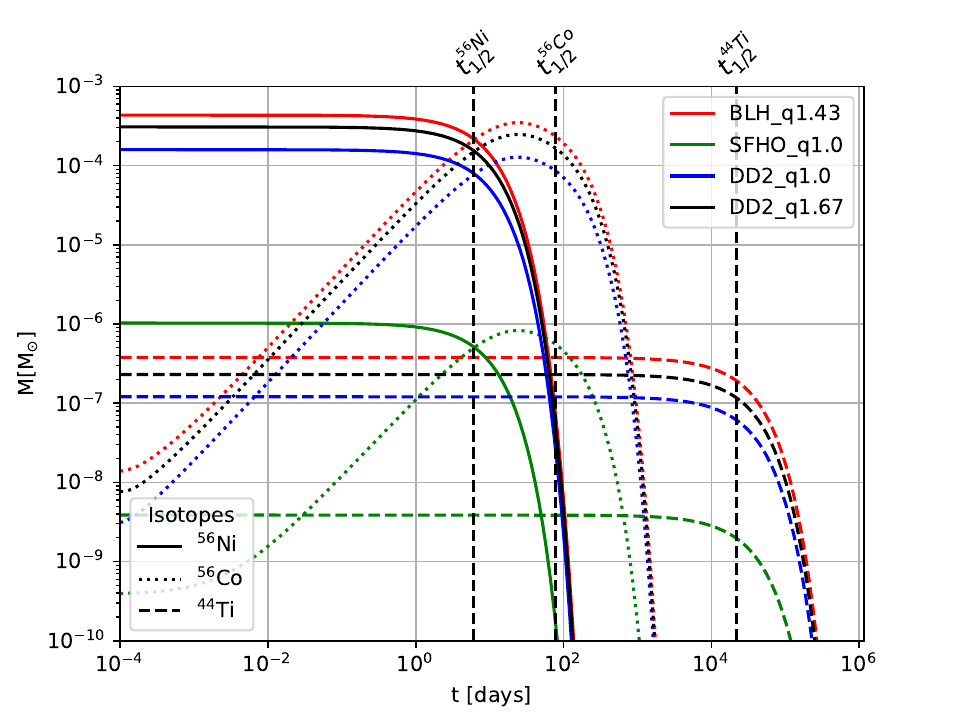}
	\caption{Evolution of $^{56}$Ni, $^{56}$Co, and $^{44}$Ti for our models. Among them, the one with the highest (lowest) initial nickel mass is BLh\_q1.43 (SFHo\_q1.0). The negligible amount of nickel displayed by SFHo\_q1.0 reflects the early collapse of its remnant into a BH and the consequent lack of a neutrino-driven wind at the poles. The values of the initial masses of the individual models are listed in Tab.~\ref{tab:EjectaMasses}. Unlike $^{56}$Co, $^{44}$Ti is not a by-product of nickel decay, but it is an important isotope in supernova physics \cite{Hanover:2025,Kosakowski:2022odv,The:2006iu}. We find no significant amount of this isotope in any of our models. }
    \label{fig:NiEvolve}
\end{figure}

These effects can be more easily illustrated by comparing the sky location of the elements for the 1~s and 150~ms data-sets of DD2\_q1.67.
This comparison is made for some selected nuclei in Fig.~\ref{fig:Redistribution}.
While the 150~ms data shows a tidal arm constrained to $\theta < 15^{\circ} $, the longer hydrodynamic evolution allows the tidal arm to expand and reach $\theta\sim{30^{\circ}}$.
This analysis agrees with the non-radial motion found at Fig.~\ref{fig:Vfdd2a}.

\begin{figure*}[t]
    \centering
        \includegraphics[width=.49\linewidth]{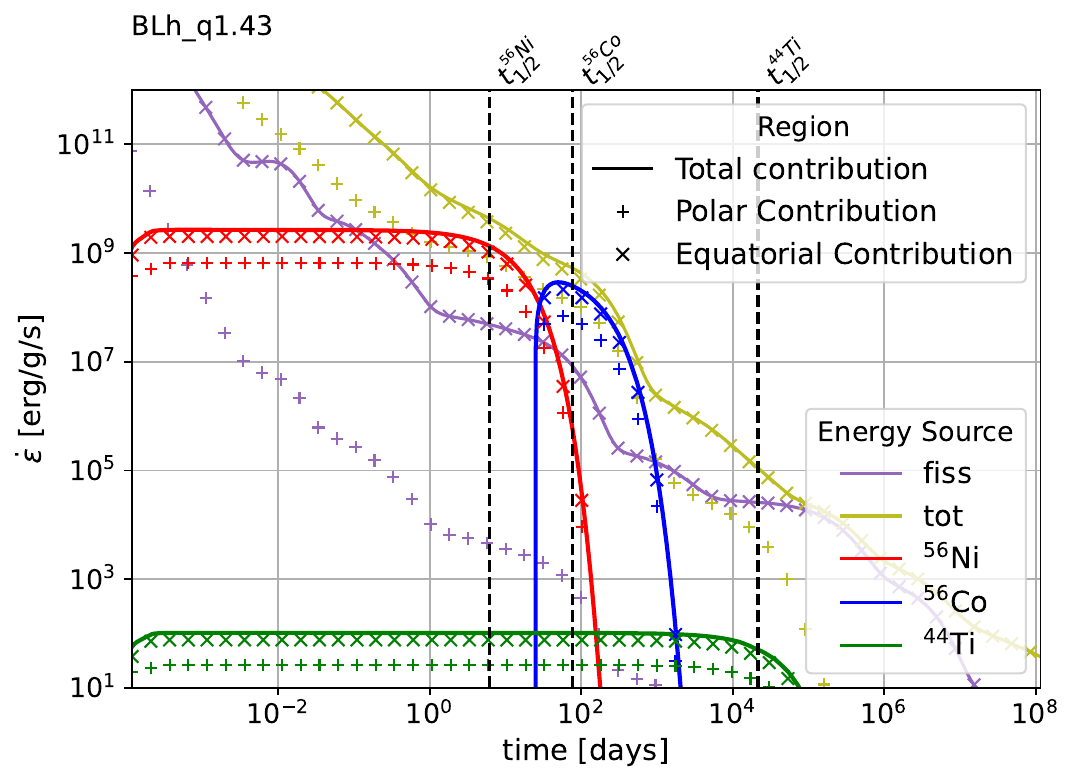}
        \includegraphics[width=.49\linewidth]{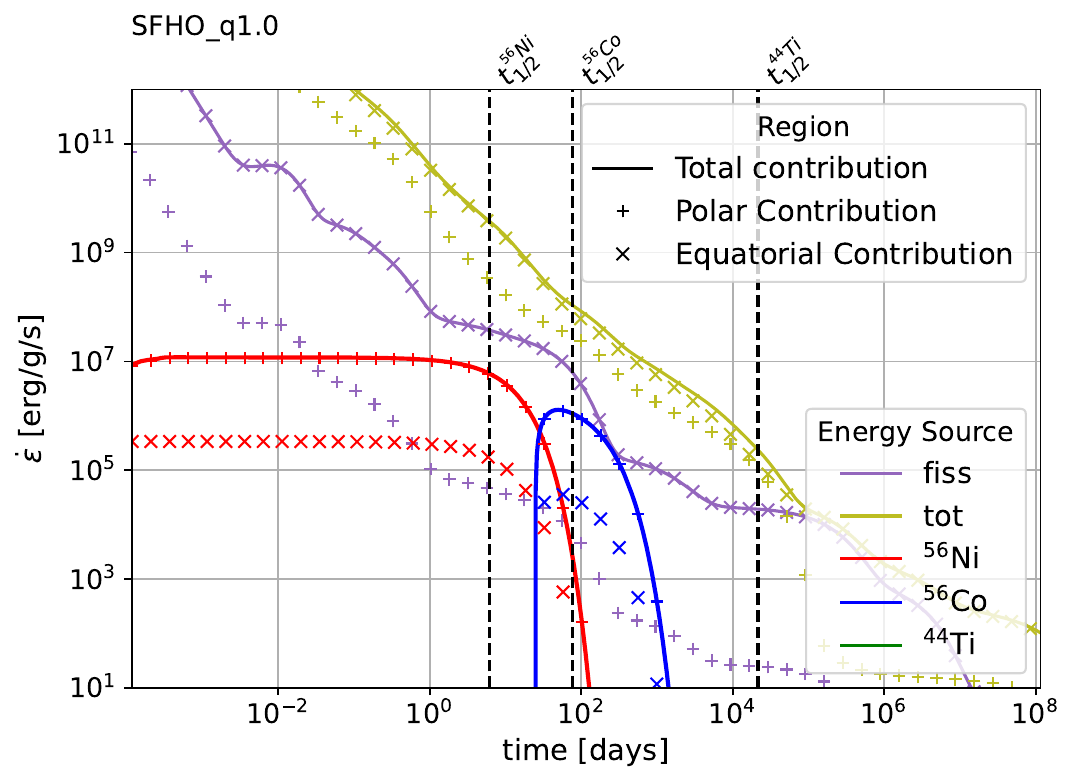}
        \includegraphics[width=.49\linewidth]{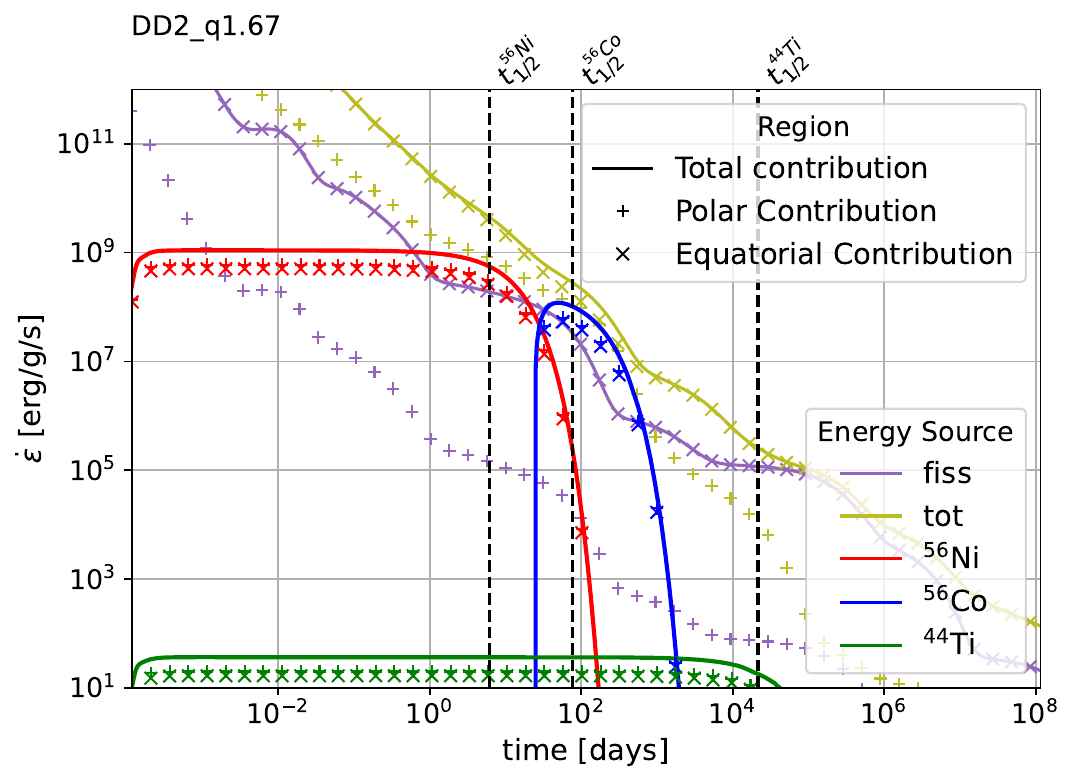}
        \includegraphics[width=.49\linewidth]{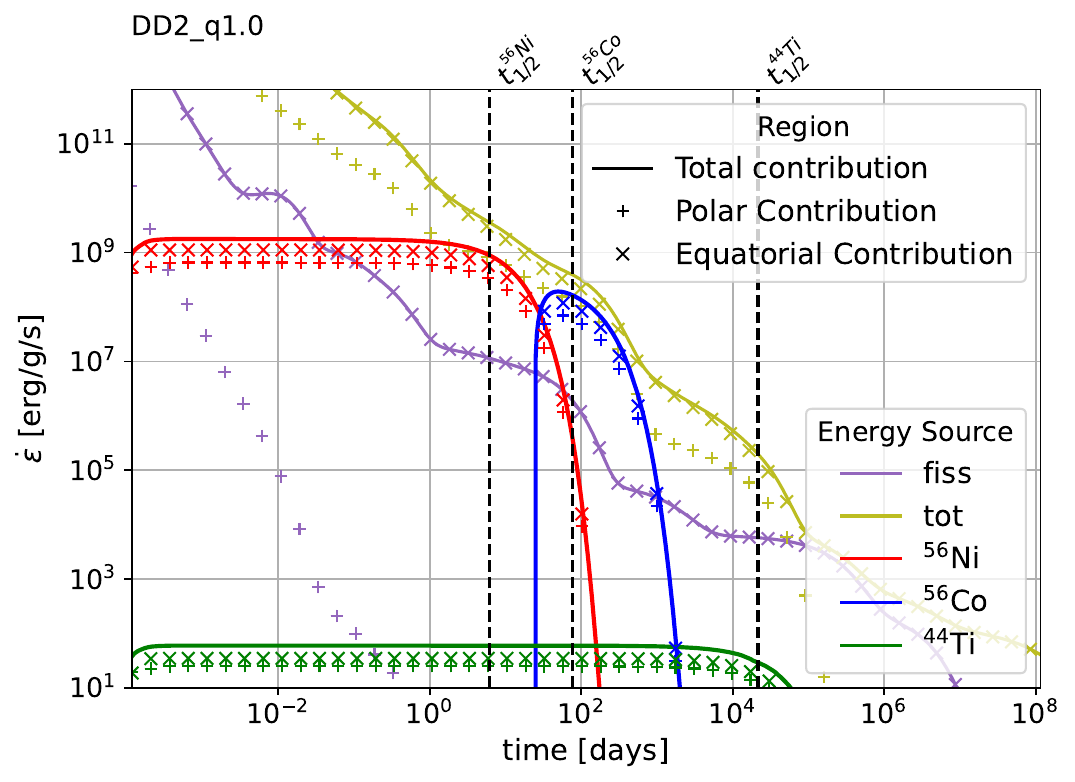}
	\caption{Total nuclear heating and its different sources.
	         The right captions are insets of the left plots for better comparison.
		 At around $t\sim{10}^{7}$~s all non-collapsing models show a clear dominance of the heating originating from the $^{56}$Ni decay chain.
		 The separation in polar ($\theta>45^{\circ}$) and equatorial ($\theta<45^{\circ}$) is performed.
		 We find that the impact of $^{56}$Co decay is stronger in the equatorial plane (in absolute terms; see below for its relative observational importance in the polar direction).
		 The underlying explanation is that the bulk of this decay chain is taking place at $30^{\circ}<\theta<45^{\circ}$,
		 as $^{56}$Fe is seen in these regions in Fig.~\ref{fig:SkyMaps}.
		 This effect is weaker in SFHo\_q1.0 and DD2\_q1.0 due to their symmetry and their lower initial $^{56}$Ni mass.
		 }
        \label{fig:EnergiesWinNet}
\end{figure*}

In light of a recent paper \cite{Jacobi:2025eak}, where yields of $10^{-3}M_{\odot}$ of $^{56}$Ni were found for BNS binaries with a NS remnant, we also analyze the production of such elements and their impact on observations. While this previous work's analysis relies on the same \thc data as our present work, it performs the nuclear evolution on profiles averaged over the azimuthal angle. The previous work performs a 2D ray-by-ray evolution of the averaged \thc data as extracted at $R_{0}=300M_{\odot}$, contrary to our three dimensional set-up. While previous results seem to overestimate the total masses, for reasons discussed in previous sections, the ratio of $^{56}$Ni to total ejecta mass remains roughly the same: 1.7-7.7\% for \cite{Jacobi:2025eak} and 2.3-5.5\% in our present analysis. SFHo\_q1.0 is the notable exception to this range of nickel masses. Due to its early collapse to a BH, this model does not present appreciable formation of $^{56}$Ni with the mass fraction staying below 0.03\% in our data.
Although this is not consistent with the mass fraction found in \cite{Jacobi:2025eak}, both analyses agree that Ni production by the SFHo\_q1.0 is of negligible effect.
Another difference between our current work and \cite{Jacobi:2025eak} is that this previous work used an online NN for their nucleosynthesis calculation.
In \cite{Magistrelli:2025xja}, it was shown that the $^{56}$Ni total mass can be affected by $\mathcal{O}$(10\%) when an online NN is used compared to post-processing of tracer particles.
Under the assumption of a fully resolved simulation by our set of tracers, this $\mathcal{O}$(10\%) error budget is not enough to explain the differences found in this work, which strengthens our confidence in the explanation based on the aforementioned total mass difference.

In Fig.~\ref{fig:NiEvolve} we show the mass evolution of $^{56}$Ni, $^{56}$Co, and $^{44}$Ti.
While the former two isotopes have appreciable masses for the first $\lesssim$3 years ($\lesssim{10}^{3}$ days) for the non-collapsing models, none of our models display a production above $4\times10^{-7}M_{\odot}$ of the latter isotope.
As $^{44}$Ti originates from the innermost, highest entropy ejecta layers via alpha-rich freeze-out it is a sensitive probe of the extreme conditions faced by the innermost fluid elements.
Our analysis suggests that $^{44}$Ti is not as relevant as in supernova events \cite{Hanover:2025,Kosakowski:2022odv,The:2006iu}.
These results suggest that our ejecta models do not reach the peak entropies required to form $^{44}$Ti.
Observationally, our work shows that we can expect a lack of the 67.87 and 78.32~keV gamma-ray lines associated with the decay of $^{44}$Ti~\cite{Chen:2023abc}.

To investigate the impact of $^{56}$Ni, $^{56}$Co, and $^{44}$Ti on the KN energetics, we studied their contribution to the overall heating energy as computed by \winnet.
In Fig.~\ref{fig:EnergiesWinNet} we show some of the main energy sources between 0.1 and 10$^{8}$~days, such as selected element decays, and fission separated in polar ($\theta>45^{\circ}$) and equatorial contributions ($\theta<45^{\circ}$).

As expected from the respective decay time scales, the heating at $ 30\lesssim t \lesssim 5\times 10^{2}$~days (1 to 17 months) for models with a NS remnant is dominated by $^{56}$Co decay rather than by $^{56}$Ni itself.
We stress this distinction explicitly, as it is not always made in the literature, \eg \cite{Jacobi:2025eak}.
Having separated the heating contribution in the polar ($\theta> 45^{\circ}$) and equatorial ($\theta< 45^{\circ}$) regions, we see that the regions most affected by this heating, for all models but SFHo\_q1.0, are the equatorial ones.
This can be understood by examining the sky location of the $A=56$ elements, which traces the final location of the $^{56}$Ni decay chain.
Both for BLh\_q1.43 and DD2\_q1.0, the $A=56$ elements have the bulk of their mass distribution at $30^{\circ}\lesssim \theta \lesssim 45^{\circ}$, see Fig.~\ref{fig:SkyMaps}.
On the other hand, DD2\_q1.67 displays a more homogeneous distribution for $A=56$, with the exception of an eye-shaped $A=56$-depleted region co-located with the tidal arm.

The contribution from the decay of $^{44}$Ti shows that, contrary to young supernova remnants ($100-1000$~years) \cite{Jerkstrand:2011,Grebenev:2012,Hanover:2025}, this decay route does not play a major role for our models.
We also show the contribution of fission in Figure~\ref{fig:EnergiesWinNet}. These contributions, which come from heavy and long-lived isotopes (225$\lesssim A \lesssim$250),  are subdominant except at late times of $10^{4}-10^{5}$~days ($30-300$~years) for all models.
Note that the energy coming from fission processes is mostly made available in the equatorial plane, in agreement with the results shown in Fig.~\ref{fig:AbsAngAlltheta}.

The properties of energy evolution discussed above are likely to leave imprints in the observed EM emission.
As the $^{56}$Co decay is dominant around 100 days and it decays by emitting gamma rays of 846.77~keV and 1238.288~keV in a ratio of 3:2 \cite{Huo:2011A56}, these lines should be stronger for binaries observed in a face-on line-of-sight. While the absolute $^{56}$Co heating is largest in the equatorial plane, where the bulk of the $A=56$ ejecta mass resides, its relative importance for the observed emission is largest for polar (face-on) observers, since polar ejecta contains proportionally less lanthanide-rich ($A\sim197$) material to outshine it. This can be seen directly in the isotope ratios: taking BLh\_q1.43 as an example, although most of the $A=56$ elements are formed at $\theta \lesssim 36^{\circ}$ (see Figure~\ref{fig:AbsAngAlltheta}), $Y_{56}/Y_{197} \sim 0.23$ for $\theta \lesssim 18^{\circ}$ (near-equatorial), increasing to $Y_{56}/Y_{197} \sim 1.6\times 10^{6}$ ($1.4\times 10^{2}$) at $36^{\circ} \lesssim \theta \lesssim 54^{\circ}$ ($72^{\circ} \lesssim \theta \lesssim 90^{\circ}$), where $Y_{i}$ is the nuclear abundance of $A=i$ elements.

\begin{table*}[t]
\centering
\caption{Total mass of selected light isotopes and of $^{88}$Sr for our BNS merger models, compared to the dynamical-ejecta values of \cite{Perego:2020evn} (bottom block, labelled ``P20''). The EOS and mass-ratio ($q=M_{1}/M_{2}$) columns make explicit which of our models is closest in setup to each P20 model; note however that P20 masses account only for the dynamical ejecta unbound within the first few milliseconds, while ours integrate over the full evolution (see text). P20 do not report Li or K masses, only that their number abundances are negligible ($Y\lesssim10^{-5}$), indicated by ``--''.}
\label{tab:lightelements}
\begin{tabular}{|c| c| c| c| c| c| c| c|}
\hline\hline
Model & EOS & $q$ & $M^{^{1}\mathrm{H}}[M_{\odot}]$ & $M^{^{4}\mathrm{He}}[M_{\odot}]$ & $M^{^{7}\mathrm{Li}}[M_{\odot}]$ & $M^{^{39}\mathrm{K}}[M_{\odot}]$ & $M^{^{88}\mathrm{Sr}}[M_{\odot}]$ \\
\hline\hline
BLh\_q1.43 & BLh  & 1.43 & $1.845\times 10^{-4}$ & $1.933\times 10^{-3}$ & $4.195\times 10^{-13}$ & $1.126\times 10^{-7}$ & $9.6\times 10^{-5}$ \\
\hline
SFHo\_q1.0 & SFHo & 1.0  & $3.586\times 10^{-7}$ & $9.667\times 10^{-5}$ & $5.074\times 10^{-13}$ & $4.559\times 10^{-10}$ & $6.6\times 10^{-5}$ \\
\hline
DD2\_q1.67 & DD2  & 1.67 & $1.167\times 10^{-4}$ & $1.301\times 10^{-3}$ & $3.767\times 10^{-12}$ & $7.388\times 10^{-8}$ & $1.3\times 10^{-4}$ \\
\hline
DD2\_q1.0  & DD2  & 1.0  & $6.421\times 10^{-5}$ & $7.754\times 10^{-4}$ & $2.186\times 10^{-13}$ & $4.155\times 10^{-8}$ & $1.1\times 10^{-4}$ \\
\hline\hline
\multicolumn{8}{|c|}{Reference: \cite{Perego:2020evn} (P20), dynamical ejecta only} \\
\hline\hline
BLh\_equal (P20)   & BLh  & 1.0  & $1.52\times 10^{-6}$ & $3.87\times 10^{-6}$ & -- & -- & $3.01\times 10^{-5}$ \\
\hline
BLh\_unequal (P20) & BLh  & 1.82 & $0.78\times 10^{-6}$ & $9.25\times 10^{-6}$ & -- & -- & $0.25\times 10^{-5}$ \\
\hline
DD2\_equal (P20)   & DD2  & 1.0  & $1.70\times 10^{-6}$ & $3.12\times 10^{-6}$ & -- & -- & $2.91\times 10^{-5}$ \\
\hline\hline
\end{tabular}
\end{table*}

We compare our light-element yields with those of \cite{Perego:2020evn}, who computed H, He, and Li--K abundances from the dynamical ejecta of BLh and DD2 both for symmetric and asymmetric binaries.
Our total masses of $^{1}$H and $^{4}$He are systematically larger than theirs by one to two orders of magnitude, for instance for hydrogen in our DD2\_q1.0 model compared to their DD2\_equal model, see Tab.~\ref{tab:lightelements}.
However, this is expected, as their masses account only for the dynamical ejecta, while ours consider over $\sim$100~ms of \thc data, and therefore include additional H/He-producing material, most notably from the neutrino-driven wind.
Despite this difference, the relative trend across EOS is consistent with their picture.
The softer BLh EOS produces more H and He than the stiffer DD2 EOS.
This agrees with the finding that the high-entropy channel is more prominent for softer EOS.
We further find that SFHo\_q1.0 exhibits a dramatic suppression of H, He, and K relative to all other models, up to $\sim$500$\times$ for hydrogen.
This aligns with the fact that SFHo\_q1.0 has the weakest neutrino-driven wind among our models.
In contrast, the production of $^{7}$Li does not follow this trend.
It is largest for DD2\_q1.67, roughly an order of magnitude above the other three models, suggesting it is not governed by the same high-entropy/neutrino-wind channel responsible for H, He, and K production.
The authors of \cite{Perego:2020evn} also kept track of strontium in their models, noticing that they are able to produce enough $^{88}$Sr to explain the inferred mass of this element for AT2017gfo \cite{Watson:2019xjv}.
Our models seem to predict larger $^{88}$Sr mass compared to observations, which was determined to be 1-5$\times10^{-5}M_{\odot}$.
Nevertheless, one should take this comparison against observations with caution, as the total Sr mass for AT2017gfo is still under debate \cite{Perego:2020evn,Domoto:2021xfq,Gillanders:2022opm,Tarumi:2023apl,Sneppen:2024ros,Arya:2026jvs}.

\subsection{Kilonova lightcurves}
\label{sec:res:lc}

\begin{figure}[t]
    \centering
    \includegraphics[width=\linewidth]{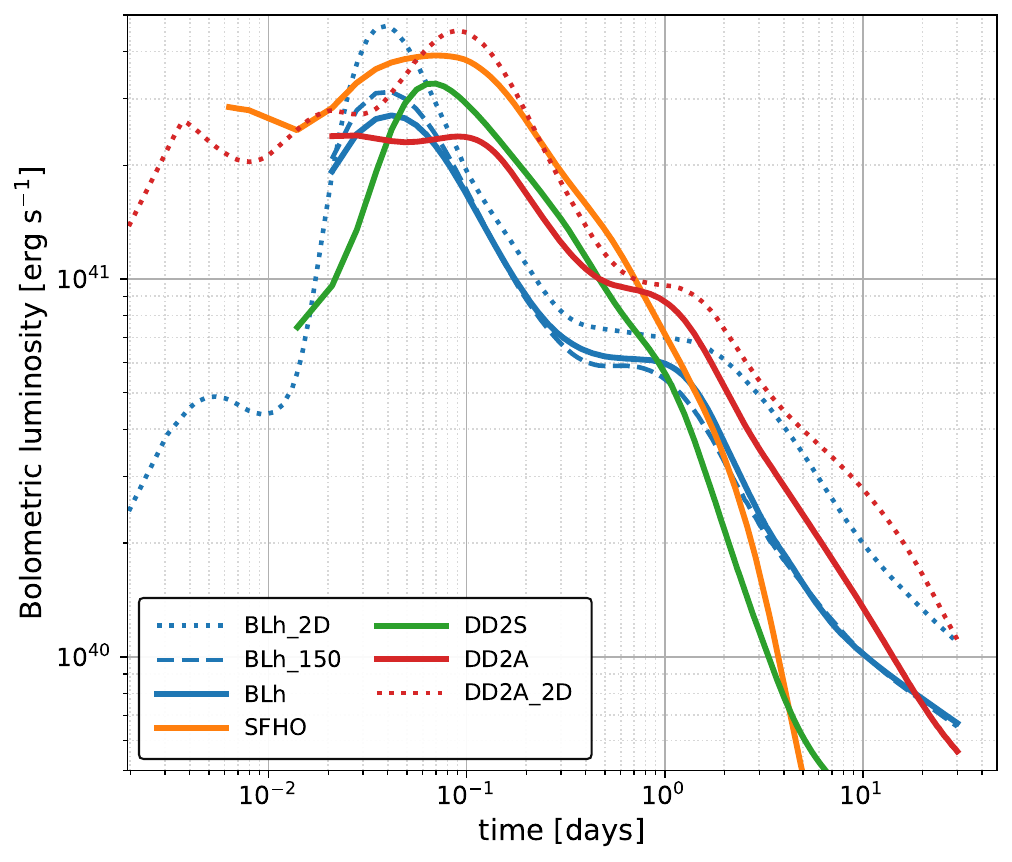}
    \caption{ Bolometric luminosity evolution of the models considered in this work.
	      The 2D \knec\ simulations of BLh\_q1.43 and DD2\_q1.67 are also shown and display enhanced luminosities compared to their corresponding 3D counterparts.
	      We further include the BLh\_q1.43 model evolved with \athena for only 150 ms to assess the impact of the shorter evolution time compared to the fiducial 1 s simulation.
	     }
    \label{fig:LumBOL}
\end{figure}

As discussed in previous sections, the $r$-process taking place in the ejecta material will create large amounts ($\sim 10^{-3}M_{\odot}$) of unstable isotopes.
Associated with the decay of these unstable elements, EM radiation of different frequencies will be emitted over time-scales ranging from hours to centuries.
We evolve the EM emission of our models using the \knec~code as described in Section~\ref{sec:method:LC}.
In this Section, we discuss the light curves of the KN of our models, but for the sake of brevity, we use BLh\_q1.43 as our fiducial model, as its mass ratio and chirp mass are the closest, among our models, to the estimated values for GW170817 \cite{Abbott:2018wiz}.
In this work, we focus on the 3D morphology of the BNS ejecta and the properties of the associated KN light curves.
To assess the impact of the 3D modeling of the BNS ejecta evolution we also evolve our models with \knec~after a mass-averaged $\phi$ marginalization of our ejecta profiles.
The $\phi$ marginalization of our models is performed directly on the \thc data, without the 3D extension performed with \athena.
For this reason, the 2D ray-by-ray results are also representative of an earlier homology assumption.

\begin{figure*}[t]
    \centering
    \includegraphics[width=\linewidth]{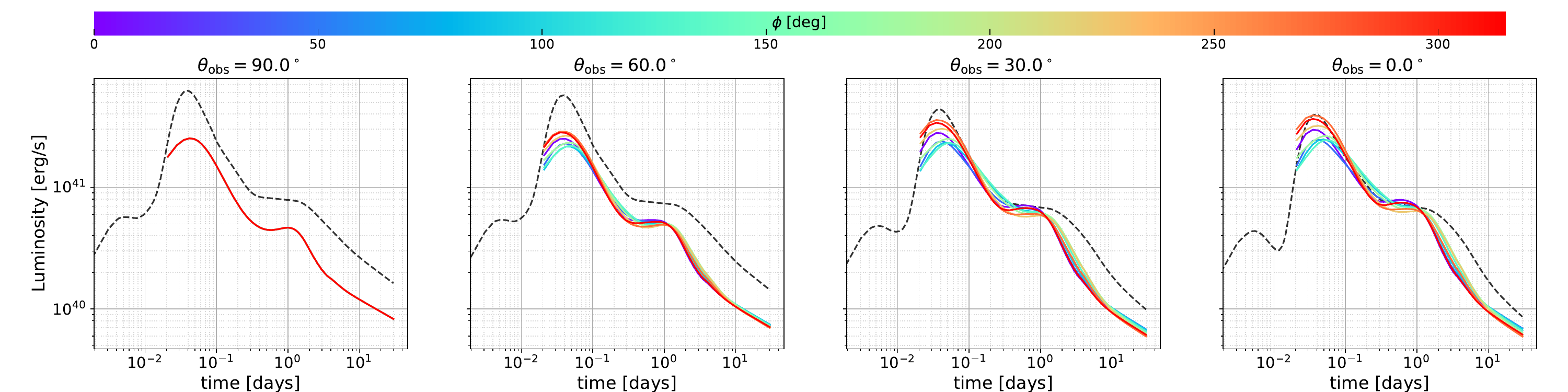}
	\caption{Observed luminosities as a function of time and angles of view with colored lines representing different azimuthal angles and dashed lines their 2D ray-by-ray counterpart. Here we show the results for BLh\_q1.43 as a fiducial model. As for the symmetric mass ratio binaries, no strong dependence with respect to $\phi$ was found, with exception to DD2\_q1.0 at time of first peak, where a factor $\sim$2.5 variation was found. The 2D ray-by-ray scheme tends to overestimate the peak luminosity and the late-time tail.}
    \label{fig:LumOBS}
\end{figure*}

We first consider the bolometric luminosity, shown in Fig.~\ref{fig:LumBOL}, before turning to its dependence on viewing angle.
BLh\_q1.43 and DD2\_q1.67 (SFHo\_q1.0 and DD2\_q1.0) display a double(single)-peaked bolometric light curve.
The properties of both peaks of the bolometric and observed luminosities are summarized in Tab.~\ref{tab:PeakLumBol} and Tab.~\ref{tab:PeakLumObs}, respectively.
As symmetric binaries have no genuine second local maximum, their tabulated ``late peak'' value reports the maximum luminosity reached within the same time window used to identify the late peak of the other three models.
This is made to allow a like-for-like comparison.
SFHo\_q1.0 reaches the highest early-time bolometric luminosity among our models ($3.9\times10^{41}\,\mathrm{erg\,s^{-1}}$ at $t\sim7.0\times10^{-2}$~days), while DD2\_q1.67 reaches the lowest ($2.4\times10^{41}\,\mathrm{erg\,s^{-1}}$ at $t\sim2.1\times10^{-2}$~days).
At the later peak the ordering is similar, with SFHo\_q1.0 remaining the brightest ($1.9\times10^{41}\,\mathrm{erg\,s^{-1}}$) and DD2\_q1.67 the dimmest ($1.3\times10^{41}\,\mathrm{erg\,s^{-1}}$), both occurring at $t\sim3\times10^{-1}$~days.
One should be aware that, although SFHo\_q1.0 and DD2\_q1.0 are the brightest models at early times, their bolometric luminosity decays faster at later times than for the asymmetric binaries, reflecting their relatively low production of lanthanides.
Figure~\ref{fig:LumBOL} also includes the 2D \knec evolutions of BLh\_q1.43 and DD2\_q1.67, which display enhanced luminosities compared to their 3D counterparts.
This is consistent with the discussed ejecta-mass overestimation, but we cannot rule out other effects, such as differences in opacity and its morphology, and the consequent evolution of the photosphere, which could have a comparable  impact on the luminosity and light curves.
Figure~\ref{fig:LumBOL} additionally includes a version of BLh\_q1.43 evolved with \athena\ for only 150\,ms, allowing us to isolate the impact of the shorter evolution time relative to our fiducial 1\,s simulation.

\begin{table*}[t]
\centering
\caption{In this table we summarize the simulated peak bolometric luminosities of our models and their time of occurrence.}
\label{tab:PeakLumBol}
\begin{tabular}{|c | c| c| c| c| }
\hline\hline
Model
& $L^{early}_{\rm peak}$ [$\mathrm{erg\,s^{-1}}$]
& $t_{L^{early}_{\rm peak}}$ [days]
& $L^{late}_{\rm peak}$ [$\mathrm{erg\,s^{-1}}$]
& $t_{L^{late}_{\rm peak}}$ [days] \\
\hline\hline
\hline
BLh\_q1.43         & 2.7 $\times 10^{41}$ & 4.2 $\times 10^{-2}$ & 6.2 $\times 10^{40}$ & $6.8\times10^{-1}$  \\
\hline
SFHo\_q1.0         & 3.9 $\times 10^{41}$ & 7.0 $\times 10^{-2}$ & 1.9 $\times 10^{41}$ & $3.1\times10^{-1}$  \\
\hline
DD2\_q1.0          & 3.3 $\times 10^{41}$ & 6.3 $\times 10^{-2}$ & 1.4 $\times 10^{41}$ & $3.1\times10^{-1}$  \\
\hline
DD2\_q1.67         & 2.4 $\times 10^{41}$ & 2.1 $\times 10^{-2}$ & 1.3 $\times 10^{41}$ & $3.0\times10^{-1}$  \\ [1ex]
\hline\hline
\end{tabular}
\end{table*}

\begin{table*}[t]
\centering
\caption{In this table we summarize the simulated double peak observed luminosities of our models and their time and angle of occurrence.}
\label{tab:PeakLumObs}
\begin{tabular}{|c | c| c| c| c| c| c| c| }
\hline\hline
Model
& $L^{early}_{\rm peak}$ [$\mathrm{erg\,s^{-1}}$]
& $(\theta,\phi)_{L^{early}_{\rm peak}}$
& $t_{L^{early}_{\rm peak}}$ [days]
& $L^{late}_{\rm peak}$ [$\mathrm{erg\,s^{-1}}$]
& $(\theta,\phi)_{L^{late}_{\rm peak}}$
& $t_{L^{late}_{\rm peak}}$ [days] \\
\hline\hline
BLh\_q1.43         & 3.9 $\times 10^{41}$ & $(0^{\circ},270^{\circ})$ & 3.5 $\times 10^{-2}$ & 7.9 $\times 10^{40}$ & $(0^{\circ},0^{\circ})$   & $5.8\times10^{-1}$  \\
\hline
SFHo\_q1.0         & 5.0 $\times 10^{41}$ & $(90^{\circ},0^{\circ}) $ & 7.6 $\times 10^{-2}$ & 2.2 $\times 10^{41}$ & $(90^{\circ},0^{\circ})$  & $3.1\times10^{-1}$  \\
\hline
DD2\_q1.0          & 5.4 $\times 10^{41}$ & $(0^{\circ},90^{\circ}) $ & 6.3 $\times 10^{-2}$ & 2.0 $\times 10^{41}$ & $(0^{\circ},135^{\circ})$ & $3.1\times10^{-1}$  \\
\hline
DD2\_q1.67         & 2.9 $\times 10^{41}$ & $(0^{\circ},0^{\circ})  $ & 8.4 $\times 10^{-2}$ & 1.6 $\times 10^{41}$ & $(0^{\circ},45^{\circ})$  & $3.0\times10^{-1}$  \\ [1ex]
\hline\hline
\end{tabular}
\end{table*}

We now turn to the observed luminosities, shown in Fig.~\ref{fig:LumOBS}, which retain the full angular dependence averaged out of the bolometric picture above.
By investigating the properties of the photosphere for our models, we see that this structure is the result of a complicated interplay between the decay of $^{56}$Ni, and light and $r$-process elements.
While the second peak is mostly powered by Ni and $r$-process decays, the dominant source of luminosity for the first peak is very model dependent.
As expected, axisymmetric models (DD2\_q1.0 and SFHo\_q1.0) display light curves that depend only weakly on the azimuthal angle.
The largest departure from axisymmetry for these models is seen around DD2\_q1.0's first peak, where the luminosity varies by a factor of $\sim2.5$ across different $\phi$ angles for an edge-on angle of view.
DD2\_q1.0 (SFHo\_q1.0) has its brightest early peak at $\theta=0^{\circ}$, $\phi=90^{\circ}$ ($\theta=90^{\circ}$, $\phi=0^{\circ}$), correlating with the slightly higher concentrations of light elements in these regions, as seen in Fig.~\ref{fig:AbsAngAllphi}.
This figure also explains why DD2\_q1.0 has a larger variability around the earlier peak when compared to SFHo\_q1.0, as the distribution of $80\lesssim A\lesssim120$ elements is more strongly $\phi$ dependent.
The later, dimmer peak instead traces the diffusion of photons through the high-opacity lanthanide curtain, whose larger optical depth delays and reddens the emission it powers.
DD2\_q1.0 has its brightest late peak at $\theta=0^{\circ}$, $\phi=135^{\circ}$, close to but distinct from its early-peak location, consistent with a shift in the dominant emitting region as the lanthanide-rich material becomes optically thin. For both asymmetric binaries, the brighter regions are found at $\theta=0^{\circ}$ at both peaks (Tab.~\ref{tab:PeakLumObs}).
BLh\_q1.43 (DD2\_q1.67) displays a slower decay of the light curve for $\phi=210^{\circ}(30^{\circ})$, regions with larger abundances of lanthanides, consistent with this second, lanthanide-driven component of the emission.

\begin{figure*}[t]
    \centering
    \includegraphics[width=\linewidth]{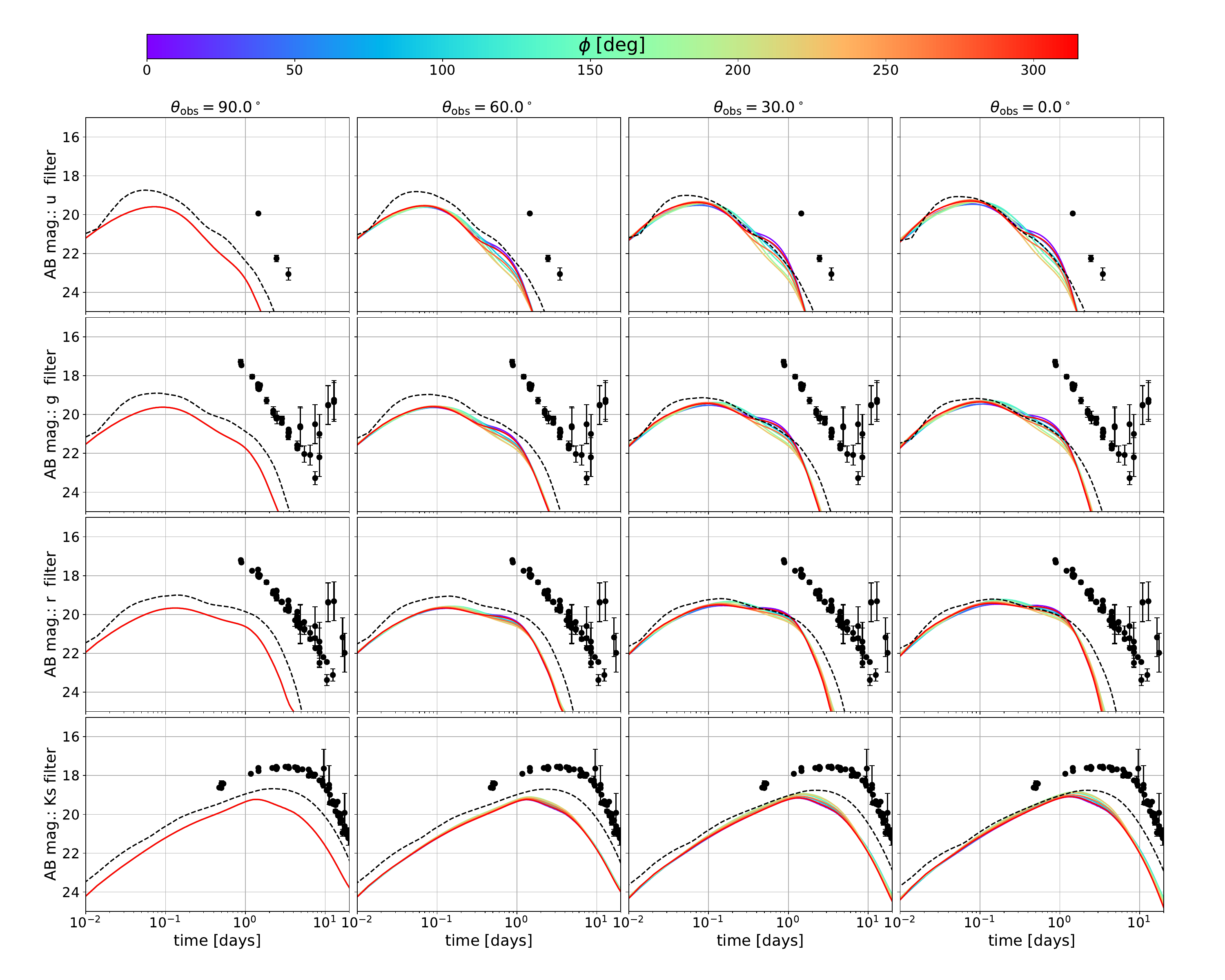}
    \caption{Here we present the AB magnitudes at 40~Mpc obtained when feeding our 3D ray-by-ray version of the \knec{} code with our BLh\_q1.43 ejecta profile at one second. While the colors represent different angles of observation ($\phi_{\rm obs}$), the dashed lines represent the $\phi$-averaged equivalent as obtained by the original 2D \knec{} set-up. Black dots and their error bars represent the collection of observational data points for the filters of interest from various telescopes \cite{Villar:2017wcc}. We note that the 2D ray-by-ray evolution tends to underestimate the peak magnitudes for asymmetric binaries.
      The 3D ray-by-ray runs do not move the theoretical models closer to observations implying that dimensionality is not a dominant factor when searching for better agreement with data.}
    \label{fig:GeminiBLH}
\end{figure*}

For a more detailed comparison, we show in Fig.~\ref{fig:GeminiBLH} the calculated AB magnitudes for different filters at 40~Mpc for our fiducial model.
Our findings indicate that changing the ray-by-ray scheme from 2D to 3D does not bridge the gap between observations \cite{Coulter:2017wya,Arcavi:2017a,Cowperthwaite:2017dyu,Diaz:2017uch,Drout:2017ijr,Evans:2017mmy,Kasliwal:2017ngb,Pian:2017gtc,Smartt:2017fuw,Troja:2017nqp,Utsumi:2017cti,Valenti:2017ngx,Villar:2017wcc} and ab-initio simulations, see some comparisons at \cite{Collins:2022ocl,Groenewegen:2025ezj,Breschi:2021tbm}. More precisely, for most of our models and filters, the AB magnitudes are increased when the dimensionality was increased.
We understand this by a combination of two sources of overestimation of ejecta mass on the part of the 2D ray-by-ray model.
The first effect is described on Appendix~\ref{app:mflux},
 where we discuss the differences found by using only fluid elements with positive mass fluxes to construct an ejecta profile versus populating ghost zones with all the fluid elements and how the former tends to estimate higher masses, especially for asymmetric binaries. The second source of a mass overestimation is the average implemented to perform the marginalization over $\phi$. In order to achieve a more representative profile, all quantities (including density) are mass-weighted averaged, which has the drawback of not preserving the total mass. These effects are more prominent for DD2\_q1.67 due to its large asymmetry. However, the difference in AB magnitude never surpasses unity at peak time for individual filters.

Our analysis suggests that the axisymmetry assumption is not responsible for the gap between simulations and observations.
To account for these differences in AB magnitudes, one needs not only more mass but also a specific composition for this new ejecta component.
The ejecta originating from the disk were also shown to extend to much lower $Y_{e}$ than in our models, reaching values as low as 0.02 \cite{Fujibayashi:2022ftg}, see Fig.~\ref{fig:Corner}.
The viscous ejecta missing from our simulations are likely responsible for this discrepancy.
Assuming that the disk neutronizes~\cite{Beloborodov:2002af,Siegel:2017nub,Sprouse:2023cdm}, viscous ejecta are expected to undergo strong $r$-process and contribute to the KN ``red'' component~\cite{Barnes:2013wka,Kasen:2013xka,Tanaka:2017qxj,Even:2019nbj,Tanaka:2019iqp}. This does not require a central NS: pure BH-disk simulations (i.e.\ with no NS present) reach similar conclusions, though the degree of neutronization is sensitive to disk spin, magnetization, and neutrino-transport fidelity, with simpler leakage-scheme models finding robust neutron-rich, red-component-producing outflows~\cite{Siegel:2017nub} while full-transport treatments of comparable systems find higher, more polar-concentrated $Y_{e}$ and a correspondingly weaker/partial $r$-process outcome~\cite{Miller:2019dpt,Curtis:2023xzm,Sprouse:2023cdm}.
Therefore the $r$ and $K_{s}$ filters are expected to be significantly impacted.
Longer simulations that account for the viscous ejection of the disk mass are necessary to serve as input to our set-up before making any quantitative statement about its impact.

\section{Conclusion}
\label{sec:con}

In this work we presented second-long 3D simulations of BNS ejecta, obtained by extending NR merger simulations with a general-relativistic hydrodynamics evolution that incorporates a non-NSE equation of state and an effective nuclear-heating prescription. This setup allows us to self-consistently continue the hydrodynamical evolution of NR ejecta profiles and to provide more reliable input to nuclear-network and radiative-transport codes.
We considered four binary models of varying EOS and mass ratio, and investigated the impact of nuclear heating on the ejecta dynamics, nucleosynthesis, and element distribution, as well as the differences between our 3D light curves and previous 2D ray-by-ray results.

Regarding the ejecta dynamics, the most asymmetric model, DD2\_q1.67, is the most strongly affected by nuclear heating, which sources a non-radial motion in the $\theta$ direction that persists up to the second time-scale (Fig.~\ref{fig:Homology}). This motion is strongly correlated with the nuclear heating energy (Fig.~\ref{fig:Energies}).

Regarding nucleosynthesis, we find that the final ejecta composition of our models is robust with respect to the time at which we terminate the tracers' evolution. The dynamics, however, can still redistribute the elements on the sky: for DD2\_q1.67, the lanthanide curtain and other heavy elements are pushed to larger polar angles when the final tracer time is extended from 150~ms to 1~s, from a collimated $\theta\lesssim15^{\circ}$ to $\theta\lesssim30^{\circ}$. This is consistent with the recent finding of \cite{Groenewegen:2025ezj} that homologous expansion only sets in after 2.5~s. Our nucleosynthesis data further confirm the result of \cite{Jacobi:2025eak} that the $^{56}$Ni decay route dominates the heating at $\sim$100 days.
Nevertheless, the total $^{56}$Ni  mass produced in our 3D models is systematically lower than for the 2D models presented in  \cite{Jacobi:2025eak}, by a factor in the range $\sim 1.3-2.6$ depending on the model.
We also find fallback masses of 1--4$\times10^{-3}M_{\odot}$ for all models except DD2\_q1.0, for which the fallback mass is $\sim8\times10^{-4}M_{\odot}$. 
For SFHo\_q1.0, we argue that this fallback mass may be sufficient to suppress the later ejection of disk material.

Turning to the KN light curves, our 3D ray-by-ray results agree qualitatively with the 2D ray-by-ray method, with the largest differences found for the asymmetric models BLh\_q1.43 and DD2\_q1.67. In these two models, the 3D light curves peak later and at a lower luminosity than their 2D counterparts, widening the gap with the AT2017gfo observations. Analytic scaling relations show that both a lower peak luminosity is a signature of a less massive ejecta profile \cite{Arnett:1980}.
We therefore interpret this difference as evidence that the 2D ray-by-ray method overestimates the ejecta mass relative to our 3D calculation.
This overestimation has two distinct origins.
The first is the injection method used by the 2D pipeline: because its radial profile is reconstructed from only the positive mass fluxes through the injection surface, it does not capture the fluid-element crossings resolved by our 1D Lagrangian evolution, and therefore predicts a slightly higher ejecta mass (see Fig.~\ref{fig:MassInjection}).
In our model, potential fluid-element crossings are intrinsically taken into account up to the 1 second extension of the NR data.
The second is the averaging used to construct the 2D profile: obtaining a representative azimuthally-averaged profile requires mass-weighted averages for all quantities of interest, including density; this preserves the microphysical state of the fluid but not the total mass.
Together, these two methodological choices account for the differences we find between the 2D and 3D light curves.
We conclude that the 2D ray-by-ray approach remains a robust but upper-limit estimate of the KN light curve, and that increased dimensionality alone is not sufficient to bridge the gap between current simulations and observations.
Differently from \cite{Neuweiler:2022eum}, we do find differences in the bolometric luminosity when the homologous expansion is relaxed (see Fig.~\ref{fig:LumBOL}), although these differences are a subdominant source of error when compared to the low dimensionality modeling.

In the near future, we plan to extend our analysis to the effect of magnetic fields on the ejecta evolution. Magnetic fields can inject energy into a fluid element, altering both its dynamics and thermodynamic state; magnetohydrodynamic turbulence developing in the ejecta interior, e.g.~\cite{Cook:2025frw}, can trap charged particles and enhance their thermalization, which could quantitatively affect the nucleosynthesis and the resulting KN light curves. We also plan to investigate jet-launching and jet-ejecta interaction with our setup.
 Our approach can further be extended to follow the ejecta evolution up to the nebular phase, providing data for 3D radiation-transport codes to compute KN light curves and spectra self-consistently.
 The interaction of the KN remnant with the surrounding interstellar medium, leading to the afterglow phase, is of particular interest, since homologous expansion (and consequently the validity of the ray-by-ray treatment) will eventually break down, requiring a fully three-dimensional scheme.

\begin{acknowledgements}
We thank David Radice for discussions and input on the project.
LFLM, MJ and SB acknowledge funding from the EU Horizon under ERC Consolidator
Grant, no. InspiReM-101043372.
MJ and SB acknowledge funding from the Deutsche Forschungsgemeinschaft (DFG) project MEMI number BE 6301/2-2 (project number 443239082).
FM acknowledges support from the Deutsche Forschungsgemeinschaft
(DFG) under Grant No. 406116891 within the Research Training Group
RTG 2522/1 .

The computations were performed on the ARA cluster at Friedrich
Schiller University Jena, on the supercomputer SuperMUC-NG at the Leibniz-
Rechenzentrum (LRZ, \url{www.lrz.de}) Munich.
ARA is a resource of Friedrich-Schiller-Universit\"at Jena
supported in part by DFG grants INST 275/334-1 FUGG, INST 275/363-1
FUGG and EU H2020 BinGraSp-714626. 
The authors acknowledge the Gauss Centre for Supercomputing
e.V. (\url{www.gauss-centre.eu}) for funding this project by providing
computing time on the GCS Supercomputer SuperMUC-NG at LRZ
(allocations {\tt pn36ge} and {\tt pn36jo}).
\end{acknowledgements}

\appendix

\section{Consistency of ejecta mass with NR data}
\label{app:mflux}

\begin{figure*}[t]
  \centering
  \includegraphics[width=.49\linewidth]{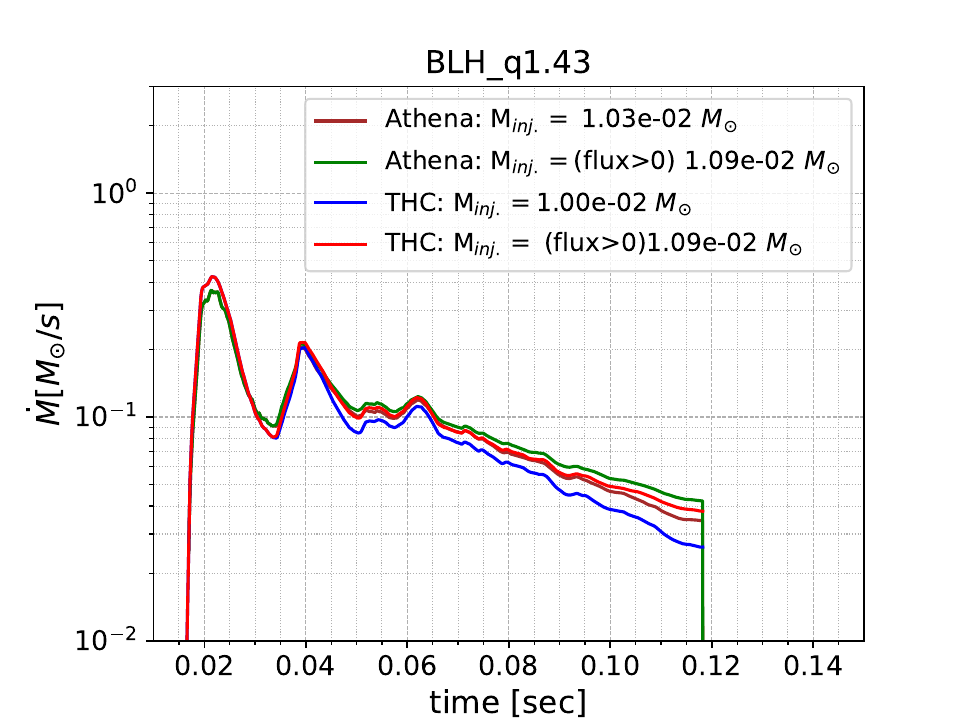}
  \includegraphics[width=.49\linewidth]{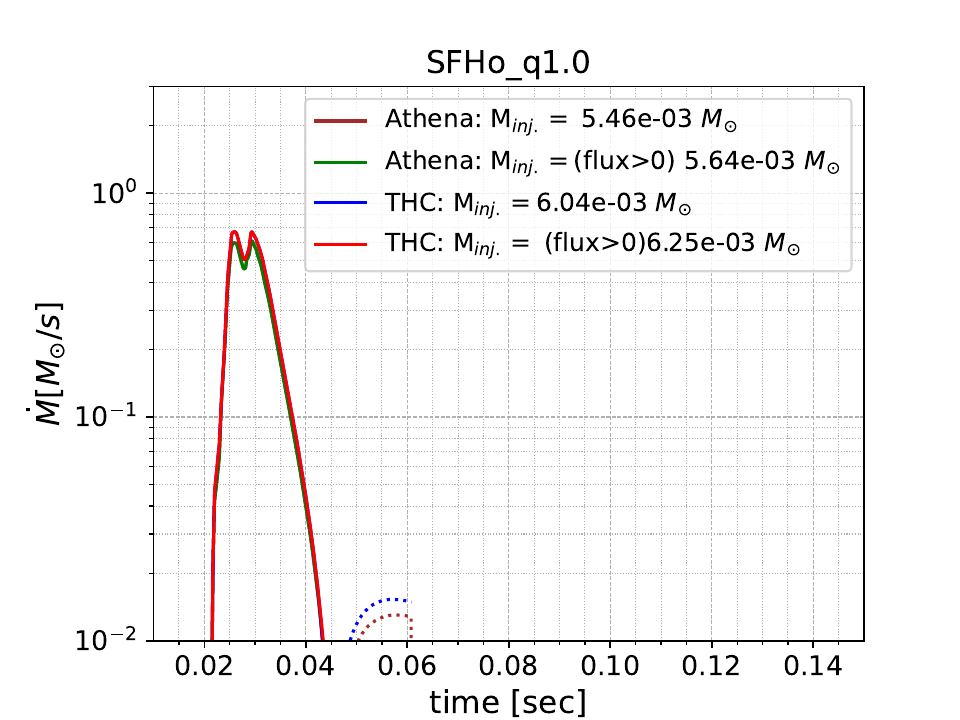}
  \includegraphics[width=.49\linewidth]{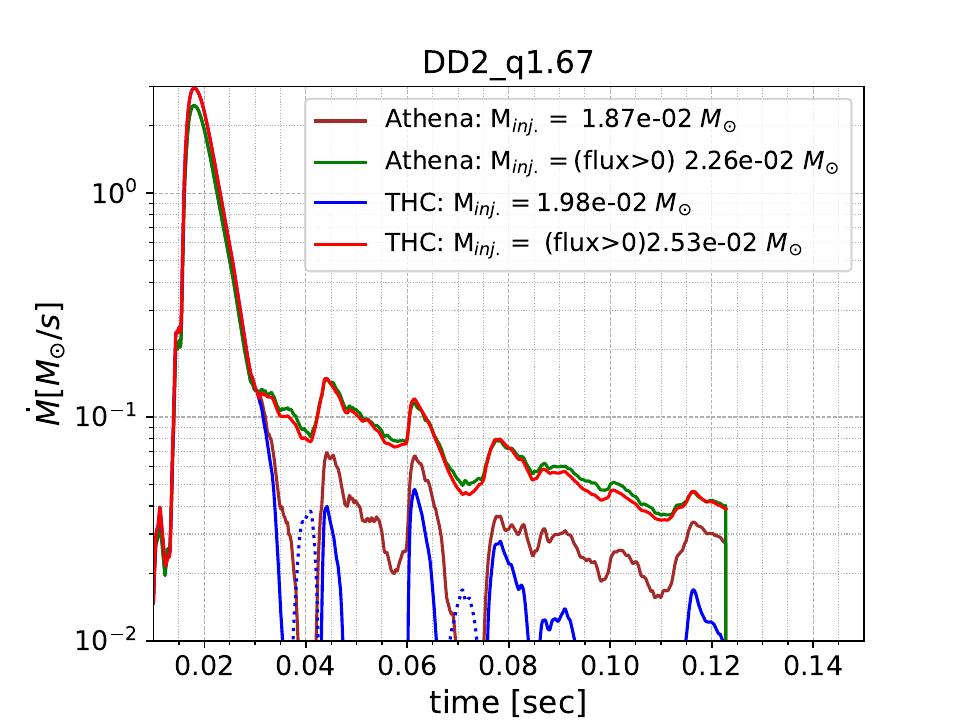}
  \includegraphics[width=.49\linewidth]{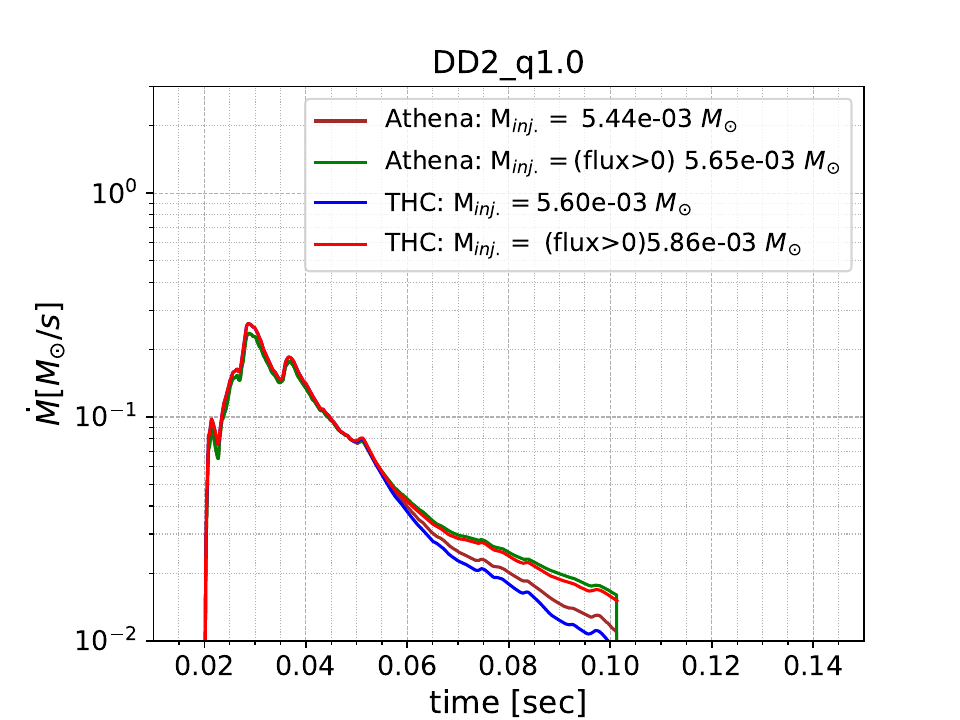}
  \caption{Here we show the comparison of the mass flux at $r=300M_{\odot}$ between the injected data in our grid and one extracted from \thc simulations.
	The integral over the injection/extraction 2-sphere is performed by considering either all the points on the surface or only the points where the mass flux is positive. We understand that our injection method closely reproduces the input data. The biggest discrepancy is seen for the DD2\_q1.67 when negative fluxes are accounted for.  This model displays the biggest value of fallback material, see Fig.~\ref{fig:EjectaMassTime}, which may explain such differences.  }
    \label{fig:MassInjection}
\end{figure*}

Figure \ref{fig:MassInjection} shows the mass flux over the innermost radial ghost zone where our injection takes place. We perform a comparison between the mass flux obtained in our simulations and the one in the pre-existing \thc data.
We also performed the same comparison by restraining our integration to grid cells with positive mass flux, as previous studies have relied on this subset of the data \cite{Magistrelli:2024zmk,Jacobi:2025eak,Magistrelli:2025xja}. Our runs systematically achieve lower ejecta mass in comparison to the \knec{} data from~\cite{Magistrelli:2024zmk,Jacobi:2025eak,Magistrelli:2025xja} that used the same NR data.
These differences in ejecta masses have two main origins: first, \knec{} simulations used  only data points with positive mass flux from the \thc data. Second, some matter is falling back in the \athena simulations.

Among our models, the only one for which the \thc data reach a time where we can guarantee that the spiral-arm ejection has ceased is SFHo\_q1.0, due to its collapse into a BH at $t\sim{8}$~ms. For this model we argue that our estimated fallback mass has a physical origin. For the other models, ejecta that would cross the ejection radius at times beyond the end of the original \thc simulation could provide support to the outer layers of the ejecta, pushing them and not allowing the fallback that we report. Nevertheless, DD2\_q1.67 displays the largest amount of fallback material, implying that a larger number of fluid elements with negative mass flux at the inner radial boundary may explain the larger discrepancies seen for this model in Fig.~\ref{fig:MassInjection} when compared to the remaining models.

\section*{Data Availability}

We will release tracer data on Zenodo/NR-GW OpenData upon final peer-review.
Other data will be shared upon reasonable request to the corresponding author.

\bibliographystyle{apsrev4-1}

\begin{thebibliography}{130}%
\makeatletter
\providecommand \@ifxundefined [1]{%
 \@ifx{#1\undefined}
}%
\providecommand \@ifnum [1]{%
 \ifnum #1\expandafter \@firstoftwo
 \else \expandafter \@secondoftwo
 \fi
}%
\providecommand \@ifx [1]{%
 \ifx #1\expandafter \@firstoftwo
 \else \expandafter \@secondoftwo
 \fi
}%
\providecommand \natexlab [1]{#1}%
\providecommand \enquote  [1]{``#1''}%
\providecommand \bibnamefont  [1]{#1}%
\providecommand \bibfnamefont [1]{#1}%
\providecommand \citenamefont [1]{#1}%
\providecommand \href@noop [0]{\@secondoftwo}%
\providecommand \href [0]{\begingroup \@sanitize@url \@href}%
\providecommand \@href[1]{\@@startlink{#1}\@@href}%
\providecommand \@@href[1]{\endgroup#1\@@endlink}%
\providecommand \@sanitize@url [0]{\catcode `\\12\catcode `\$12\catcode
  `\&12\catcode `\#12\catcode `\^12\catcode `\_12\catcode `\%12\relax}%
\providecommand \@@startlink[1]{}%
\providecommand \@@endlink[0]{}%
\providecommand \url  [0]{\begingroup\@sanitize@url \@url }%
\providecommand \@url [1]{\endgroup\@href {#1}{\urlprefix }}%
\providecommand \urlprefix  [0]{URL }%
\providecommand \Eprint [0]{\href }%
\providecommand \doibase [0]{http://dx.doi.org/}%
\providecommand \selectlanguage [0]{\@gobble}%
\providecommand \bibinfo  [0]{\@secondoftwo}%
\providecommand \bibfield  [0]{\@secondoftwo}%
\providecommand \translation [1]{[#1]}%
\providecommand \BibitemOpen [0]{}%
\providecommand \bibitemStop [0]{}%
\providecommand \bibitemNoStop [0]{.\EOS\space}%
\providecommand \EOS [0]{\spacefactor3000\relax}%
\providecommand \BibitemShut  [1]{\csname bibitem#1\endcsname}%
\let\auto@bib@innerbib\@empty
\bibitem [{\citenamefont {Radice}\ \emph {et~al.}(2020)\citenamefont {Radice},
  \citenamefont {Bernuzzi},\ and\ \citenamefont {Perego}}]{Radice:2020ddv}%
  \BibitemOpen
  \bibfield  {author} {\bibinfo {author} {\bibfnamefont {D.}~\bibnamefont
  {Radice}}, \bibinfo {author} {\bibfnamefont {S.}~\bibnamefont {Bernuzzi}}, \
  and\ \bibinfo {author} {\bibfnamefont {A.}~\bibnamefont {Perego}},\ }\href
  {\doibase 10.1146/annurev-nucl-013120-114541} {\bibfield  {journal} {\bibinfo
   {journal} {Ann. Rev. Nucl. Part. Sci.}\ }\textbf {\bibinfo {volume} {70}}
  (\bibinfo {year} {2020}),\ 10.1146/annurev-nucl-013120-114541},\ \Eprint
  {http://arxiv.org/abs/2002.03863} {arXiv:2002.03863 [astro-ph.HE]}
  \BibitemShut {NoStop}%
\bibitem [{\citenamefont {Li}\ and\ \citenamefont
  {Paczynski}(1998)}]{Li:1998bw}%
  \BibitemOpen
  \bibfield  {author} {\bibinfo {author} {\bibfnamefont {L.-X.}\ \bibnamefont
  {Li}}\ and\ \bibinfo {author} {\bibfnamefont {B.}~\bibnamefont {Paczynski}},\
  }\href {\doibase 10.1086/311680} {\bibfield  {journal} {\bibinfo  {journal}
  {Astrophys.J.}\ }\textbf {\bibinfo {volume} {507}},\ \bibinfo {pages} {L59}
  (\bibinfo {year} {1998})},\ \Eprint {http://arxiv.org/abs/astro-ph/9807272}
  {arXiv:astro-ph/9807272 [astro-ph]} \BibitemShut {NoStop}%
\bibitem [{\citenamefont {Lattimer}\ and\ \citenamefont
  {Schramm}(1974)}]{Lattimer:1974slx}%
  \BibitemOpen
  \bibfield  {author} {\bibinfo {author} {\bibfnamefont {J.~M.}\ \bibnamefont
  {Lattimer}}\ and\ \bibinfo {author} {\bibfnamefont {D.~N.}\ \bibnamefont
  {Schramm}},\ }\href {\doibase 10.1086/181612} {\bibfield  {journal} {\bibinfo
   {journal} {Astrophys. J.}\ }\textbf {\bibinfo {volume} {192}},\ \bibinfo
  {pages} {L145} (\bibinfo {year} {1974})}\BibitemShut {NoStop}%
\bibitem [{\citenamefont {Metzger}\ \emph {et~al.}(2010)\citenamefont
  {Metzger}, \citenamefont {Martinez-Pinedo}, \citenamefont {Darbha},
  \citenamefont {Quataert}, \citenamefont {Arcones} \emph
  {et~al.}}]{Metzger:2010sy}%
  \BibitemOpen
  \bibfield  {author} {\bibinfo {author} {\bibfnamefont {B.}~\bibnamefont
  {Metzger}}, \bibinfo {author} {\bibfnamefont {G.}~\bibnamefont
  {Martinez-Pinedo}}, \bibinfo {author} {\bibfnamefont {S.}~\bibnamefont
  {Darbha}}, \bibinfo {author} {\bibfnamefont {E.}~\bibnamefont {Quataert}},
  \bibinfo {author} {\bibfnamefont {A.}~\bibnamefont {Arcones}},  \emph
  {et~al.},\ }\href {\doibase 10.1111/j.1365-2966.2010.16864.x} {\bibfield
  {journal} {\bibinfo  {journal} {Mon.Not.Roy.Astron.Soc.}\ }\textbf {\bibinfo
  {volume} {406}},\ \bibinfo {pages} {2650} (\bibinfo {year} {2010})},\ \Eprint
  {http://arxiv.org/abs/1001.5029} {arXiv:1001.5029 [astro-ph.HE]} \BibitemShut
  {NoStop}%
\bibitem [{\citenamefont {Abbott}\ \emph {et~al.}(2018)\citenamefont {Abbott}
  \emph {et~al.}}]{Abbott:2018hgk}%
  \BibitemOpen
  \bibfield  {author} {\bibinfo {author} {\bibfnamefont {B.~P.}\ \bibnamefont
  {Abbott}} \emph {et~al.} (\bibinfo {collaboration} {LIGO Scientific,
  Virgo}),\ }\href {\doibase 10.3847/1538-4357/ab0f3d} {\  (\bibinfo {year}
  {2018}),\ 10.3847/1538-4357/ab0f3d},\ \Eprint
  {http://arxiv.org/abs/1810.02581} {arXiv:1810.02581 [gr-qc]} \BibitemShut
  {NoStop}%
\bibitem [{\citenamefont {Abbott}\ \emph {et~al.}(2019)\citenamefont {Abbott}
  \emph {et~al.}}]{Abbott:2018wiz}%
  \BibitemOpen
  \bibfield  {author} {\bibinfo {author} {\bibfnamefont {B.~P.}\ \bibnamefont
  {Abbott}} \emph {et~al.} (\bibinfo {collaboration} {LIGO Scientific,
  Virgo}),\ }\href {\doibase 10.1103/PhysRevX.9.011001} {\bibfield  {journal}
  {\bibinfo  {journal} {Phys. Rev.}\ }\textbf {\bibinfo {volume} {X9}},\
  \bibinfo {pages} {011001} (\bibinfo {year} {2019})},\ \Eprint
  {http://arxiv.org/abs/1805.11579} {arXiv:1805.11579 [gr-qc]} \BibitemShut
  {NoStop}%
\bibitem [{\citenamefont {Abbott}\ \emph {et~al.}(2017)\citenamefont {Abbott}
  \emph {et~al.}}]{GBM:2017lvd}%
  \BibitemOpen
  \bibfield  {author} {\bibinfo {author} {\bibfnamefont {B.~P.}\ \bibnamefont
  {Abbott}} \emph {et~al.} (\bibinfo {collaboration} {GROND, SALT Group,
  OzGrav, DFN, INTEGRAL, Virgo, Insight-Hxmt, MAXI Team, Fermi-LAT, J-GEM,
  RATIR, IceCube, CAASTRO, LWA, ePESSTO, GRAWITA, RIMAS, SKA South
  Africa/MeerKAT, H.E.S.S., 1M2H Team, IKI-GW Follow-up, Fermi GBM, Pi of Sky,
  DWF (Deeper Wider Faster Program), Dark Energy Survey, MASTER, AstroSat
  Cadmium Zinc Telluride Imager Team, Swift, Pierre Auger, ASKAP, VINROUGE,
  JAGWAR, Chandra Team at McGill University, TTU-NRAO, GROWTH, AGILE Team, MWA,
  ATCA, AST3, TOROS, Pan-STARRS, NuSTAR, ATLAS Telescopes, BOOTES, CaltechNRAO,
  LIGO Scientific, High Time Resolution Universe Survey, Nordic Optical
  Telescope, Las Cumbres Observatory Group, TZAC Consortium, LOFAR, IPN, DLT40,
  Texas Tech University, HAWC, ANTARES, KU, Dark Energy Camera GW-EM, CALET,
  Euro VLBI Team, ALMA}),\ }\href {\doibase 10.3847/2041-8213/aa91c9}
  {\bibfield  {journal} {\bibinfo  {journal} {Astrophys. J.}\ }\textbf
  {\bibinfo {volume} {848}},\ \bibinfo {pages} {L12} (\bibinfo {year}
  {2017})},\ \Eprint {http://arxiv.org/abs/1710.05833} {arXiv:1710.05833
  [astro-ph.HE]} \BibitemShut {NoStop}%
\bibitem [{\citenamefont {Coulter}\ \emph {et~al.}(2017)\citenamefont {Coulter}
  \emph {et~al.}}]{Coulter:2017wya}%
  \BibitemOpen
  \bibfield  {author} {\bibinfo {author} {\bibfnamefont {D.~A.}\ \bibnamefont
  {Coulter}} \emph {et~al.},\ }\href {\doibase 10.1126/science.aap9811}
  {\bibfield  {journal} {\bibinfo  {journal} {Science}\ }\textbf {\bibinfo
  {volume} {358}},\ \bibinfo {pages} {1556} (\bibinfo {year} {2017})},\ \Eprint
  {http://arxiv.org/abs/1710.05452} {arXiv:1710.05452 [astro-ph.HE]}
  \BibitemShut {NoStop}%
\bibitem [{\citenamefont {Arcavi}\ \emph {et~al.}(2017)\citenamefont {Arcavi},
  \citenamefont {McCully}, \citenamefont {Hosseinzadeh}, \citenamefont
  {Howell}, \citenamefont {Vasylyev}, \citenamefont {Poznanski}, \citenamefont
  {Zaltzman}, \citenamefont {Maoz}, \citenamefont {Singer}, \citenamefont
  {Valenti}, \citenamefont {Kasen}, \citenamefont {Barnes}, \citenamefont
  {Piran},\ and\ \citenamefont {fai Fong}}]{Arcavi:2017a}%
  \BibitemOpen
  \bibfield  {author} {\bibinfo {author} {\bibfnamefont {I.}~\bibnamefont
  {Arcavi}}, \bibinfo {author} {\bibfnamefont {C.}~\bibnamefont {McCully}},
  \bibinfo {author} {\bibfnamefont {G.}~\bibnamefont {Hosseinzadeh}}, \bibinfo
  {author} {\bibfnamefont {D.~A.}\ \bibnamefont {Howell}}, \bibinfo {author}
  {\bibfnamefont {S.}~\bibnamefont {Vasylyev}}, \bibinfo {author}
  {\bibfnamefont {D.}~\bibnamefont {Poznanski}}, \bibinfo {author}
  {\bibfnamefont {M.}~\bibnamefont {Zaltzman}}, \bibinfo {author}
  {\bibfnamefont {D.}~\bibnamefont {Maoz}}, \bibinfo {author} {\bibfnamefont
  {L.}~\bibnamefont {Singer}}, \bibinfo {author} {\bibfnamefont
  {S.}~\bibnamefont {Valenti}}, \bibinfo {author} {\bibfnamefont
  {D.}~\bibnamefont {Kasen}}, \bibinfo {author} {\bibfnamefont
  {J.}~\bibnamefont {Barnes}}, \bibinfo {author} {\bibfnamefont
  {T.}~\bibnamefont {Piran}}, \ and\ \bibinfo {author} {\bibfnamefont
  {W.}~\bibnamefont {fai Fong}},\ }\href {\doibase 10.3847/2041-8213/aa910f}
  {\bibfield  {journal} {\bibinfo  {journal} {The Astrophysical Journal}\
  }\textbf {\bibinfo {volume} {848}},\ \bibinfo {pages} {L33} (\bibinfo {year}
  {2017})}\BibitemShut {NoStop}%
\bibitem [{\citenamefont {Cowperthwaite}\ \emph {et~al.}(2017)\citenamefont
  {Cowperthwaite} \emph {et~al.}}]{Cowperthwaite:2017dyu}%
  \BibitemOpen
  \bibfield  {author} {\bibinfo {author} {\bibfnamefont {P.~S.}\ \bibnamefont
  {Cowperthwaite}} \emph {et~al.},\ }\href {\doibase 10.3847/2041-8213/aa8fc7}
  {\bibfield  {journal} {\bibinfo  {journal} {Astrophys. J.}\ }\textbf
  {\bibinfo {volume} {848}},\ \bibinfo {pages} {L17} (\bibinfo {year}
  {2017})},\ \Eprint {http://arxiv.org/abs/1710.05840} {arXiv:1710.05840
  [astro-ph.HE]} \BibitemShut {NoStop}%
\bibitem [{\citenamefont {Díaz}\ \emph {et~al.}(2017)\citenamefont {Díaz}
  \emph {et~al.}}]{Diaz:2017uch}%
  \BibitemOpen
  \bibfield  {author} {\bibinfo {author} {\bibfnamefont {M.~C.}\ \bibnamefont
  {Díaz}} \emph {et~al.} (\bibinfo {collaboration} {TOROS}),\ }\href@noop {}
  {\bibfield  {journal} {\bibinfo  {journal} {Submitted to: Astrophys. J.}\ }
  (\bibinfo {year} {2017})},\ \Eprint {http://arxiv.org/abs/1710.05844}
  {arXiv:1710.05844 [astro-ph.HE]} \BibitemShut {NoStop}%
\bibitem [{\citenamefont {Drout}\ \emph {et~al.}(2017)\citenamefont {Drout}
  \emph {et~al.}}]{Drout:2017ijr}%
  \BibitemOpen
  \bibfield  {author} {\bibinfo {author} {\bibfnamefont {M.~R.}\ \bibnamefont
  {Drout}} \emph {et~al.},\ }\href {\doibase 10.1126/science.aaq0049}
  {\bibfield  {journal} {\bibinfo  {journal} {Science}\ }\textbf {\bibinfo
  {volume} {358}},\ \bibinfo {pages} {1570} (\bibinfo {year} {2017})},\ \Eprint
  {http://arxiv.org/abs/1710.05443} {arXiv:1710.05443 [astro-ph.HE]}
  \BibitemShut {NoStop}%
\bibitem [{\citenamefont {Evans}\ \emph {et~al.}(2017)\citenamefont {Evans}
  \emph {et~al.}}]{Evans:2017mmy}%
  \BibitemOpen
  \bibfield  {author} {\bibinfo {author} {\bibfnamefont {P.~A.}\ \bibnamefont
  {Evans}} \emph {et~al.},\ }\href {\doibase 10.1126/science.aap9580}
  {\bibfield  {journal} {\bibinfo  {journal} {Science}\ }\textbf {\bibinfo
  {volume} {358}},\ \bibinfo {pages} {1565} (\bibinfo {year} {2017})},\ \Eprint
  {http://arxiv.org/abs/1710.05437} {arXiv:1710.05437 [astro-ph.HE]}
  \BibitemShut {NoStop}%
\bibitem [{\citenamefont {Kasliwal}\ \emph {et~al.}(2017)\citenamefont
  {Kasliwal} \emph {et~al.}}]{Kasliwal:2017ngb}%
  \BibitemOpen
  \bibfield  {author} {\bibinfo {author} {\bibfnamefont {M.~M.}\ \bibnamefont
  {Kasliwal}} \emph {et~al.},\ }\href {\doibase 10.1126/science.aap9455}
  {\bibfield  {journal} {\bibinfo  {journal} {Science}\ }\textbf {\bibinfo
  {volume} {358}},\ \bibinfo {pages} {1559} (\bibinfo {year} {2017})},\ \Eprint
  {http://arxiv.org/abs/1710.05436} {arXiv:1710.05436 [astro-ph.HE]}
  \BibitemShut {NoStop}%
\bibitem [{\citenamefont {Pian}\ \emph {et~al.}(2017)\citenamefont {Pian} \emph
  {et~al.}}]{Pian:2017gtc}%
  \BibitemOpen
  \bibfield  {author} {\bibinfo {author} {\bibfnamefont {E.}~\bibnamefont
  {Pian}} \emph {et~al.},\ }\href {\doibase 10.1038/nature24298} {\bibfield
  {journal} {\bibinfo  {journal} {Nature}\ } (\bibinfo {year} {2017}),\
  10.1038/nature24298},\ \Eprint {http://arxiv.org/abs/1710.05858}
  {arXiv:1710.05858 [astro-ph.HE]} \BibitemShut {NoStop}%
\bibitem [{\citenamefont {Smartt}\ \emph {et~al.}(2017)\citenamefont {Smartt}
  \emph {et~al.}}]{Smartt:2017fuw}%
  \BibitemOpen
  \bibfield  {author} {\bibinfo {author} {\bibfnamefont {S.~J.}\ \bibnamefont
  {Smartt}} \emph {et~al.},\ }\href {\doibase 10.1038/nature24303} {\bibfield
  {journal} {\bibinfo  {journal} {Nature}\ } (\bibinfo {year} {2017}),\
  10.1038/nature24303},\ \Eprint {http://arxiv.org/abs/1710.05841}
  {arXiv:1710.05841 [astro-ph.HE]} \BibitemShut {NoStop}%
\bibitem [{\citenamefont {Troja}\ \emph {et~al.}(2017)\citenamefont {Troja}
  \emph {et~al.}}]{Troja:2017nqp}%
  \BibitemOpen
  \bibfield  {author} {\bibinfo {author} {\bibfnamefont {E.}~\bibnamefont
  {Troja}} \emph {et~al.},\ }\href {\doibase 10.1038/nature24290} {\bibfield
  {journal} {\bibinfo  {journal} {Nature}\ } (\bibinfo {year} {2017}),\
  10.1038/nature24290},\ \Eprint {http://arxiv.org/abs/1710.05433}
  {arXiv:1710.05433 [astro-ph.HE]} \BibitemShut {NoStop}%
\bibitem [{\citenamefont {Utsumi}\ \emph {et~al.}(2017)\citenamefont {Utsumi}
  \emph {et~al.}}]{Utsumi:2017cti}%
  \BibitemOpen
  \bibfield  {author} {\bibinfo {author} {\bibfnamefont {Y.}~\bibnamefont
  {Utsumi}} \emph {et~al.} (\bibinfo {collaboration} {J-GEM}),\ }\href
  {\doibase 10.1093/pasj/psx118} {\bibfield  {journal} {\bibinfo  {journal}
  {Publ.\ Astron.\ Soc.\ Jap.}\ }\textbf {\bibinfo {volume} {69}},\ \bibinfo
  {pages} {101} (\bibinfo {year} {2017})},\ \Eprint
  {http://arxiv.org/abs/1710.05848} {arXiv:1710.05848 [astro-ph.HE]}
  \BibitemShut {NoStop}%
\bibitem [{\citenamefont {Valenti}\ \emph {et~al.}(2017)\citenamefont
  {Valenti}, \citenamefont {Sand}, \citenamefont {Yang}, \citenamefont
  {Cappellaro}, \citenamefont {Tartaglia}, \citenamefont {Corsi}, \citenamefont
  {Jha}, \citenamefont {Reichart}, \citenamefont {Haislip},\ and\ \citenamefont
  {Kouprianov}}]{Valenti:2017ngx}%
  \BibitemOpen
  \bibfield  {author} {\bibinfo {author} {\bibfnamefont {S.}~\bibnamefont
  {Valenti}}, \bibinfo {author} {\bibfnamefont {D.~J.}\ \bibnamefont {Sand}},
  \bibinfo {author} {\bibfnamefont {S.}~\bibnamefont {Yang}}, \bibinfo {author}
  {\bibfnamefont {E.}~\bibnamefont {Cappellaro}}, \bibinfo {author}
  {\bibfnamefont {L.}~\bibnamefont {Tartaglia}}, \bibinfo {author}
  {\bibfnamefont {A.}~\bibnamefont {Corsi}}, \bibinfo {author} {\bibfnamefont
  {S.~W.}\ \bibnamefont {Jha}}, \bibinfo {author} {\bibfnamefont {D.~E.}\
  \bibnamefont {Reichart}}, \bibinfo {author} {\bibfnamefont {J.}~\bibnamefont
  {Haislip}}, \ and\ \bibinfo {author} {\bibfnamefont {V.}~\bibnamefont
  {Kouprianov}},\ }\href {\doibase 10.3847/2041-8213/aa8edf} {\bibfield
  {journal} {\bibinfo  {journal} {Astrophys. J.}\ }\textbf {\bibinfo {volume}
  {848}},\ \bibinfo {pages} {L24} (\bibinfo {year} {2017})},\ \Eprint
  {http://arxiv.org/abs/1710.05854} {arXiv:1710.05854 [astro-ph.HE]}
  \BibitemShut {NoStop}%
\bibitem [{\citenamefont {Villar}\ \emph {et~al.}(2017)\citenamefont {Villar}
  \emph {et~al.}}]{Villar:2017wcc}%
  \BibitemOpen
  \bibfield  {author} {\bibinfo {author} {\bibfnamefont {V.~A.}\ \bibnamefont
  {Villar}} \emph {et~al.},\ }\href {\doibase 10.3847/2041-8213/aa9c84}
  {\bibfield  {journal} {\bibinfo  {journal} {Astrophys. J.}\ }\textbf
  {\bibinfo {volume} {851}},\ \bibinfo {pages} {L21} (\bibinfo {year}
  {2017})},\ \Eprint {http://arxiv.org/abs/1710.11576} {arXiv:1710.11576
  [astro-ph.HE]} \BibitemShut {NoStop}%
\bibitem [{\citenamefont {Bernuzzi}(2020)}]{Bernuzzi:2020tgt}%
  \BibitemOpen
  \bibfield  {author} {\bibinfo {author} {\bibfnamefont {S.}~\bibnamefont
  {Bernuzzi}},\ }\href {\doibase 10.1007/s10714-020-02752-5} {\bibfield
  {journal} {\bibinfo  {journal} {Gen. Rel. Grav.}\ }\textbf {\bibinfo {volume}
  {52}},\ \bibinfo {pages} {108} (\bibinfo {year} {2020})},\ \Eprint
  {http://arxiv.org/abs/2004.06419} {arXiv:2004.06419 [astro-ph.HE]}
  \BibitemShut {NoStop}%
\bibitem [{\citenamefont {Radice}\ \emph {et~al.}(2018)\citenamefont {Radice},
  \citenamefont {Perego}, \citenamefont {Hotokezaka}, \citenamefont {Fromm},
  \citenamefont {Bernuzzi},\ and\ \citenamefont {Roberts}}]{Radice:2018pdn}%
  \BibitemOpen
  \bibfield  {author} {\bibinfo {author} {\bibfnamefont {D.}~\bibnamefont
  {Radice}}, \bibinfo {author} {\bibfnamefont {A.}~\bibnamefont {Perego}},
  \bibinfo {author} {\bibfnamefont {K.}~\bibnamefont {Hotokezaka}}, \bibinfo
  {author} {\bibfnamefont {S.~A.}\ \bibnamefont {Fromm}}, \bibinfo {author}
  {\bibfnamefont {S.}~\bibnamefont {Bernuzzi}}, \ and\ \bibinfo {author}
  {\bibfnamefont {L.~F.}\ \bibnamefont {Roberts}},\ }\href {\doibase
  10.3847/1538-4357/aaf054} {\bibfield  {journal} {\bibinfo  {journal}
  {Astrophys. J.}\ }\textbf {\bibinfo {volume} {869}},\ \bibinfo {pages} {130}
  (\bibinfo {year} {2018})},\ \Eprint {http://arxiv.org/abs/1809.11161}
  {arXiv:1809.11161 [astro-ph.HE]} \BibitemShut {NoStop}%
\bibitem [{\citenamefont {Rosswog}\ \emph {et~al.}(1999)\citenamefont
  {Rosswog}, \citenamefont {Liebendoerfer}, \citenamefont {Thielemann},
  \citenamefont {Davies}, \citenamefont {Benz} \emph
  {et~al.}}]{Rosswog:1998hy}%
  \BibitemOpen
  \bibfield  {author} {\bibinfo {author} {\bibfnamefont {S.}~\bibnamefont
  {Rosswog}}, \bibinfo {author} {\bibfnamefont {M.}~\bibnamefont
  {Liebendoerfer}}, \bibinfo {author} {\bibfnamefont {F.}~\bibnamefont
  {Thielemann}}, \bibinfo {author} {\bibfnamefont {M.}~\bibnamefont {Davies}},
  \bibinfo {author} {\bibfnamefont {W.}~\bibnamefont {Benz}},  \emph {et~al.},\
  }\href@noop {} {\bibfield  {journal} {\bibinfo  {journal}
  {Astron.Astrophys.}\ }\textbf {\bibinfo {volume} {341}},\ \bibinfo {pages}
  {499} (\bibinfo {year} {1999})},\ \Eprint
  {http://arxiv.org/abs/astro-ph/9811367} {arXiv:astro-ph/9811367 [astro-ph]}
  \BibitemShut {NoStop}%
\bibitem [{\citenamefont {Rosswog}(2013)}]{Rosswog:2013}%
  \BibitemOpen
  \bibfield  {author} {\bibinfo {author} {\bibfnamefont {S.}~\bibnamefont
  {Rosswog}},\ }\href@noop {} {\bibfield  {journal} {\bibinfo  {journal}
  {Philosophical Transactions A}\ }\textbf {\bibinfo {volume} {373}},\ \bibinfo
  {pages} {2036} (\bibinfo {year} {2013})}\BibitemShut {NoStop}%
\bibitem [{\citenamefont {Perego}\ \emph {et~al.}(2019)\citenamefont {Perego},
  \citenamefont {Bernuzzi},\ and\ \citenamefont {Radice}}]{Perego:2019adq}%
  \BibitemOpen
  \bibfield  {author} {\bibinfo {author} {\bibfnamefont {A.}~\bibnamefont
  {Perego}}, \bibinfo {author} {\bibfnamefont {S.}~\bibnamefont {Bernuzzi}}, \
  and\ \bibinfo {author} {\bibfnamefont {D.}~\bibnamefont {Radice}},\ }\href
  {\doibase 10.1140/epja/i2019-12810-7} {\bibfield  {journal} {\bibinfo
  {journal} {Eur. Phys. J.}\ }\textbf {\bibinfo {volume} {A55}},\ \bibinfo
  {pages} {124} (\bibinfo {year} {2019})},\ \Eprint
  {http://arxiv.org/abs/1903.07898} {arXiv:1903.07898 [gr-qc]} \BibitemShut
  {NoStop}%
\bibitem [{\citenamefont {Nedora}\ \emph {et~al.}(2019)\citenamefont {Nedora},
  \citenamefont {Bernuzzi}, \citenamefont {Radice}, \citenamefont {Perego},
  \citenamefont {Endrizzi},\ and\ \citenamefont {Ortiz}}]{Nedora:2019jhl}%
  \BibitemOpen
  \bibfield  {author} {\bibinfo {author} {\bibfnamefont {V.}~\bibnamefont
  {Nedora}}, \bibinfo {author} {\bibfnamefont {S.}~\bibnamefont {Bernuzzi}},
  \bibinfo {author} {\bibfnamefont {D.}~\bibnamefont {Radice}}, \bibinfo
  {author} {\bibfnamefont {A.}~\bibnamefont {Perego}}, \bibinfo {author}
  {\bibfnamefont {A.}~\bibnamefont {Endrizzi}}, \ and\ \bibinfo {author}
  {\bibfnamefont {N.}~\bibnamefont {Ortiz}},\ }\href {\doibase
  10.3847/2041-8213/ab5794} {\bibfield  {journal} {\bibinfo  {journal}
  {Astrophys. J.}\ }\textbf {\bibinfo {volume} {886}},\ \bibinfo {pages} {L30}
  (\bibinfo {year} {2019})},\ \Eprint {http://arxiv.org/abs/1907.04872}
  {arXiv:1907.04872 [astro-ph.HE]} \BibitemShut {NoStop}%
\bibitem [{\citenamefont {Radice}\ and\ \citenamefont
  {Bernuzzi}(2024)}]{Radice:2023xxn}%
  \BibitemOpen
  \bibfield  {author} {\bibinfo {author} {\bibfnamefont {D.}~\bibnamefont
  {Radice}}\ and\ \bibinfo {author} {\bibfnamefont {S.}~\bibnamefont
  {Bernuzzi}},\ }\href {\doibase 10.1088/1742-6596/2742/1/012009} {\bibfield
  {journal} {\bibinfo  {journal} {J. Phys. Conf. Ser.}\ }\textbf {\bibinfo
  {volume} {2742}},\ \bibinfo {pages} {012009} (\bibinfo {year} {2024})},\
  \Eprint {http://arxiv.org/abs/2310.09934} {arXiv:2310.09934 [astro-ph.HE]}
  \BibitemShut {NoStop}%
\bibitem [{\citenamefont {Combi}\ and\ \citenamefont
  {Siegel}(2023)}]{Combi:2022nhg}%
  \BibitemOpen
  \bibfield  {author} {\bibinfo {author} {\bibfnamefont {L.}~\bibnamefont
  {Combi}}\ and\ \bibinfo {author} {\bibfnamefont {D.~M.}\ \bibnamefont
  {Siegel}},\ }\href {\doibase 10.3847/1538-4357/acac29} {\bibfield  {journal}
  {\bibinfo  {journal} {Astrophys. J.}\ }\textbf {\bibinfo {volume} {944}},\
  \bibinfo {pages} {28} (\bibinfo {year} {2023})},\ \Eprint
  {http://arxiv.org/abs/2206.03618} {arXiv:2206.03618 [astro-ph.HE]}
  \BibitemShut {NoStop}%
\bibitem [{\citenamefont {Dessart}\ \emph {et~al.}(2009)\citenamefont
  {Dessart}, \citenamefont {Ott}, \citenamefont {Burrows}, \citenamefont
  {Rosswog},\ and\ \citenamefont {Livne}}]{Dessart:2008zd}%
  \BibitemOpen
  \bibfield  {author} {\bibinfo {author} {\bibfnamefont {L.}~\bibnamefont
  {Dessart}}, \bibinfo {author} {\bibfnamefont {C.}~\bibnamefont {Ott}},
  \bibinfo {author} {\bibfnamefont {A.}~\bibnamefont {Burrows}}, \bibinfo
  {author} {\bibfnamefont {S.}~\bibnamefont {Rosswog}}, \ and\ \bibinfo
  {author} {\bibfnamefont {E.}~\bibnamefont {Livne}},\ }\href {\doibase
  10.1088/0004-637X/690/2/1681} {\bibfield  {journal} {\bibinfo  {journal}
  {Astrophys.J.}\ }\textbf {\bibinfo {volume} {690}},\ \bibinfo {pages} {1681}
  (\bibinfo {year} {2009})},\ \Eprint {http://arxiv.org/abs/0806.4380}
  {arXiv:0806.4380 [astro-ph]} \BibitemShut {NoStop}%
\bibitem [{\citenamefont {Perego}\ \emph {et~al.}(2014)\citenamefont {Perego},
  \citenamefont {Rosswog}, \citenamefont {Cabezon}, \citenamefont {Korobkin},
  \citenamefont {Kaeppeli} \emph {et~al.}}]{Perego:2014fma}%
  \BibitemOpen
  \bibfield  {author} {\bibinfo {author} {\bibfnamefont {A.}~\bibnamefont
  {Perego}}, \bibinfo {author} {\bibfnamefont {S.}~\bibnamefont {Rosswog}},
  \bibinfo {author} {\bibfnamefont {R.}~\bibnamefont {Cabezon}}, \bibinfo
  {author} {\bibfnamefont {O.}~\bibnamefont {Korobkin}}, \bibinfo {author}
  {\bibfnamefont {R.}~\bibnamefont {Kaeppeli}},  \emph {et~al.},\ }\href
  {\doibase 10.1093/mnras/stu1352} {\bibfield  {journal} {\bibinfo  {journal}
  {Mon.Not.Roy.Astron.Soc.}\ }\textbf {\bibinfo {volume} {443}},\ \bibinfo
  {pages} {3134} (\bibinfo {year} {2014})},\ \Eprint
  {http://arxiv.org/abs/1405.6730} {arXiv:1405.6730 [astro-ph.HE]} \BibitemShut
  {NoStop}%
\bibitem [{\citenamefont {Martin}\ \emph {et~al.}(2015)\citenamefont {Martin},
  \citenamefont {Perego}, \citenamefont {Arcones}, \citenamefont {Thielemann},
  \citenamefont {Korobkin},\ and\ \citenamefont {Rosswog}}]{Martin:2015hxa}%
  \BibitemOpen
  \bibfield  {author} {\bibinfo {author} {\bibfnamefont {D.}~\bibnamefont
  {Martin}}, \bibinfo {author} {\bibfnamefont {A.}~\bibnamefont {Perego}},
  \bibinfo {author} {\bibfnamefont {A.}~\bibnamefont {Arcones}}, \bibinfo
  {author} {\bibfnamefont {F.-K.}\ \bibnamefont {Thielemann}}, \bibinfo
  {author} {\bibfnamefont {O.}~\bibnamefont {Korobkin}}, \ and\ \bibinfo
  {author} {\bibfnamefont {S.}~\bibnamefont {Rosswog}},\ }\href {\doibase
  10.1088/0004-637X/813/1/2} {\bibfield  {journal} {\bibinfo  {journal}
  {Astrophys. J.}\ }\textbf {\bibinfo {volume} {813}},\ \bibinfo {pages} {2}
  (\bibinfo {year} {2015})},\ \Eprint {http://arxiv.org/abs/1506.05048}
  {arXiv:1506.05048 [astro-ph.SR]} \BibitemShut {NoStop}%
\bibitem [{\citenamefont {Kiuchi}\ \emph {et~al.}(2023)\citenamefont {Kiuchi},
  \citenamefont {Fujibayashi}, \citenamefont {Hayashi}, \citenamefont
  {Kyutoku}, \citenamefont {Sekiguchi},\ and\ \citenamefont
  {Shibata}}]{Kiuchi:2022nin}%
  \BibitemOpen
  \bibfield  {author} {\bibinfo {author} {\bibfnamefont {K.}~\bibnamefont
  {Kiuchi}}, \bibinfo {author} {\bibfnamefont {S.}~\bibnamefont {Fujibayashi}},
  \bibinfo {author} {\bibfnamefont {K.}~\bibnamefont {Hayashi}}, \bibinfo
  {author} {\bibfnamefont {K.}~\bibnamefont {Kyutoku}}, \bibinfo {author}
  {\bibfnamefont {Y.}~\bibnamefont {Sekiguchi}}, \ and\ \bibinfo {author}
  {\bibfnamefont {M.}~\bibnamefont {Shibata}},\ }\href {\doibase
  10.1103/PhysRevLett.131.011401} {\bibfield  {journal} {\bibinfo  {journal}
  {Phys. Rev. Lett.}\ }\textbf {\bibinfo {volume} {131}},\ \bibinfo {pages}
  {011401} (\bibinfo {year} {2023})},\ \Eprint
  {http://arxiv.org/abs/2211.07637} {arXiv:2211.07637 [astro-ph.HE]}
  \BibitemShut {NoStop}%
\bibitem [{\citenamefont {Goriely}\ \emph {et~al.}(2011)\citenamefont
  {Goriely}, \citenamefont {Bauswein},\ and\ \citenamefont
  {Janka}}]{Goriely:2011vg}%
  \BibitemOpen
  \bibfield  {author} {\bibinfo {author} {\bibfnamefont {S.}~\bibnamefont
  {Goriely}}, \bibinfo {author} {\bibfnamefont {A.}~\bibnamefont {Bauswein}}, \
  and\ \bibinfo {author} {\bibfnamefont {H.-T.}\ \bibnamefont {Janka}},\ }\href
  {\doibase 10.1088/2041-8205/738/2/L32} {\bibfield  {journal} {\bibinfo
  {journal} {Astrophys.J.}\ }\textbf {\bibinfo {volume} {738}},\ \bibinfo
  {pages} {L32} (\bibinfo {year} {2011})},\ \Eprint
  {http://arxiv.org/abs/1107.0899} {arXiv:1107.0899 [astro-ph.SR]} \BibitemShut
  {NoStop}%
\bibitem [{\citenamefont {Korobkin}\ \emph {et~al.}(2012)\citenamefont
  {Korobkin}, \citenamefont {Rosswog}, \citenamefont {Arcones},\ and\
  \citenamefont {Winteler}}]{Korobkin:2012uy}%
  \BibitemOpen
  \bibfield  {author} {\bibinfo {author} {\bibfnamefont {O.}~\bibnamefont
  {Korobkin}}, \bibinfo {author} {\bibfnamefont {S.}~\bibnamefont {Rosswog}},
  \bibinfo {author} {\bibfnamefont {A.}~\bibnamefont {Arcones}}, \ and\
  \bibinfo {author} {\bibfnamefont {C.}~\bibnamefont {Winteler}},\ }\href
  {\doibase 10.1111/j.1365-2966.2012.21859.x} {\bibfield  {journal} {\bibinfo
  {journal} {Mon. Not. Roy. Astron. Soc.}\ }\textbf {\bibinfo {volume} {426}},\
  \bibinfo {pages} {1940} (\bibinfo {year} {2012})},\ \Eprint
  {http://arxiv.org/abs/1206.2379} {arXiv:1206.2379 [astro-ph.SR]} \BibitemShut
  {NoStop}%
\bibitem [{\citenamefont {Jacobi}\ \emph {et~al.}(2026)\citenamefont {Jacobi},
  \citenamefont {Magistrelli}, \citenamefont {Loffredo}, \citenamefont
  {Ricigliano}, \citenamefont {Chiesa}, \citenamefont {Bernuzzi}, \citenamefont
  {Perego},\ and\ \citenamefont {Arcones}}]{Jacobi:2025eak}%
  \BibitemOpen
  \bibfield  {author} {\bibinfo {author} {\bibfnamefont {M.}~\bibnamefont
  {Jacobi}}, \bibinfo {author} {\bibfnamefont {F.}~\bibnamefont {Magistrelli}},
  \bibinfo {author} {\bibfnamefont {E.}~\bibnamefont {Loffredo}}, \bibinfo
  {author} {\bibfnamefont {G.}~\bibnamefont {Ricigliano}}, \bibinfo {author}
  {\bibfnamefont {L.}~\bibnamefont {Chiesa}}, \bibinfo {author} {\bibfnamefont
  {S.}~\bibnamefont {Bernuzzi}}, \bibinfo {author} {\bibfnamefont
  {A.}~\bibnamefont {Perego}}, \ and\ \bibinfo {author} {\bibfnamefont
  {A.}~\bibnamefont {Arcones}},\ }\href {\doibase 10.3847/2041-8213/ae4104}
  {\bibfield  {journal} {\bibinfo  {journal} {Astrophys. J. Lett.}\ }\textbf
  {\bibinfo {volume} {999}},\ \bibinfo {pages} {L16} (\bibinfo {year}
  {2026})},\ \Eprint {http://arxiv.org/abs/2503.17445} {arXiv:2503.17445
  [astro-ph.HE]} \BibitemShut {NoStop}%
\bibitem [{\citenamefont {Beloborodov}(2003)}]{Beloborodov:2002af}%
  \BibitemOpen
  \bibfield  {author} {\bibinfo {author} {\bibfnamefont {A.~M.}\ \bibnamefont
  {Beloborodov}},\ }\href {\doibase 10.1086/374217} {\bibfield  {journal}
  {\bibinfo  {journal} {Astrophys. J.}\ }\textbf {\bibinfo {volume} {588}},\
  \bibinfo {pages} {931} (\bibinfo {year} {2003})},\ \Eprint
  {http://arxiv.org/abs/astro-ph/0210522} {arXiv:astro-ph/0210522} \BibitemShut
  {NoStop}%
\bibitem [{\citenamefont {Siegel}\ and\ \citenamefont
  {Metzger}(2017)}]{Siegel:2017nub}%
  \BibitemOpen
  \bibfield  {author} {\bibinfo {author} {\bibfnamefont {D.~M.}\ \bibnamefont
  {Siegel}}\ and\ \bibinfo {author} {\bibfnamefont {B.~D.}\ \bibnamefont
  {Metzger}},\ }\href {\doibase 10.1103/PhysRevLett.119.231102} {\bibfield
  {journal} {\bibinfo  {journal} {Phys. Rev. Lett.}\ }\textbf {\bibinfo
  {volume} {119}},\ \bibinfo {pages} {231102} (\bibinfo {year} {2017})},\
  \Eprint {http://arxiv.org/abs/1705.05473} {arXiv:1705.05473 [astro-ph.HE]}
  \BibitemShut {NoStop}%
\bibitem [{\citenamefont {Sprouse}\ \emph {et~al.}(2024)\citenamefont
  {Sprouse}, \citenamefont {Lund}, \citenamefont {Miller}, \citenamefont
  {McLaughlin},\ and\ \citenamefont {Mumpower}}]{Sprouse:2023cdm}%
  \BibitemOpen
  \bibfield  {author} {\bibinfo {author} {\bibfnamefont {T.~M.}\ \bibnamefont
  {Sprouse}}, \bibinfo {author} {\bibfnamefont {K.~A.}\ \bibnamefont {Lund}},
  \bibinfo {author} {\bibfnamefont {J.~M.}\ \bibnamefont {Miller}}, \bibinfo
  {author} {\bibfnamefont {G.~C.}\ \bibnamefont {McLaughlin}}, \ and\ \bibinfo
  {author} {\bibfnamefont {M.~R.}\ \bibnamefont {Mumpower}},\ }\href {\doibase
  10.3847/1538-4357/ad1819} {\bibfield  {journal} {\bibinfo  {journal}
  {Astrophys. J.}\ }\textbf {\bibinfo {volume} {962}},\ \bibinfo {pages} {79}
  (\bibinfo {year} {2024})},\ \Eprint {http://arxiv.org/abs/2309.07966}
  {arXiv:2309.07966 [astro-ph.HE]} \BibitemShut {NoStop}%
\bibitem [{\citenamefont {Kasen}\ \emph {et~al.}(2015)\citenamefont {Kasen},
  \citenamefont {Fern{\'a}ndez},\ and\ \citenamefont
  {Metzger}}]{Kasen:2014toa}%
  \BibitemOpen
  \bibfield  {author} {\bibinfo {author} {\bibfnamefont {D.}~\bibnamefont
  {Kasen}}, \bibinfo {author} {\bibfnamefont {R.}~\bibnamefont
  {Fern{\'a}ndez}}, \ and\ \bibinfo {author} {\bibfnamefont {B.}~\bibnamefont
  {Metzger}},\ }\href {\doibase 10.1093/mnras/stv721} {\bibfield  {journal}
  {\bibinfo  {journal} {Mon. Not. Roy. Astron. Soc.}\ }\textbf {\bibinfo
  {volume} {450}},\ \bibinfo {pages} {1777} (\bibinfo {year} {2015})},\ \Eprint
  {http://arxiv.org/abs/1411.3726} {arXiv:1411.3726 [astro-ph.HE]} \BibitemShut
  {NoStop}%
\bibitem [{\citenamefont {Tanaka}\ \emph {et~al.}(2017)\citenamefont {Tanaka}
  \emph {et~al.}}]{Tanaka:2017qxj}%
  \BibitemOpen
  \bibfield  {author} {\bibinfo {author} {\bibfnamefont {M.}~\bibnamefont
  {Tanaka}} \emph {et~al.},\ }\href {\doibase 10.1093/pasj/psx121} {\bibfield
  {journal} {\bibinfo  {journal} {Publ. Astron. Soc. Jap.}\ } (\bibinfo {year}
  {2017}),\ 10.1093/pasj/psx121},\ \Eprint {http://arxiv.org/abs/1710.05850}
  {arXiv:1710.05850 [astro-ph.HE]} \BibitemShut {NoStop}%
\bibitem [{\citenamefont {Perego}\ \emph {et~al.}(2017)\citenamefont {Perego},
  \citenamefont {Radice},\ and\ \citenamefont {Bernuzzi}}]{Perego:2017wtu}%
  \BibitemOpen
  \bibfield  {author} {\bibinfo {author} {\bibfnamefont {A.}~\bibnamefont
  {Perego}}, \bibinfo {author} {\bibfnamefont {D.}~\bibnamefont {Radice}}, \
  and\ \bibinfo {author} {\bibfnamefont {S.}~\bibnamefont {Bernuzzi}},\ }\href
  {\doibase 10.3847/2041-8213/aa9ab9} {\bibfield  {journal} {\bibinfo
  {journal} {Astrophys. J.}\ }\textbf {\bibinfo {volume} {850}},\ \bibinfo
  {pages} {L37} (\bibinfo {year} {2017})},\ \Eprint
  {http://arxiv.org/abs/1711.03982} {arXiv:1711.03982 [astro-ph.HE]}
  \BibitemShut {NoStop}%
\bibitem [{\citenamefont {Nedora}\ \emph {et~al.}(2021)\citenamefont {Nedora},
  \citenamefont {Bernuzzi}, \citenamefont {Radice}, \citenamefont {Daszuta},
  \citenamefont {Endrizzi}, \citenamefont {Perego}, \citenamefont {Prakash},
  \citenamefont {Safarzadeh}, \citenamefont {Schianchi},\ and\ \citenamefont
  {Logoteta}}]{Nedora:2020hxc}%
  \BibitemOpen
  \bibfield  {author} {\bibinfo {author} {\bibfnamefont {V.}~\bibnamefont
  {Nedora}}, \bibinfo {author} {\bibfnamefont {S.}~\bibnamefont {Bernuzzi}},
  \bibinfo {author} {\bibfnamefont {D.}~\bibnamefont {Radice}}, \bibinfo
  {author} {\bibfnamefont {B.}~\bibnamefont {Daszuta}}, \bibinfo {author}
  {\bibfnamefont {A.}~\bibnamefont {Endrizzi}}, \bibinfo {author}
  {\bibfnamefont {A.}~\bibnamefont {Perego}}, \bibinfo {author} {\bibfnamefont
  {A.}~\bibnamefont {Prakash}}, \bibinfo {author} {\bibfnamefont
  {M.}~\bibnamefont {Safarzadeh}}, \bibinfo {author} {\bibfnamefont
  {F.}~\bibnamefont {Schianchi}}, \ and\ \bibinfo {author} {\bibfnamefont
  {D.}~\bibnamefont {Logoteta}},\ }\href {\doibase 10.3847/1538-4357/abc9be}
  {\bibfield  {journal} {\bibinfo  {journal} {Astrophys. J.}\ }\textbf
  {\bibinfo {volume} {906}},\ \bibinfo {pages} {98} (\bibinfo {year} {2021})},\
  \Eprint {http://arxiv.org/abs/2008.04333} {arXiv:2008.04333 [astro-ph.HE]}
  \BibitemShut {NoStop}%
\bibitem [{\citenamefont {Lippuner}\ and\ \citenamefont
  {Roberts}(2017)}]{Lippuner:2017tyn}%
  \BibitemOpen
  \bibfield  {author} {\bibinfo {author} {\bibfnamefont {J.}~\bibnamefont
  {Lippuner}}\ and\ \bibinfo {author} {\bibfnamefont {L.~F.}\ \bibnamefont
  {Roberts}},\ }\href {\doibase 10.3847/1538-4365/aa94cb} {\bibfield  {journal}
  {\bibinfo  {journal} {Astrophys. J. Suppl.}\ }\textbf {\bibinfo {volume}
  {233}},\ \bibinfo {pages} {18} (\bibinfo {year} {2017})},\ \Eprint
  {http://arxiv.org/abs/1706.06198} {arXiv:1706.06198 [astro-ph.HE]}
  \BibitemShut {NoStop}%
\bibitem [{\citenamefont {Reichert}\ \emph {et~al.}(2023)\citenamefont
  {Reichert} \emph {et~al.}}]{Reichert:2023xqy}%
  \BibitemOpen
  \bibfield  {author} {\bibinfo {author} {\bibfnamefont {M.}~\bibnamefont
  {Reichert}} \emph {et~al.},\ }\href {\doibase 10.3847/1538-4365/acf033}
  {\bibfield  {journal} {\bibinfo  {journal} {Astrophys. J. Suppl.}\ }\textbf
  {\bibinfo {volume} {268}},\ \bibinfo {pages} {66} (\bibinfo {year} {2023})},\
  \Eprint {http://arxiv.org/abs/2305.07048} {arXiv:2305.07048 [astro-ph.IM]}
  \BibitemShut {NoStop}%
\bibitem [{\citenamefont {Magistrelli}\ \emph {et~al.}(2024)\citenamefont
  {Magistrelli}, \citenamefont {Bernuzzi}, \citenamefont {Perego},\ and\
  \citenamefont {Radice}}]{Magistrelli:2024zmk}%
  \BibitemOpen
  \bibfield  {author} {\bibinfo {author} {\bibfnamefont {F.}~\bibnamefont
  {Magistrelli}}, \bibinfo {author} {\bibfnamefont {S.}~\bibnamefont
  {Bernuzzi}}, \bibinfo {author} {\bibfnamefont {A.}~\bibnamefont {Perego}}, \
  and\ \bibinfo {author} {\bibfnamefont {D.}~\bibnamefont {Radice}},\ }\href
  {\doibase 10.3847/2041-8213/ad74e0} {\bibfield  {journal} {\bibinfo
  {journal} {Astrophys. J. Lett.}\ }\textbf {\bibinfo {volume} {974}},\
  \bibinfo {pages} {L5} (\bibinfo {year} {2024})},\ \Eprint
  {http://arxiv.org/abs/2403.13883} {arXiv:2403.13883 [astro-ph.HE]}
  \BibitemShut {NoStop}%
\bibitem [{\citenamefont {Magistrelli}\ \emph {et~al.}(2026)\citenamefont
  {Magistrelli}, \citenamefont {Bernuzzi}, \citenamefont {Perego},
  \citenamefont {Jacobi},\ and\ \citenamefont {Fontes}}]{Magistrelli:2025xja}%
  \BibitemOpen
  \bibfield  {author} {\bibinfo {author} {\bibfnamefont {F.}~\bibnamefont
  {Magistrelli}}, \bibinfo {author} {\bibfnamefont {S.}~\bibnamefont
  {Bernuzzi}}, \bibinfo {author} {\bibfnamefont {A.}~\bibnamefont {Perego}},
  \bibinfo {author} {\bibfnamefont {M.}~\bibnamefont {Jacobi}}, \ and\ \bibinfo
  {author} {\bibfnamefont {C.~J.}\ \bibnamefont {Fontes}},\ }\href {\doibase
  10.1051/0004-6361/202558682} {\bibfield  {journal} {\bibinfo  {journal}
  {Astron. Astrophys.}\ }\textbf {\bibinfo {volume} {708}},\ \bibinfo {pages}
  {A126} (\bibinfo {year} {2026})},\ \Eprint {http://arxiv.org/abs/2512.11032}
  {arXiv:2512.11032 [astro-ph.HE]} \BibitemShut {NoStop}%
\bibitem [{\citenamefont {Radice}\ \emph {et~al.}(2016)\citenamefont {Radice},
  \citenamefont {Galeazzi}, \citenamefont {Lippuner}, \citenamefont {Roberts},
  \citenamefont {Ott},\ and\ \citenamefont {Rezzolla}}]{Radice:2016dwd}%
  \BibitemOpen
  \bibfield  {author} {\bibinfo {author} {\bibfnamefont {D.}~\bibnamefont
  {Radice}}, \bibinfo {author} {\bibfnamefont {F.}~\bibnamefont {Galeazzi}},
  \bibinfo {author} {\bibfnamefont {J.}~\bibnamefont {Lippuner}}, \bibinfo
  {author} {\bibfnamefont {L.~F.}\ \bibnamefont {Roberts}}, \bibinfo {author}
  {\bibfnamefont {C.~D.}\ \bibnamefont {Ott}}, \ and\ \bibinfo {author}
  {\bibfnamefont {L.}~\bibnamefont {Rezzolla}},\ }\href {\doibase
  10.1093/mnras/stw1227} {\bibfield  {journal} {\bibinfo  {journal} {Mon. Not.
  Roy. Astron. Soc.}\ }\textbf {\bibinfo {volume} {460}},\ \bibinfo {pages}
  {3255} (\bibinfo {year} {2016})},\ \Eprint {http://arxiv.org/abs/1601.02426}
  {arXiv:1601.02426 [astro-ph.HE]} \BibitemShut {NoStop}%
\bibitem [{\citenamefont {Radice}\ and\ \citenamefont
  {Dai}(2019)}]{Radice:2018ozg}%
  \BibitemOpen
  \bibfield  {author} {\bibinfo {author} {\bibfnamefont {D.}~\bibnamefont
  {Radice}}\ and\ \bibinfo {author} {\bibfnamefont {L.}~\bibnamefont {Dai}},\
  }\href {\doibase 10.1140/epja/i2019-12716-4} {\bibfield  {journal} {\bibinfo
  {journal} {Eur. Phys. J.}\ }\textbf {\bibinfo {volume} {A55}},\ \bibinfo
  {pages} {50} (\bibinfo {year} {2019})},\ \Eprint
  {http://arxiv.org/abs/1810.12917} {arXiv:1810.12917 [astro-ph.HE]}
  \BibitemShut {NoStop}%
\bibitem [{\citenamefont {Camilletti}\ \emph {et~al.}(2022)\citenamefont
  {Camilletti}, \citenamefont {Chiesa}, \citenamefont {Ricigliano},
  \citenamefont {Perego}, \citenamefont {Lippold}, \citenamefont {Padamata},
  \citenamefont {Bernuzzi}, \citenamefont {Radice}, \citenamefont {Logoteta},\
  and\ \citenamefont {Guercilena}}]{Camilletti:2022jms}%
  \BibitemOpen
  \bibfield  {author} {\bibinfo {author} {\bibfnamefont {A.}~\bibnamefont
  {Camilletti}}, \bibinfo {author} {\bibfnamefont {L.}~\bibnamefont {Chiesa}},
  \bibinfo {author} {\bibfnamefont {G.}~\bibnamefont {Ricigliano}}, \bibinfo
  {author} {\bibfnamefont {A.}~\bibnamefont {Perego}}, \bibinfo {author}
  {\bibfnamefont {L.~C.}\ \bibnamefont {Lippold}}, \bibinfo {author}
  {\bibfnamefont {S.}~\bibnamefont {Padamata}}, \bibinfo {author}
  {\bibfnamefont {S.}~\bibnamefont {Bernuzzi}}, \bibinfo {author}
  {\bibfnamefont {D.}~\bibnamefont {Radice}}, \bibinfo {author} {\bibfnamefont
  {D.}~\bibnamefont {Logoteta}}, \ and\ \bibinfo {author} {\bibfnamefont
  {F.~M.}\ \bibnamefont {Guercilena}},\ }\href {\doibase
  10.1093/mnras/stac2333} {\bibfield  {journal} {\bibinfo  {journal} {Mon. Not.
  Roy. Astron. Soc.}\ }\textbf {\bibinfo {volume} {516}},\ \bibinfo {pages}
  {4760} (\bibinfo {year} {2022})},\ \Eprint {http://arxiv.org/abs/2204.05336}
  {arXiv:2204.05336 [astro-ph.HE]} \BibitemShut {NoStop}%
\bibitem [{\citenamefont {Neuweiler}\ \emph {et~al.}(2023)\citenamefont
  {Neuweiler}, \citenamefont {Dietrich}, \citenamefont {Bulla}, \citenamefont
  {Chaurasia}, \citenamefont {Rosswog},\ and\ \citenamefont
  {Ujevic}}]{Neuweiler:2022eum}%
  \BibitemOpen
  \bibfield  {author} {\bibinfo {author} {\bibfnamefont {A.}~\bibnamefont
  {Neuweiler}}, \bibinfo {author} {\bibfnamefont {T.}~\bibnamefont {Dietrich}},
  \bibinfo {author} {\bibfnamefont {M.}~\bibnamefont {Bulla}}, \bibinfo
  {author} {\bibfnamefont {S.~V.}\ \bibnamefont {Chaurasia}}, \bibinfo {author}
  {\bibfnamefont {S.}~\bibnamefont {Rosswog}}, \ and\ \bibinfo {author}
  {\bibfnamefont {M.}~\bibnamefont {Ujevic}},\ }\href {\doibase
  10.1103/PhysRevD.107.023016} {\bibfield  {journal} {\bibinfo  {journal}
  {Phys. Rev. D}\ }\textbf {\bibinfo {volume} {107}},\ \bibinfo {pages}
  {023016} (\bibinfo {year} {2023})},\ \Eprint
  {http://arxiv.org/abs/2208.13460} {arXiv:2208.13460 [astro-ph.HE]}
  \BibitemShut {NoStop}%
\bibitem [{\citenamefont {Neuweiler}\ \emph {et~al.}(2026)\citenamefont
  {Neuweiler} \emph {et~al.}}]{Neuweiler:2025klw}%
  \BibitemOpen
  \bibfield  {author} {\bibinfo {author} {\bibfnamefont {A.}~\bibnamefont
  {Neuweiler}} \emph {et~al.},\ }\href {\doibase 10.1103/mxlf-8sbm} {\bibfield
  {journal} {\bibinfo  {journal} {Phys. Rev. D}\ }\textbf {\bibinfo {volume}
  {113}},\ \bibinfo {pages} {043038} (\bibinfo {year} {2026})},\ \Eprint
  {http://arxiv.org/abs/2510.14850} {arXiv:2510.14850 [astro-ph.HE]}
  \BibitemShut {NoStop}%
\bibitem [{\citenamefont {Rosswog}\ \emph {et~al.}(2014)\citenamefont
  {Rosswog}, \citenamefont {Korobkin}, \citenamefont {Arcones}, \citenamefont
  {Thielemann},\ and\ \citenamefont {Piran}}]{Rosswog:2013kqa}%
  \BibitemOpen
  \bibfield  {author} {\bibinfo {author} {\bibfnamefont {S.}~\bibnamefont
  {Rosswog}}, \bibinfo {author} {\bibfnamefont {O.}~\bibnamefont {Korobkin}},
  \bibinfo {author} {\bibfnamefont {A.}~\bibnamefont {Arcones}}, \bibinfo
  {author} {\bibfnamefont {F.~K.}\ \bibnamefont {Thielemann}}, \ and\ \bibinfo
  {author} {\bibfnamefont {T.}~\bibnamefont {Piran}},\ }\href {\doibase
  10.1093/mnras/stt2502} {\bibfield  {journal} {\bibinfo  {journal} {Mon. Not.
  Roy. Astron. Soc.}\ }\textbf {\bibinfo {volume} {439}},\ \bibinfo {pages}
  {744} (\bibinfo {year} {2014})},\ \Eprint {http://arxiv.org/abs/1307.2939}
  {arXiv:1307.2939 [astro-ph.HE]} \BibitemShut {NoStop}%
\bibitem [{\citenamefont {Kawaguchi}\ \emph {et~al.}(2024)\citenamefont
  {Kawaguchi}, \citenamefont {Domoto}, \citenamefont {Fujibayashi},
  \citenamefont {Hamidani}, \citenamefont {Hayashi}, \citenamefont {Shibata},
  \citenamefont {Tanaka},\ and\ \citenamefont {Wanajo}}]{Kawaguchi:2024hdk}%
  \BibitemOpen
  \bibfield  {author} {\bibinfo {author} {\bibfnamefont {K.}~\bibnamefont
  {Kawaguchi}}, \bibinfo {author} {\bibfnamefont {N.}~\bibnamefont {Domoto}},
  \bibinfo {author} {\bibfnamefont {S.}~\bibnamefont {Fujibayashi}}, \bibinfo
  {author} {\bibfnamefont {H.}~\bibnamefont {Hamidani}}, \bibinfo {author}
  {\bibfnamefont {K.}~\bibnamefont {Hayashi}}, \bibinfo {author} {\bibfnamefont
  {M.}~\bibnamefont {Shibata}}, \bibinfo {author} {\bibfnamefont
  {M.}~\bibnamefont {Tanaka}}, \ and\ \bibinfo {author} {\bibfnamefont
  {S.}~\bibnamefont {Wanajo}},\ }\href {\doibase 10.1093/mnras/stae2594}
  {\bibfield  {journal} {\bibinfo  {journal} {Mon. Not. Roy. Astron. Soc.}\
  }\textbf {\bibinfo {volume} {535}},\ \bibinfo {pages} {3711} (\bibinfo {year}
  {2024})},\ \Eprint {http://arxiv.org/abs/2404.15027} {arXiv:2404.15027
  [astro-ph.HE]} \BibitemShut {NoStop}%
\bibitem [{\citenamefont {Morozova}\ \emph {et~al.}(2015)\citenamefont
  {Morozova}, \citenamefont {Piro}, \citenamefont {Renzo}, \citenamefont {Ott},
  \citenamefont {Clausen}, \citenamefont {Couch}, \citenamefont {Ellis},\ and\
  \citenamefont {Roberts}}]{Morozova:2015bla}%
  \BibitemOpen
  \bibfield  {author} {\bibinfo {author} {\bibfnamefont {V.}~\bibnamefont
  {Morozova}}, \bibinfo {author} {\bibfnamefont {A.~L.}\ \bibnamefont {Piro}},
  \bibinfo {author} {\bibfnamefont {M.}~\bibnamefont {Renzo}}, \bibinfo
  {author} {\bibfnamefont {C.~D.}\ \bibnamefont {Ott}}, \bibinfo {author}
  {\bibfnamefont {D.}~\bibnamefont {Clausen}}, \bibinfo {author} {\bibfnamefont
  {S.~M.}\ \bibnamefont {Couch}}, \bibinfo {author} {\bibfnamefont
  {J.}~\bibnamefont {Ellis}}, \ and\ \bibinfo {author} {\bibfnamefont {L.~F.}\
  \bibnamefont {Roberts}},\ }\href {\doibase 10.1088/0004-637X/814/1/63}
  {\bibfield  {journal} {\bibinfo  {journal} {Astrophys. J.}\ }\textbf
  {\bibinfo {volume} {814}},\ \bibinfo {pages} {63} (\bibinfo {year} {2015})},\
  \Eprint {http://arxiv.org/abs/1505.06746} {arXiv:1505.06746 [astro-ph.HE]}
  \BibitemShut {NoStop}%
\bibitem [{\citenamefont {Bulla}(2019)}]{Bulla:2019muo}%
  \BibitemOpen
  \bibfield  {author} {\bibinfo {author} {\bibfnamefont {M.}~\bibnamefont
  {Bulla}},\ }\href {\doibase 10.1093/mnras/stz2495} {\bibfield  {journal}
  {\bibinfo  {journal} {Mon. Not. Roy. Astron. Soc.}\ }\textbf {\bibinfo
  {volume} {489}},\ \bibinfo {pages} {5037} (\bibinfo {year} {2019})},\ \Eprint
  {http://arxiv.org/abs/1906.04205} {arXiv:1906.04205 [astro-ph.HE]}
  \BibitemShut {NoStop}%
\bibitem [{\citenamefont {Collins}\ \emph {et~al.}(2023)\citenamefont
  {Collins}, \citenamefont {Bauswein}, \citenamefont {Sim}, \citenamefont
  {Vijayan}, \citenamefont {Mart\'\i{}nez-Pinedo}, \citenamefont {Just},
  \citenamefont {Shingles},\ and\ \citenamefont {Kromer}}]{Collins:2022ocl}%
  \BibitemOpen
  \bibfield  {author} {\bibinfo {author} {\bibfnamefont {C.~E.}\ \bibnamefont
  {Collins}}, \bibinfo {author} {\bibfnamefont {A.}~\bibnamefont {Bauswein}},
  \bibinfo {author} {\bibfnamefont {S.~A.}\ \bibnamefont {Sim}}, \bibinfo
  {author} {\bibfnamefont {V.}~\bibnamefont {Vijayan}}, \bibinfo {author}
  {\bibfnamefont {G.}~\bibnamefont {Mart\'\i{}nez-Pinedo}}, \bibinfo {author}
  {\bibfnamefont {O.}~\bibnamefont {Just}}, \bibinfo {author} {\bibfnamefont
  {L.~J.}\ \bibnamefont {Shingles}}, \ and\ \bibinfo {author} {\bibfnamefont
  {M.}~\bibnamefont {Kromer}},\ }\href {\doibase 10.1093/mnras/stad606}
  {\bibfield  {journal} {\bibinfo  {journal} {Mon. Not. Roy. Astron. Soc.}\
  }\textbf {\bibinfo {volume} {521}},\ \bibinfo {pages} {1858} (\bibinfo {year}
  {2023})},\ \Eprint {http://arxiv.org/abs/2209.05246} {arXiv:2209.05246
  [astro-ph.HE]} \BibitemShut {NoStop}%
\bibitem [{\citenamefont {Groenewegen}\ \emph {et~al.}(2025)\citenamefont
  {Groenewegen}, \citenamefont {Curtis}, \citenamefont {M{\"o}sta},
  \citenamefont {Kasen},\ and\ \citenamefont
  {Brethauer}}]{Groenewegen:2025ezj}%
  \BibitemOpen
  \bibfield  {author} {\bibinfo {author} {\bibfnamefont {L.~S.}\ \bibnamefont
  {Groenewegen}}, \bibinfo {author} {\bibfnamefont {S.}~\bibnamefont {Curtis}},
  \bibinfo {author} {\bibfnamefont {P.}~\bibnamefont {M{\"o}sta}}, \bibinfo
  {author} {\bibfnamefont {D.}~\bibnamefont {Kasen}}, \ and\ \bibinfo {author}
  {\bibfnamefont {D.}~\bibnamefont {Brethauer}},\ }\href {\doibase
  10.1093/mnras/staf1529} {\bibfield  {journal} {\bibinfo  {journal} {Mon. Not.
  Roy. Astron. Soc.}\ }\textbf {\bibinfo {volume} {543}},\ \bibinfo {pages}
  {2836} (\bibinfo {year} {2025})},\ \Eprint {http://arxiv.org/abs/2508.00062}
  {arXiv:2508.00062 [astro-ph.HE]} \BibitemShut {NoStop}%
\bibitem [{\citenamefont {Breschi}\ \emph {et~al.}(2021)\citenamefont
  {Breschi}, \citenamefont {Perego}, \citenamefont {Bernuzzi}, \citenamefont
  {Del~Pozzo}, \citenamefont {Nedora}, \citenamefont {Radice},\ and\
  \citenamefont {Vescovi}}]{Breschi:2021tbm}%
  \BibitemOpen
  \bibfield  {author} {\bibinfo {author} {\bibfnamefont {M.}~\bibnamefont
  {Breschi}}, \bibinfo {author} {\bibfnamefont {A.}~\bibnamefont {Perego}},
  \bibinfo {author} {\bibfnamefont {S.}~\bibnamefont {Bernuzzi}}, \bibinfo
  {author} {\bibfnamefont {W.}~\bibnamefont {Del~Pozzo}}, \bibinfo {author}
  {\bibfnamefont {V.}~\bibnamefont {Nedora}}, \bibinfo {author} {\bibfnamefont
  {D.}~\bibnamefont {Radice}}, \ and\ \bibinfo {author} {\bibfnamefont
  {D.}~\bibnamefont {Vescovi}},\ }\href {\doibase 10.1093/mnras/stab1287}
  {\bibfield  {journal} {\bibinfo  {journal} {Mon. Not. Roy. Astron. Soc.}\
  }\textbf {\bibinfo {volume} {505}},\ \bibinfo {pages} {1661} (\bibinfo {year}
  {2021})},\ \Eprint {http://arxiv.org/abs/2101.01201} {arXiv:2101.01201
  [astro-ph.HE]} \BibitemShut {NoStop}%
\bibitem [{\citenamefont {Kasen}\ \emph {et~al.}(2017)\citenamefont {Kasen},
  \citenamefont {Metzger}, \citenamefont {Barnes}, \citenamefont {Quataert},\
  and\ \citenamefont {Ramirez-Ruiz}}]{Kasen:2017sxr}%
  \BibitemOpen
  \bibfield  {author} {\bibinfo {author} {\bibfnamefont {D.}~\bibnamefont
  {Kasen}}, \bibinfo {author} {\bibfnamefont {B.}~\bibnamefont {Metzger}},
  \bibinfo {author} {\bibfnamefont {J.}~\bibnamefont {Barnes}}, \bibinfo
  {author} {\bibfnamefont {E.}~\bibnamefont {Quataert}}, \ and\ \bibinfo
  {author} {\bibfnamefont {E.}~\bibnamefont {Ramirez-Ruiz}},\ }\href {\doibase
  10.1038/nature24453} {\bibfield  {journal} {\bibinfo  {journal} {Nature}\ }
  (\bibinfo {year} {2017}),\ 10.1038/nature24453},\ \bibinfo {note}
  {[Nature551,80(2017)]},\ \Eprint {http://arxiv.org/abs/1710.05463}
  {arXiv:1710.05463 [astro-ph.HE]} \BibitemShut {NoStop}%
\bibitem [{\citenamefont {Metzger}(2020)}]{Metzger:2019zeh}%
  \BibitemOpen
  \bibfield  {author} {\bibinfo {author} {\bibfnamefont {B.~D.}\ \bibnamefont
  {Metzger}},\ }\href {\doibase 10.1007/s41114-019-0024-0} {\bibfield
  {journal} {\bibinfo  {journal} {Living Rev. Rel.}\ }\textbf {\bibinfo
  {volume} {23}},\ \bibinfo {pages} {1} (\bibinfo {year} {2020})},\ \Eprint
  {http://arxiv.org/abs/1910.01617} {arXiv:1910.01617 [astro-ph.HE]}
  \BibitemShut {NoStop}%
\bibitem [{\citenamefont {Nicholl}\ \emph {et~al.}(2021)\citenamefont
  {Nicholl}, \citenamefont {Margalit}, \citenamefont {Schmidt}, \citenamefont
  {Smith}, \citenamefont {Ridley},\ and\ \citenamefont
  {Nuttall}}]{Nicholl:2021abc}%
  \BibitemOpen
  \bibfield  {author} {\bibinfo {author} {\bibfnamefont {M.}~\bibnamefont
  {Nicholl}}, \bibinfo {author} {\bibfnamefont {B.}~\bibnamefont {Margalit}},
  \bibinfo {author} {\bibfnamefont {P.}~\bibnamefont {Schmidt}}, \bibinfo
  {author} {\bibfnamefont {G.~P.}\ \bibnamefont {Smith}}, \bibinfo {author}
  {\bibfnamefont {E.~J.}\ \bibnamefont {Ridley}}, \ and\ \bibinfo {author}
  {\bibfnamefont {J.}~\bibnamefont {Nuttall}},\ }\href@noop {} {\  (\bibinfo
  {year} {2021})},\ \Eprint {http://arxiv.org/abs/2102.02229} {arXiv:2102.02229
  [astro-ph.HE]} \BibitemShut {NoStop}%
\bibitem [{\citenamefont {Just}\ \emph {et~al.}(2023)\citenamefont {Just},
  \citenamefont {Vijayan}, \citenamefont {Xiong}, \citenamefont {Goriely},
  \citenamefont {Soultanis}, \citenamefont {Bauswein}, \citenamefont {Guilet},
  \citenamefont {Janka}, \citenamefont {Janka},\ and\ \citenamefont
  {Mart\'\i{}nez-Pinedo}}]{Just:2023wtj}%
  \BibitemOpen
  \bibfield  {author} {\bibinfo {author} {\bibfnamefont {O.}~\bibnamefont
  {Just}}, \bibinfo {author} {\bibfnamefont {V.}~\bibnamefont {Vijayan}},
  \bibinfo {author} {\bibfnamefont {Z.}~\bibnamefont {Xiong}}, \bibinfo
  {author} {\bibfnamefont {S.}~\bibnamefont {Goriely}}, \bibinfo {author}
  {\bibfnamefont {T.}~\bibnamefont {Soultanis}}, \bibinfo {author}
  {\bibfnamefont {A.}~\bibnamefont {Bauswein}}, \bibinfo {author}
  {\bibfnamefont {J.}~\bibnamefont {Guilet}}, \bibinfo {author} {\bibfnamefont
  {H.~T.}\ \bibnamefont {Janka}}, \bibinfo {author} {\bibfnamefont {H.-T.}\
  \bibnamefont {Janka}}, \ and\ \bibinfo {author} {\bibfnamefont
  {G.}~\bibnamefont {Mart\'\i{}nez-Pinedo}},\ }\href {\doibase
  10.3847/2041-8213/acdad2} {\bibfield  {journal} {\bibinfo  {journal}
  {Astrophys. J. Lett.}\ }\textbf {\bibinfo {volume} {951}},\ \bibinfo {pages}
  {L12} (\bibinfo {year} {2023})},\ \Eprint {http://arxiv.org/abs/2302.10928}
  {arXiv:2302.10928 [astro-ph.HE]} \BibitemShut {NoStop}%
\bibitem [{\citenamefont {Just}\ \emph {et~al.}(2015)\citenamefont {Just},
  \citenamefont {Obergaulinger},\ and\ \citenamefont {Janka}}]{Just:2015fda}%
  \BibitemOpen
  \bibfield  {author} {\bibinfo {author} {\bibfnamefont {O.}~\bibnamefont
  {Just}}, \bibinfo {author} {\bibfnamefont {M.}~\bibnamefont {Obergaulinger}},
  \ and\ \bibinfo {author} {\bibfnamefont {H.~T.}\ \bibnamefont {Janka}},\
  }\href {\doibase 10.1093/mnras/stv1892} {\bibfield  {journal} {\bibinfo
  {journal} {Mon. Not. Roy. Astron. Soc.}\ }\textbf {\bibinfo {volume} {453}},\
  \bibinfo {pages} {3386} (\bibinfo {year} {2015})},\ \Eprint
  {http://arxiv.org/abs/1501.02999} {arXiv:1501.02999 [astro-ph.HE]}
  \BibitemShut {NoStop}%
\bibitem [{\citenamefont {{Obergaulinger}}(2008)}]{Obergaulinger:Thesis:2008}%
  \BibitemOpen
  \bibfield  {author} {\bibinfo {author} {\bibfnamefont {M.}~\bibnamefont
  {{Obergaulinger}}},\ }\emph {\bibinfo {title} {{Astrophysical
  magnetohydrodynamics and radiative transfer: numerical methods and
  applications}}},\ \href@noop {} {Ph.D. thesis},\ \bibinfo  {school}
  {Max-Planck-Institute for Astrophysics, Garching} (\bibinfo {year}
  {2008})\BibitemShut {NoStop}%
\bibitem [{\citenamefont {{Stone}}\ \emph {et~al.}(2020)\citenamefont
  {{Stone}}, \citenamefont {{Tomida}}, \citenamefont {{White}},\ and\
  \citenamefont {{Felker}}}]{Stone:2020}%
  \BibitemOpen
  \bibfield  {author} {\bibinfo {author} {\bibfnamefont {J.~M.}\ \bibnamefont
  {{Stone}}}, \bibinfo {author} {\bibfnamefont {K.}~\bibnamefont {{Tomida}}},
  \bibinfo {author} {\bibfnamefont {C.~J.}\ \bibnamefont {{White}}}, \ and\
  \bibinfo {author} {\bibfnamefont {K.~G.}\ \bibnamefont {{Felker}}},\ }\href
  {\doibase 10.3847/1538-4365/ab929b} {\bibfield  {journal} {\bibinfo
  {journal} {Astrophys.\ J.\ Suppl.}\ }\textbf {\bibinfo {volume} {249}},\
  \bibinfo {eid} {4} (\bibinfo {year} {2020})},\ \Eprint
  {http://arxiv.org/abs/2005.06651} {arXiv:2005.06651 [astro-ph.IM]}
  \BibitemShut {NoStop}%
\bibitem [{\citenamefont {Radice}\ and\ \citenamefont
  {Bernuzzi}(2023)}]{Radice:2023zlw}%
  \BibitemOpen
  \bibfield  {author} {\bibinfo {author} {\bibfnamefont {D.}~\bibnamefont
  {Radice}}\ and\ \bibinfo {author} {\bibfnamefont {S.}~\bibnamefont
  {Bernuzzi}},\ }\href {\doibase 10.3847/1538-4357/ad0235} {\bibfield
  {journal} {\bibinfo  {journal} {Astrophys. J.}\ }\textbf {\bibinfo {volume}
  {959}},\ \bibinfo {pages} {46} (\bibinfo {year} {2023})},\ \Eprint
  {http://arxiv.org/abs/2306.13709} {arXiv:2306.13709 [astro-ph.HE]}
  \BibitemShut {NoStop}%
\bibitem [{\citenamefont {Bernuzzi}\ \emph {et~al.}(2025)\citenamefont
  {Bernuzzi}, \citenamefont {Magistrelli}, \citenamefont {Jacobi},
  \citenamefont {Logoteta}, \citenamefont {Perego},\ and\ \citenamefont
  {Radice}}]{Bernuzzi:2024mfx}%
  \BibitemOpen
  \bibfield  {author} {\bibinfo {author} {\bibfnamefont {S.}~\bibnamefont
  {Bernuzzi}}, \bibinfo {author} {\bibfnamefont {F.}~\bibnamefont
  {Magistrelli}}, \bibinfo {author} {\bibfnamefont {M.}~\bibnamefont {Jacobi}},
  \bibinfo {author} {\bibfnamefont {D.}~\bibnamefont {Logoteta}}, \bibinfo
  {author} {\bibfnamefont {A.}~\bibnamefont {Perego}}, \ and\ \bibinfo {author}
  {\bibfnamefont {D.}~\bibnamefont {Radice}},\ }\href {\doibase
  10.1093/mnras/staf1147} {\bibfield  {journal} {\bibinfo  {journal} {Mon. Not.
  Roy. Astron. Soc.}\ }\textbf {\bibinfo {volume} {256}},\ \bibinfo {pages}
  {271} (\bibinfo {year} {2025})},\ \Eprint {http://arxiv.org/abs/2409.18185}
  {arXiv:2409.18185 [astro-ph.HE]} \BibitemShut {NoStop}%
\bibitem [{\citenamefont {Gutierrez}\ \emph {et~al.}(2025)\citenamefont
  {Gutierrez}, \citenamefont {Bhattacharya}, \citenamefont {Radice},
  \citenamefont {Murase},\ and\ \citenamefont {Bernuzzi}}]{Gutierrez:2024pch}%
  \BibitemOpen
  \bibfield  {author} {\bibinfo {author} {\bibfnamefont {E.~M.}\ \bibnamefont
  {Gutierrez}}, \bibinfo {author} {\bibfnamefont {M.}~\bibnamefont
  {Bhattacharya}}, \bibinfo {author} {\bibfnamefont {D.}~\bibnamefont
  {Radice}}, \bibinfo {author} {\bibfnamefont {K.}~\bibnamefont {Murase}}, \
  and\ \bibinfo {author} {\bibfnamefont {S.}~\bibnamefont {Bernuzzi}},\ }\href
  {\doibase 10.1103/PhysRevD.111.063031} {\bibfield  {journal} {\bibinfo
  {journal} {Phys. Rev. D}\ }\textbf {\bibinfo {volume} {111}},\ \bibinfo
  {pages} {063031} (\bibinfo {year} {2025})},\ \Eprint
  {http://arxiv.org/abs/2408.15973} {arXiv:2408.15973 [astro-ph.HE]}
  \BibitemShut {NoStop}%
\bibitem [{\citenamefont {{Timmes}}\ and\ \citenamefont
  {{Swesty}}(2000)}]{Timmes:2000a}%
  \BibitemOpen
  \bibfield  {author} {\bibinfo {author} {\bibfnamefont {F.~X.}\ \bibnamefont
  {{Timmes}}}\ and\ \bibinfo {author} {\bibfnamefont {F.~D.}\ \bibnamefont
  {{Swesty}}},\ }\href {\doibase 10.1086/313304} {\bibfield  {journal}
  {\bibinfo  {journal} {Astrophys.\ J.\ Suppl.}\ }\textbf {\bibinfo {volume}
  {126}},\ \bibinfo {pages} {501} (\bibinfo {year} {2000})}\BibitemShut
  {NoStop}%
\bibitem [{\citenamefont {Wu}\ \emph {et~al.}(2022)\citenamefont {Wu},
  \citenamefont {Ricigliano}, \citenamefont {Kashyap}, \citenamefont {Perego},\
  and\ \citenamefont {Radice}}]{Wu:2021ibi}%
  \BibitemOpen
  \bibfield  {author} {\bibinfo {author} {\bibfnamefont {Z.}~\bibnamefont
  {Wu}}, \bibinfo {author} {\bibfnamefont {G.}~\bibnamefont {Ricigliano}},
  \bibinfo {author} {\bibfnamefont {R.}~\bibnamefont {Kashyap}}, \bibinfo
  {author} {\bibfnamefont {A.}~\bibnamefont {Perego}}, \ and\ \bibinfo {author}
  {\bibfnamefont {D.}~\bibnamefont {Radice}},\ }\href {\doibase
  10.1093/mnras/stac399} {\bibfield  {journal} {\bibinfo  {journal} {Mon. Not.
  Roy. Astron. Soc.}\ }\textbf {\bibinfo {volume} {512}},\ \bibinfo {pages}
  {328} (\bibinfo {year} {2022})},\ \Eprint {http://arxiv.org/abs/2111.06870}
  {arXiv:2111.06870 [astro-ph.HE]} \BibitemShut {NoStop}%
\bibitem [{\citenamefont {Banyuls}\ \emph {et~al.}(1997)\citenamefont
  {Banyuls}, \citenamefont {Font}, \citenamefont {Ibanez}, \citenamefont
  {Marti},\ and\ \citenamefont {Miralles}}]{Banyuls:1997zz}%
  \BibitemOpen
  \bibfield  {author} {\bibinfo {author} {\bibfnamefont {F.}~\bibnamefont
  {Banyuls}}, \bibinfo {author} {\bibfnamefont {J.~A.}\ \bibnamefont {Font}},
  \bibinfo {author} {\bibfnamefont {J.~M.~A.}\ \bibnamefont {Ibanez}}, \bibinfo
  {author} {\bibfnamefont {J.~M.~A.}\ \bibnamefont {Marti}}, \ and\ \bibinfo
  {author} {\bibfnamefont {J.~A.}\ \bibnamefont {Miralles}},\ }\href@noop {}
  {\bibfield  {journal} {\bibinfo  {journal} {Astrophys. J.}\ }\textbf
  {\bibinfo {volume} {476}},\ \bibinfo {pages} {221} (\bibinfo {year}
  {1997})}\BibitemShut {NoStop}%
\bibitem [{\citenamefont {Bombaci}\ and\ \citenamefont
  {Logoteta}(2018)}]{Bombaci:2018ksa}%
  \BibitemOpen
  \bibfield  {author} {\bibinfo {author} {\bibfnamefont {I.}~\bibnamefont
  {Bombaci}}\ and\ \bibinfo {author} {\bibfnamefont {D.}~\bibnamefont
  {Logoteta}},\ }\href {\doibase 10.1051/0004-6361/201731604} {\bibfield
  {journal} {\bibinfo  {journal} {Astron. Astrophys.}\ }\textbf {\bibinfo
  {volume} {609}},\ \bibinfo {pages} {A128} (\bibinfo {year} {2018})},\ \Eprint
  {http://arxiv.org/abs/1805.11846} {arXiv:1805.11846 [astro-ph.HE]}
  \BibitemShut {NoStop}%
\bibitem [{\citenamefont {Logoteta}\ \emph {et~al.}(2021)\citenamefont
  {Logoteta}, \citenamefont {Perego},\ and\ \citenamefont
  {Bombaci}}]{Logoteta:2020yxf}%
  \BibitemOpen
  \bibfield  {author} {\bibinfo {author} {\bibfnamefont {D.}~\bibnamefont
  {Logoteta}}, \bibinfo {author} {\bibfnamefont {A.}~\bibnamefont {Perego}}, \
  and\ \bibinfo {author} {\bibfnamefont {I.}~\bibnamefont {Bombaci}},\ }\href
  {\doibase 10.1051/0004-6361/202039457} {\bibfield  {journal} {\bibinfo
  {journal} {Astron. Astrophys.}\ }\textbf {\bibinfo {volume} {646}},\ \bibinfo
  {pages} {A55} (\bibinfo {year} {2021})},\ \Eprint
  {http://arxiv.org/abs/2012.03599} {arXiv:2012.03599 [nucl-th]} \BibitemShut
  {NoStop}%
\bibitem [{\citenamefont {Steiner}\ \emph {et~al.}(2010)\citenamefont
  {Steiner}, \citenamefont {Lattimer},\ and\ \citenamefont
  {Brown}}]{Steiner:2010fz}%
  \BibitemOpen
  \bibfield  {author} {\bibinfo {author} {\bibfnamefont {A.~W.}\ \bibnamefont
  {Steiner}}, \bibinfo {author} {\bibfnamefont {J.~M.}\ \bibnamefont
  {Lattimer}}, \ and\ \bibinfo {author} {\bibfnamefont {E.~F.}\ \bibnamefont
  {Brown}},\ }\href {\doibase 10.1088/0004-637X/722/1/33} {\bibfield  {journal}
  {\bibinfo  {journal} {Astrophys. J.}\ }\textbf {\bibinfo {volume} {722}},\
  \bibinfo {pages} {33} (\bibinfo {year} {2010})},\ \Eprint
  {http://arxiv.org/abs/1005.0811} {arXiv:1005.0811 [astro-ph.HE]} \BibitemShut
  {NoStop}%
\bibitem [{\citenamefont {Typel}\ \emph {et~al.}(2010)\citenamefont {Typel},
  \citenamefont {Ropke}, \citenamefont {Klahn}, \citenamefont {Blaschke},\ and\
  \citenamefont {Wolter}}]{Typel:2009sy}%
  \BibitemOpen
  \bibfield  {author} {\bibinfo {author} {\bibfnamefont {S.}~\bibnamefont
  {Typel}}, \bibinfo {author} {\bibfnamefont {G.}~\bibnamefont {Ropke}},
  \bibinfo {author} {\bibfnamefont {T.}~\bibnamefont {Klahn}}, \bibinfo
  {author} {\bibfnamefont {D.}~\bibnamefont {Blaschke}}, \ and\ \bibinfo
  {author} {\bibfnamefont {H.~H.}\ \bibnamefont {Wolter}},\ }\href {\doibase
  10.1103/PhysRevC.81.015803} {\bibfield  {journal} {\bibinfo  {journal} {Phys.
  Rev.}\ }\textbf {\bibinfo {volume} {C81}},\ \bibinfo {pages} {015803}
  (\bibinfo {year} {2010})},\ \Eprint {http://arxiv.org/abs/0908.2344}
  {arXiv:0908.2344 [nucl-th]} \BibitemShut {NoStop}%
\bibitem [{\citenamefont {Hempel}\ and\ \citenamefont
  {Schaffner-Bielich}(2010)}]{Hempel:2009mc}%
  \BibitemOpen
  \bibfield  {author} {\bibinfo {author} {\bibfnamefont {M.}~\bibnamefont
  {Hempel}}\ and\ \bibinfo {author} {\bibfnamefont {J.}~\bibnamefont
  {Schaffner-Bielich}},\ }\href {\doibase 10.1016/j.nuclphysa.2010.02.010}
  {\bibfield  {journal} {\bibinfo  {journal} {Nucl. Phys.}\ }\textbf {\bibinfo
  {volume} {A837}},\ \bibinfo {pages} {210} (\bibinfo {year} {2010})},\ \Eprint
  {http://arxiv.org/abs/0911.4073} {arXiv:0911.4073 [nucl-th]} \BibitemShut
  {NoStop}%
\bibitem [{\citenamefont {Radice}\ and\ \citenamefont
  {Rezzolla}(2012)}]{Radice:2012cu}%
  \BibitemOpen
  \bibfield  {author} {\bibinfo {author} {\bibfnamefont {D.}~\bibnamefont
  {Radice}}\ and\ \bibinfo {author} {\bibfnamefont {L.}~\bibnamefont
  {Rezzolla}},\ }\href {\doibase 10.1051/0004-6361/201219735} {\bibfield
  {journal} {\bibinfo  {journal} {Astron. Astrophys.}\ }\textbf {\bibinfo
  {volume} {547}},\ \bibinfo {pages} {A26} (\bibinfo {year} {2012})},\ \Eprint
  {http://arxiv.org/abs/1206.6502} {arXiv:1206.6502 [astro-ph.IM]} \BibitemShut
  {NoStop}%
\bibitem [{\citenamefont {Radice}\ \emph
  {et~al.}(2014{\natexlab{a}})\citenamefont {Radice}, \citenamefont
  {Rezzolla},\ and\ \citenamefont {Galeazzi}}]{Radice:2013hxh}%
  \BibitemOpen
  \bibfield  {author} {\bibinfo {author} {\bibfnamefont {D.}~\bibnamefont
  {Radice}}, \bibinfo {author} {\bibfnamefont {L.}~\bibnamefont {Rezzolla}}, \
  and\ \bibinfo {author} {\bibfnamefont {F.}~\bibnamefont {Galeazzi}},\ }\href
  {\doibase 10.1093/mnrasl/slt137} {\bibfield  {journal} {\bibinfo  {journal}
  {Mon.Not.Roy.Astron.Soc.}\ }\textbf {\bibinfo {volume} {437}},\ \bibinfo
  {pages} {L46} (\bibinfo {year} {2014}{\natexlab{a}})},\ \Eprint
  {http://arxiv.org/abs/1306.6052} {arXiv:1306.6052 [gr-qc]} \BibitemShut
  {NoStop}%
\bibitem [{\citenamefont {Radice}\ \emph
  {et~al.}(2014{\natexlab{b}})\citenamefont {Radice}, \citenamefont
  {Rezzolla},\ and\ \citenamefont {Galeazzi}}]{Radice:2013xpa}%
  \BibitemOpen
  \bibfield  {author} {\bibinfo {author} {\bibfnamefont {D.}~\bibnamefont
  {Radice}}, \bibinfo {author} {\bibfnamefont {L.}~\bibnamefont {Rezzolla}}, \
  and\ \bibinfo {author} {\bibfnamefont {F.}~\bibnamefont {Galeazzi}},\ }\href
  {\doibase 10.1088/0264-9381/31/7/075012} {\bibfield  {journal} {\bibinfo
  {journal} {Class.Quant.Grav.}\ }\textbf {\bibinfo {volume} {31}},\ \bibinfo
  {pages} {075012} (\bibinfo {year} {2014}{\natexlab{b}})},\ \Eprint
  {http://arxiv.org/abs/1312.5004} {arXiv:1312.5004 [gr-qc]} \BibitemShut
  {NoStop}%
\bibitem [{\citenamefont {Radice}\ \emph {et~al.}(2015)\citenamefont {Radice},
  \citenamefont {Rezzolla},\ and\ \citenamefont {Galeazzi}}]{Radice:2015nva}%
  \BibitemOpen
  \bibfield  {author} {\bibinfo {author} {\bibfnamefont {D.}~\bibnamefont
  {Radice}}, \bibinfo {author} {\bibfnamefont {L.}~\bibnamefont {Rezzolla}}, \
  and\ \bibinfo {author} {\bibfnamefont {F.}~\bibnamefont {Galeazzi}},\
  }\bibfield  {booktitle} {\emph {\bibinfo {booktitle} {{Proceedings, Numerical
  Modeling of Space Plasma Flows (ASTRONUM-2014): Long Beach, CA, USA, June
  23-27, 2014}}},\ }\href@noop {} {\bibfield  {journal} {\bibinfo  {journal}
  {ASP Conf. Ser.}\ }\textbf {\bibinfo {volume} {498}},\ \bibinfo {pages} {121}
  (\bibinfo {year} {2015})},\ \Eprint {http://arxiv.org/abs/1502.00551}
  {arXiv:1502.00551 [gr-qc]} \BibitemShut {NoStop}%
\bibitem [{\citenamefont {Radice}(2017)}]{Radice:2017zta}%
  \BibitemOpen
  \bibfield  {author} {\bibinfo {author} {\bibfnamefont {D.}~\bibnamefont
  {Radice}},\ }\href {\doibase 10.3847/2041-8213/aa6483} {\bibfield  {journal}
  {\bibinfo  {journal} {Astrophys. J.}\ }\textbf {\bibinfo {volume} {838}},\
  \bibinfo {pages} {L2} (\bibinfo {year} {2017})},\ \Eprint
  {http://arxiv.org/abs/1703.02046} {arXiv:1703.02046 [astro-ph.HE]}
  \BibitemShut {NoStop}%
\bibitem [{\citenamefont {Radice}(2020)}]{Radice:2020ids}%
  \BibitemOpen
  \bibfield  {author} {\bibinfo {author} {\bibfnamefont {D.}~\bibnamefont
  {Radice}},\ }\href {\doibase 10.3390/sym12081249} {\bibfield  {journal}
  {\bibinfo  {journal} {Symmetry}\ }\textbf {\bibinfo {volume} {12}},\ \bibinfo
  {pages} {1249} (\bibinfo {year} {2020})},\ \Eprint
  {http://arxiv.org/abs/2005.09002} {arXiv:2005.09002 [astro-ph.HE]}
  \BibitemShut {NoStop}%
\bibitem [{\citenamefont {Radice}\ \emph {et~al.}(2022)\citenamefont {Radice},
  \citenamefont {Bernuzzi}, \citenamefont {Perego},\ and\ \citenamefont
  {Haas}}]{Radice:2021jtw}%
  \BibitemOpen
  \bibfield  {author} {\bibinfo {author} {\bibfnamefont {D.}~\bibnamefont
  {Radice}}, \bibinfo {author} {\bibfnamefont {S.}~\bibnamefont {Bernuzzi}},
  \bibinfo {author} {\bibfnamefont {A.}~\bibnamefont {Perego}}, \ and\ \bibinfo
  {author} {\bibfnamefont {R.}~\bibnamefont {Haas}},\ }\href {\doibase
  10.1093/mnras/stac589} {\bibfield  {journal} {\bibinfo  {journal} {Mon. Not.
  Roy. Astron. Soc.}\ }\textbf {\bibinfo {volume} {512}},\ \bibinfo {pages}
  {1499} (\bibinfo {year} {2022})},\ \Eprint {http://arxiv.org/abs/2111.14858}
  {arXiv:2111.14858 [astro-ph.HE]} \BibitemShut {NoStop}%
\bibitem [{\citenamefont {{Gong}}\ \emph {et~al.}(2023)\citenamefont {{Gong}},
  \citenamefont {{Ho}}, \citenamefont {{Stone}}, \citenamefont {{Ostriker}},
  \citenamefont {{Caselli}}, \citenamefont {{Grassi}}, \citenamefont {{Kim}},
  \citenamefont {{Kim}},\ and\ \citenamefont {{Halevi}}}]{Gong:2023abc}%
  \BibitemOpen
  \bibfield  {author} {\bibinfo {author} {\bibfnamefont {M.}~\bibnamefont
  {{Gong}}}, \bibinfo {author} {\bibfnamefont {K.~W.}\ \bibnamefont {{Ho}}},
  \bibinfo {author} {\bibfnamefont {J.~M.}\ \bibnamefont {{Stone}}}, \bibinfo
  {author} {\bibfnamefont {E.~C.}\ \bibnamefont {{Ostriker}}}, \bibinfo
  {author} {\bibfnamefont {P.}~\bibnamefont {{Caselli}}}, \bibinfo {author}
  {\bibfnamefont {T.}~\bibnamefont {{Grassi}}}, \bibinfo {author}
  {\bibfnamefont {C.-G.}\ \bibnamefont {{Kim}}}, \bibinfo {author}
  {\bibfnamefont {J.-G.}\ \bibnamefont {{Kim}}}, \ and\ \bibinfo {author}
  {\bibfnamefont {G.}~\bibnamefont {{Halevi}}},\ }\href {\doibase
  10.3847/1538-4365/aceaf9} {\bibfield  {journal} {\bibinfo  {journal}
  {Astrophys.\ J.\ Suppl.}\ }\textbf {\bibinfo {volume} {268}},\ \bibinfo {eid}
  {42} (\bibinfo {year} {2023})},\ \Eprint {http://arxiv.org/abs/2305.04965}
  {arXiv:2305.04965 [astro-ph.GA]} \BibitemShut {NoStop}%
\bibitem [{\citenamefont {Just}\ \emph {et~al.}(2026)\citenamefont {Just},
  \citenamefont {Xiong},\ and\ \citenamefont
  {Mart{\'\i}nez-Pinedo}}]{Just:2025hyy}%
  \BibitemOpen
  \bibfield  {author} {\bibinfo {author} {\bibfnamefont {O.}~\bibnamefont
  {Just}}, \bibinfo {author} {\bibfnamefont {Z.}~\bibnamefont {Xiong}}, \ and\
  \bibinfo {author} {\bibfnamefont {G.}~\bibnamefont {Mart{\'\i}nez-Pinedo}},\
  }\href {\doibase 10.1103/gl2l-7f3g} {\bibfield  {journal} {\bibinfo
  {journal} {Phys. Rev. D}\ }\textbf {\bibinfo {volume} {113}},\ \bibinfo
  {pages} {083022} (\bibinfo {year} {2026})},\ \Eprint
  {http://arxiv.org/abs/2507.09040} {arXiv:2507.09040 [astro-ph.SR]}
  \BibitemShut {NoStop}%
\bibitem [{\citenamefont {Cook}\ \emph
  {et~al.}(2025{\natexlab{a}})\citenamefont {Cook}, \citenamefont {Daszuta},
  \citenamefont {Fields}, \citenamefont {Hammond}, \citenamefont {Albanesi},
  \citenamefont {Zappa}, \citenamefont {Bernuzzi},\ and\ \citenamefont
  {Radice}}]{Cook:2023bag}%
  \BibitemOpen
  \bibfield  {author} {\bibinfo {author} {\bibfnamefont {W.}~\bibnamefont
  {Cook}}, \bibinfo {author} {\bibfnamefont {B.}~\bibnamefont {Daszuta}},
  \bibinfo {author} {\bibfnamefont {J.}~\bibnamefont {Fields}}, \bibinfo
  {author} {\bibfnamefont {P.}~\bibnamefont {Hammond}}, \bibinfo {author}
  {\bibfnamefont {S.}~\bibnamefont {Albanesi}}, \bibinfo {author}
  {\bibfnamefont {F.}~\bibnamefont {Zappa}}, \bibinfo {author} {\bibfnamefont
  {S.}~\bibnamefont {Bernuzzi}}, \ and\ \bibinfo {author} {\bibfnamefont
  {D.}~\bibnamefont {Radice}},\ }\href {\doibase 10.3847/1538-4365/ad87d4}
  {\bibfield  {journal} {\bibinfo  {journal} {Astrophys. J. Suppl.}\ }\textbf
  {\bibinfo {volume} {277}},\ \bibinfo {pages} {3} (\bibinfo {year}
  {2025}{\natexlab{a}})},\ \Eprint {http://arxiv.org/abs/2311.04989}
  {arXiv:2311.04989 [gr-qc]} \BibitemShut {NoStop}%
\bibitem [{\citenamefont {Kastaun}\ \emph {et~al.}(2021)\citenamefont
  {Kastaun}, \citenamefont {Kalinani},\ and\ \citenamefont
  {Ciolfi}}]{Kastaun:2020uxr}%
  \BibitemOpen
  \bibfield  {author} {\bibinfo {author} {\bibfnamefont {W.}~\bibnamefont
  {Kastaun}}, \bibinfo {author} {\bibfnamefont {J.~V.}\ \bibnamefont
  {Kalinani}}, \ and\ \bibinfo {author} {\bibfnamefont {R.}~\bibnamefont
  {Ciolfi}},\ }\href {\doibase 10.1103/PhysRevD.103.023018} {\bibfield
  {journal} {\bibinfo  {journal} {Phys. Rev. D}\ }\textbf {\bibinfo {volume}
  {103}},\ \bibinfo {pages} {023018} (\bibinfo {year} {2021})},\ \Eprint
  {http://arxiv.org/abs/2005.01821} {arXiv:2005.01821 [gr-qc]} \BibitemShut
  {NoStop}%
\bibitem [{\citenamefont {{Cyburt}}\ \emph {et~al.}(2010)\citenamefont
  {{Cyburt}}, \citenamefont {{Amthor}}, \citenamefont {{Ferguson}},
  \citenamefont {{Meisel}}, \citenamefont {{Smith}}, \citenamefont {{Warren}},
  \citenamefont {{Heger}}, \citenamefont {{Hoffman}}, \citenamefont
  {{Rauscher}}, \citenamefont {{Sakharuk}}, \citenamefont {{Schatz}},
  \citenamefont {{Thielemann}},\ and\ \citenamefont
  {{Wiescher}}}]{Cyburt:2010a}%
  \BibitemOpen
  \bibfield  {author} {\bibinfo {author} {\bibfnamefont {R.~H.}\ \bibnamefont
  {{Cyburt}}}, \bibinfo {author} {\bibfnamefont {A.~M.}\ \bibnamefont
  {{Amthor}}}, \bibinfo {author} {\bibfnamefont {R.}~\bibnamefont
  {{Ferguson}}}, \bibinfo {author} {\bibfnamefont {Z.}~\bibnamefont
  {{Meisel}}}, \bibinfo {author} {\bibfnamefont {K.}~\bibnamefont {{Smith}}},
  \bibinfo {author} {\bibfnamefont {S.}~\bibnamefont {{Warren}}}, \bibinfo
  {author} {\bibfnamefont {A.}~\bibnamefont {{Heger}}}, \bibinfo {author}
  {\bibfnamefont {R.~D.}\ \bibnamefont {{Hoffman}}}, \bibinfo {author}
  {\bibfnamefont {T.}~\bibnamefont {{Rauscher}}}, \bibinfo {author}
  {\bibfnamefont {A.}~\bibnamefont {{Sakharuk}}}, \bibinfo {author}
  {\bibfnamefont {H.}~\bibnamefont {{Schatz}}}, \bibinfo {author}
  {\bibfnamefont {F.~K.}\ \bibnamefont {{Thielemann}}}, \ and\ \bibinfo
  {author} {\bibfnamefont {M.}~\bibnamefont {{Wiescher}}},\ }\href {\doibase
  10.1088/0067-0049/189/1/240} {\bibfield  {journal} {\bibinfo  {journal}
  {Astrophys.\ J.\ Suppl.}\ }\textbf {\bibinfo {volume} {189}},\ \bibinfo
  {pages} {240} (\bibinfo {year} {2010})}\BibitemShut {NoStop}%
\bibitem [{\citenamefont {Fuller}\ \emph {et~al.}(1985)\citenamefont {Fuller},
  \citenamefont {Fowler},\ and\ \citenamefont {Newman}}]{Fuller:1985}%
  \BibitemOpen
  \bibfield  {author} {\bibinfo {author} {\bibfnamefont {G.~M.}\ \bibnamefont
  {Fuller}}, \bibinfo {author} {\bibfnamefont {W.~A.}\ \bibnamefont {Fowler}},
  \ and\ \bibinfo {author} {\bibfnamefont {M.~J.}\ \bibnamefont {Newman}},\
  }\href {\doibase 10.1086/163208} {\bibfield  {journal} {\bibinfo  {journal}
  {Astrophys. J.}\ }\textbf {\bibinfo {volume} {293}},\ \bibinfo {pages} {1}
  (\bibinfo {year} {1985})}\BibitemShut {NoStop}%
\bibitem [{\citenamefont {Oda}\ \emph {et~al.}(1994)\citenamefont {Oda},
  \citenamefont {Hino}, \citenamefont {Muto}, \citenamefont {Takahara},\ and\
  \citenamefont {Sato}}]{Oda:1994}%
  \BibitemOpen
  \bibfield  {author} {\bibinfo {author} {\bibfnamefont {T.}~\bibnamefont
  {Oda}}, \bibinfo {author} {\bibfnamefont {M.}~\bibnamefont {Hino}}, \bibinfo
  {author} {\bibfnamefont {K.}~\bibnamefont {Muto}}, \bibinfo {author}
  {\bibfnamefont {M.}~\bibnamefont {Takahara}}, \ and\ \bibinfo {author}
  {\bibfnamefont {K.}~\bibnamefont {Sato}},\ }\href {\doibase
  10.1006/adnd.1994.1007} {\bibfield  {journal} {\bibinfo  {journal} {Atom.
  Data Nucl. Data Tabl.}\ }\textbf {\bibinfo {volume} {56}},\ \bibinfo {pages}
  {231} (\bibinfo {year} {1994})}\BibitemShut {NoStop}%
\bibitem [{\citenamefont {Langanke}\ and\ \citenamefont
  {Mart\'{i}nez-Pinedo}(2001)}]{Langanke:2001}%
  \BibitemOpen
  \bibfield  {author} {\bibinfo {author} {\bibfnamefont {K.}~\bibnamefont
  {Langanke}}\ and\ \bibinfo {author} {\bibfnamefont {G.}~\bibnamefont
  {Mart\'{i}nez-Pinedo}},\ }\href {\doibase 10.1006/adnd.2001.0865} {\bibfield
  {journal} {\bibinfo  {journal} {Atom. Data Nucl. Data Tabl.}\ }\textbf
  {\bibinfo {volume} {79}},\ \bibinfo {pages} {1} (\bibinfo {year}
  {2001})}\BibitemShut {NoStop}%
\bibitem [{\citenamefont {Pruet}\ and\ \citenamefont
  {Fuller}(2003)}]{PruetFuller:2003}%
  \BibitemOpen
  \bibfield  {author} {\bibinfo {author} {\bibfnamefont {J.}~\bibnamefont
  {Pruet}}\ and\ \bibinfo {author} {\bibfnamefont {G.~M.}\ \bibnamefont
  {Fuller}},\ }\href {\doibase 10.1086/376753} {\bibfield  {journal} {\bibinfo
  {journal} {Astrophys. J. Suppl.}\ }\textbf {\bibinfo {volume} {149}},\
  \bibinfo {pages} {189} (\bibinfo {year} {2003})},\ \Eprint
  {http://arxiv.org/abs/astro-ph/0303235} {arXiv:astro-ph/0303235} \BibitemShut
  {NoStop}%
\bibitem [{\citenamefont {Suzuki}\ \emph {et~al.}(2016)\citenamefont {Suzuki},
  \citenamefont {Toki},\ and\ \citenamefont {Nomoto}}]{Suzuki:2016}%
  \BibitemOpen
  \bibfield  {author} {\bibinfo {author} {\bibfnamefont {T.}~\bibnamefont
  {Suzuki}}, \bibinfo {author} {\bibfnamefont {H.}~\bibnamefont {Toki}}, \ and\
  \bibinfo {author} {\bibfnamefont {K.}~\bibnamefont {Nomoto}},\ }\href
  {\doibase 10.3847/0004-637X/817/2/163} {\bibfield  {journal} {\bibinfo
  {journal} {Astrophys. J.}\ }\textbf {\bibinfo {volume} {817}},\ \bibinfo
  {pages} {163} (\bibinfo {year} {2016})},\ \Eprint
  {http://arxiv.org/abs/1512.00801} {arXiv:1512.00801 [astro-ph.SR]}
  \BibitemShut {NoStop}%
\bibitem [{\citenamefont {Mumpower}\ \emph {et~al.}(2022)\citenamefont
  {Mumpower}, \citenamefont {Kawano},\ and\ \citenamefont
  {Sprouse}}]{Mumpower:2022}%
  \BibitemOpen
  \bibfield  {author} {\bibinfo {author} {\bibfnamefont {M.~R.}\ \bibnamefont
  {Mumpower}}, \bibinfo {author} {\bibfnamefont {T.}~\bibnamefont {Kawano}}, \
  and\ \bibinfo {author} {\bibfnamefont {T.~M.}\ \bibnamefont {Sprouse}},\
  }\href {\doibase 10.1103/PhysRevC.106.065805} {\bibfield  {journal} {\bibinfo
   {journal} {Phys. Rev. C}\ }\textbf {\bibinfo {volume} {106}},\ \bibinfo
  {pages} {065805} (\bibinfo {year} {2022})},\ \Eprint
  {http://arxiv.org/abs/2201.02889} {arXiv:2201.02889 [nucl-th]} \BibitemShut
  {NoStop}%
\bibitem [{\citenamefont {Panov}\ \emph {et~al.}(2010)\citenamefont {Panov},
  \citenamefont {Korneev}, \citenamefont {Rauscher}, \citenamefont
  {Mart\'{i}nez-Pinedo}, \citenamefont {Kelic-Heil}, \citenamefont {Zinner},\
  and\ \citenamefont {Thielemann}}]{Panov:2010}%
  \BibitemOpen
  \bibfield  {author} {\bibinfo {author} {\bibfnamefont {I.~V.}\ \bibnamefont
  {Panov}}, \bibinfo {author} {\bibfnamefont {I.~Y.}\ \bibnamefont {Korneev}},
  \bibinfo {author} {\bibfnamefont {T.}~\bibnamefont {Rauscher}}, \bibinfo
  {author} {\bibfnamefont {G.}~\bibnamefont {Mart\'{i}nez-Pinedo}}, \bibinfo
  {author} {\bibfnamefont {A.}~\bibnamefont {Kelic-Heil}}, \bibinfo {author}
  {\bibfnamefont {N.~T.}\ \bibnamefont {Zinner}}, \ and\ \bibinfo {author}
  {\bibfnamefont {F.-K.}\ \bibnamefont {Thielemann}},\ }\href {\doibase
  10.1051/0004-6361/200911967} {\bibfield  {journal} {\bibinfo  {journal}
  {Astron. Astrophys.}\ }\textbf {\bibinfo {volume} {513}},\ \bibinfo {pages}
  {A61} (\bibinfo {year} {2010})},\ \Eprint {http://arxiv.org/abs/0911.2181}
  {arXiv:0911.2181 [astro-ph.SR]} \BibitemShut {NoStop}%
\bibitem [{\citenamefont {Khuyagbaatar}(2020)}]{Khuyagbaatar:2020}%
  \BibitemOpen
  \bibfield  {author} {\bibinfo {author} {\bibfnamefont {J.}~\bibnamefont
  {Khuyagbaatar}},\ }\href {\doibase 10.1016/j.nuclphysa.2020.121958}
  {\bibfield  {journal} {\bibinfo  {journal} {Nucl. Phys. A}\ }\textbf
  {\bibinfo {volume} {1002}},\ \bibinfo {pages} {121958} (\bibinfo {year}
  {2020})}\BibitemShut {NoStop}%
\bibitem [{\citenamefont {M{\"o}ller}\ \emph {et~al.}(2015)\citenamefont
  {M{\"o}ller}, \citenamefont {Sierk}, \citenamefont {Ichikawa}, \citenamefont
  {Iwamoto},\ and\ \citenamefont {Mumpower}}]{Moller:2015}%
  \BibitemOpen
  \bibfield  {author} {\bibinfo {author} {\bibfnamefont {P.}~\bibnamefont
  {M{\"o}ller}}, \bibinfo {author} {\bibfnamefont {A.~J.}\ \bibnamefont
  {Sierk}}, \bibinfo {author} {\bibfnamefont {T.}~\bibnamefont {Ichikawa}},
  \bibinfo {author} {\bibfnamefont {A.}~\bibnamefont {Iwamoto}}, \ and\
  \bibinfo {author} {\bibfnamefont {M.}~\bibnamefont {Mumpower}},\ }\href
  {\doibase 10.1103/PhysRevC.91.024310} {\bibfield  {journal} {\bibinfo
  {journal} {Phys. Rev. C}\ }\textbf {\bibinfo {volume} {91}},\ \bibinfo
  {pages} {024310} (\bibinfo {year} {2015})},\ \Eprint
  {http://arxiv.org/abs/1502.03635} {arXiv:1502.03635 [nucl-th]} \BibitemShut
  {NoStop}%
\bibitem [{\citenamefont {Panov}\ \emph {et~al.}(2001)\citenamefont {Panov},
  \citenamefont {Freiburghaus},\ and\ \citenamefont {Thielemann}}]{Panov:2001}%
  \BibitemOpen
  \bibfield  {author} {\bibinfo {author} {\bibfnamefont {I.~V.}\ \bibnamefont
  {Panov}}, \bibinfo {author} {\bibfnamefont {C.}~\bibnamefont {Freiburghaus}},
  \ and\ \bibinfo {author} {\bibfnamefont {F.-K.}\ \bibnamefont {Thielemann}},\
  }\href {\doibase 10.1016/S0375-9474(01)00797-7} {\bibfield  {journal}
  {\bibinfo  {journal} {Nucl. Phys. A}\ }\textbf {\bibinfo {volume} {688}},\
  \bibinfo {pages} {587} (\bibinfo {year} {2001})}\BibitemShut {NoStop}%
\bibitem [{\citenamefont {Dong}\ and\ \citenamefont
  {Ren}(2005)}]{DongRen:2005}%
  \BibitemOpen
  \bibfield  {author} {\bibinfo {author} {\bibfnamefont {T.}~\bibnamefont
  {Dong}}\ and\ \bibinfo {author} {\bibfnamefont {Z.}~\bibnamefont {Ren}},\
  }\href {\doibase 10.1140/epja/i2005-10142-y} {\bibfield  {journal} {\bibinfo
  {journal} {Eur. Phys. J. A}\ }\textbf {\bibinfo {volume} {26}},\ \bibinfo
  {pages} {69} (\bibinfo {year} {2005})}\BibitemShut {NoStop}%
\bibitem [{\citenamefont {Rauscher}(2003)}]{Rauscher:2003}%
  \BibitemOpen
  \bibfield  {author} {\bibinfo {author} {\bibfnamefont {T.}~\bibnamefont
  {Rauscher}},\ }\href {\doibase 10.1086/375733} {\bibfield  {journal}
  {\bibinfo  {journal} {Astrophys. J. Suppl.}\ }\textbf {\bibinfo {volume}
  {147}},\ \bibinfo {pages} {403} (\bibinfo {year} {2003})},\ \Eprint
  {http://arxiv.org/abs/astro-ph/0304022} {arXiv:astro-ph/0304022} \BibitemShut
  {NoStop}%
\bibitem [{\citenamefont {Kravchuk}\ and\ \citenamefont
  {Yakovlev}(2014)}]{Kravchuk:2014}%
  \BibitemOpen
  \bibfield  {author} {\bibinfo {author} {\bibfnamefont {P.~A.}\ \bibnamefont
  {Kravchuk}}\ and\ \bibinfo {author} {\bibfnamefont {D.~G.}\ \bibnamefont
  {Yakovlev}},\ }\href {\doibase 10.1103/PhysRevC.89.015802} {\bibfield
  {journal} {\bibinfo  {journal} {Phys. Rev. C}\ }\textbf {\bibinfo {volume}
  {89}},\ \bibinfo {pages} {015802} (\bibinfo {year} {2014})},\ \Eprint
  {http://arxiv.org/abs/1311.7005} {arXiv:1311.7005 [astro-ph.SR]} \BibitemShut
  {NoStop}%
\bibitem [{\citenamefont {Timmes}\ and\ \citenamefont
  {Arnett}(1999)}]{TimmesArnett:1999}%
  \BibitemOpen
  \bibfield  {author} {\bibinfo {author} {\bibfnamefont {F.~X.}\ \bibnamefont
  {Timmes}}\ and\ \bibinfo {author} {\bibfnamefont {D.}~\bibnamefont
  {Arnett}},\ }\href {\doibase 10.1086/313271} {\bibfield  {journal} {\bibinfo
  {journal} {Astrophys. J. Suppl.}\ }\textbf {\bibinfo {volume} {125}},\
  \bibinfo {pages} {277} (\bibinfo {year} {1999})}\BibitemShut {NoStop}%
\bibitem [{\citenamefont {Paczynski}(1986)}]{Paczynski:1986px}%
  \BibitemOpen
  \bibfield  {author} {\bibinfo {author} {\bibfnamefont {B.}~\bibnamefont
  {Paczynski}},\ }\href {\doibase 10.1086/184740} {\bibfield  {journal}
  {\bibinfo  {journal} {Astrophys. J. Lett.}\ }\textbf {\bibinfo {volume}
  {308}},\ \bibinfo {pages} {L43} (\bibinfo {year} {1986})}\BibitemShut
  {NoStop}%
\bibitem [{\citenamefont {Lippuner}\ and\ \citenamefont
  {Roberts}(2015)}]{Lippuner:2015gwa}%
  \BibitemOpen
  \bibfield  {author} {\bibinfo {author} {\bibfnamefont {J.}~\bibnamefont
  {Lippuner}}\ and\ \bibinfo {author} {\bibfnamefont {L.~F.}\ \bibnamefont
  {Roberts}},\ }\href {\doibase 10.1088/0004-637X/815/2/82} {\bibfield
  {journal} {\bibinfo  {journal} {Astrophys. J.}\ }\textbf {\bibinfo {volume}
  {815}},\ \bibinfo {pages} {82} (\bibinfo {year} {2015})},\ \Eprint
  {http://arxiv.org/abs/1508.03133} {arXiv:1508.03133 [astro-ph.HE]}
  \BibitemShut {NoStop}%
\bibitem [{\citenamefont {Perego}\ \emph {et~al.}(2022)\citenamefont {Perego}
  \emph {et~al.}}]{Perego:2020evn}%
  \BibitemOpen
  \bibfield  {author} {\bibinfo {author} {\bibfnamefont {A.}~\bibnamefont
  {Perego}} \emph {et~al.},\ }\href {\doibase 10.3847/1538-4357/ac3751}
  {\bibfield  {journal} {\bibinfo  {journal} {Astrophys. J.}\ }\textbf
  {\bibinfo {volume} {925}},\ \bibinfo {pages} {22} (\bibinfo {year} {2022})},\
  \Eprint {http://arxiv.org/abs/2009.08988} {arXiv:2009.08988 [astro-ph.HE]}
  \BibitemShut {NoStop}%
\bibitem [{\citenamefont {{Kasen}}\ and\ \citenamefont
  {{Barnes}}(2019)}]{Kasen:2018drm}%
  \BibitemOpen
  \bibfield  {author} {\bibinfo {author} {\bibfnamefont {D.}~\bibnamefont
  {{Kasen}}}\ and\ \bibinfo {author} {\bibfnamefont {J.}~\bibnamefont
  {{Barnes}}},\ }\href {\doibase 10.3847/1538-4357/ab06c2} {\bibfield
  {journal} {\bibinfo  {journal} {Astrophys.\ J.}\ }\textbf {\bibinfo {volume}
  {876}},\ \bibinfo {eid} {128} (\bibinfo {year} {2019})},\ \Eprint
  {http://arxiv.org/abs/1807.03319} {arXiv:1807.03319 [astro-ph.HE]}
  \BibitemShut {NoStop}%
\bibitem [{\citenamefont {Curtis}\ \emph {et~al.}(2022)\citenamefont {Curtis},
  \citenamefont {M\"osta}, \citenamefont {Wu}, \citenamefont {Radice},
  \citenamefont {Roberts}, \citenamefont {Ricigliano},\ and\ \citenamefont
  {Perego}}]{Curtis:2021guz}%
  \BibitemOpen
  \bibfield  {author} {\bibinfo {author} {\bibfnamefont {S.}~\bibnamefont
  {Curtis}}, \bibinfo {author} {\bibfnamefont {P.}~\bibnamefont {M\"osta}},
  \bibinfo {author} {\bibfnamefont {Z.}~\bibnamefont {Wu}}, \bibinfo {author}
  {\bibfnamefont {D.}~\bibnamefont {Radice}}, \bibinfo {author} {\bibfnamefont
  {L.}~\bibnamefont {Roberts}}, \bibinfo {author} {\bibfnamefont
  {G.}~\bibnamefont {Ricigliano}}, \ and\ \bibinfo {author} {\bibfnamefont
  {A.}~\bibnamefont {Perego}},\ }\href {\doibase 10.1093/mnras/stac3128}
  {\bibfield  {journal} {\bibinfo  {journal} {Mon. Not. Roy. Astron. Soc.}\
  }\textbf {\bibinfo {volume} {518}},\ \bibinfo {pages} {5313} (\bibinfo {year}
  {2022})},\ \Eprint {http://arxiv.org/abs/2112.00772} {arXiv:2112.00772
  [astro-ph.HE]} \BibitemShut {NoStop}%
\bibitem [{\citenamefont {Ricigliano}\ \emph {et~al.}(2024)\citenamefont
  {Ricigliano}, \citenamefont {Jacobi},\ and\ \citenamefont
  {Arcones}}]{Ricigliano:2024lwf}%
  \BibitemOpen
  \bibfield  {author} {\bibinfo {author} {\bibfnamefont {G.}~\bibnamefont
  {Ricigliano}}, \bibinfo {author} {\bibfnamefont {M.}~\bibnamefont {Jacobi}},
  \ and\ \bibinfo {author} {\bibfnamefont {A.}~\bibnamefont {Arcones}},\ }\href
  {\doibase 10.1093/mnras/stae1979} {\bibfield  {journal} {\bibinfo  {journal}
  {Mon. Not. Roy. Astron. Soc.}\ }\textbf {\bibinfo {volume} {533}},\ \bibinfo
  {pages} {2096} (\bibinfo {year} {2024})},\ \Eprint
  {http://arxiv.org/abs/2406.03649} {arXiv:2406.03649 [astro-ph.HE]}
  \BibitemShut {NoStop}%
\bibitem [{\citenamefont {{Hanover}}\ and\ \citenamefont
  {{Leising}}(2025)}]{Hanover:2025}%
  \BibitemOpen
  \bibfield  {author} {\bibinfo {author} {\bibfnamefont {T.~E.}\ \bibnamefont
  {{Hanover}}}\ and\ \bibinfo {author} {\bibfnamefont {M.~D.}\ \bibnamefont
  {{Leising}}},\ }\href {\doibase 10.3847/1538-4357/adec8b} {\bibfield
  {journal} {\bibinfo  {journal} {\apj}\ }\textbf {\bibinfo {volume} {989}},\
  \bibinfo {eid} {199} (\bibinfo {year} {2025})},\ \Eprint
  {http://arxiv.org/abs/2507.11647} {arXiv:2507.11647 [astro-ph.HE]}
  \BibitemShut {NoStop}%
\bibitem [{\citenamefont {Kosakowski}\ \emph {et~al.}(2022)\citenamefont
  {Kosakowski}, \citenamefont {Ugalino}, \citenamefont {Fisher}, \citenamefont
  {Graur}, \citenamefont {Bobrick},\ and\ \citenamefont
  {Perets}}]{Kosakowski:2022odv}%
  \BibitemOpen
  \bibfield  {author} {\bibinfo {author} {\bibfnamefont {D.}~\bibnamefont
  {Kosakowski}}, \bibinfo {author} {\bibfnamefont {M.~I.}\ \bibnamefont
  {Ugalino}}, \bibinfo {author} {\bibfnamefont {R.}~\bibnamefont {Fisher}},
  \bibinfo {author} {\bibfnamefont {O.}~\bibnamefont {Graur}}, \bibinfo
  {author} {\bibfnamefont {A.}~\bibnamefont {Bobrick}}, \ and\ \bibinfo
  {author} {\bibfnamefont {H.~B.}\ \bibnamefont {Perets}},\ }\href {\doibase
  10.1093/mnrasl/slac152} {\bibfield  {journal} {\bibinfo  {journal} {Mon. Not.
  Roy. Astron. Soc.}\ }\textbf {\bibinfo {volume} {519}},\ \bibinfo {pages}
  {L74} (\bibinfo {year} {2022})},\ \Eprint {http://arxiv.org/abs/2210.10804}
  {arXiv:2210.10804 [astro-ph.HE]} \BibitemShut {NoStop}%
\bibitem [{\citenamefont {The}\ \emph {et~al.}(2006)\citenamefont {The},
  \citenamefont {Clayton}, \citenamefont {Diehl}, \citenamefont {Hartmann},
  \citenamefont {Iyudin}, \citenamefont {Leising}, \citenamefont {Meyer},
  \citenamefont {Motizuki},\ and\ \citenamefont {Schonfelder}}]{The:2006iu}%
  \BibitemOpen
  \bibfield  {author} {\bibinfo {author} {\bibfnamefont {L.-S.}\ \bibnamefont
  {The}}, \bibinfo {author} {\bibfnamefont {D.~D.}\ \bibnamefont {Clayton}},
  \bibinfo {author} {\bibfnamefont {R.}~\bibnamefont {Diehl}}, \bibinfo
  {author} {\bibfnamefont {D.~H.}\ \bibnamefont {Hartmann}}, \bibinfo {author}
  {\bibfnamefont {A.~F.}\ \bibnamefont {Iyudin}}, \bibinfo {author}
  {\bibfnamefont {M.~D.}\ \bibnamefont {Leising}}, \bibinfo {author}
  {\bibfnamefont {B.~S.}\ \bibnamefont {Meyer}}, \bibinfo {author}
  {\bibfnamefont {Y.}~\bibnamefont {Motizuki}}, \ and\ \bibinfo {author}
  {\bibfnamefont {V.}~\bibnamefont {Schonfelder}},\ }\href {\doibase
  10.1051/0004-6361:20054626} {\bibfield  {journal} {\bibinfo  {journal}
  {Astron. Astrophys.}\ }\textbf {\bibinfo {volume} {450}},\ \bibinfo {pages}
  {1037} (\bibinfo {year} {2006})},\ \Eprint
  {http://arxiv.org/abs/astro-ph/0601039} {arXiv:astro-ph/0601039} \BibitemShut
  {NoStop}%
\bibitem [{\citenamefont {Chen}\ and\ \citenamefont
  {Singh}(2023)}]{Chen:2023abc}%
  \BibitemOpen
  \bibfield  {author} {\bibinfo {author} {\bibfnamefont {J.}~\bibnamefont
  {Chen}}\ and\ \bibinfo {author} {\bibfnamefont {B.}~\bibnamefont {Singh}},\
  }\href {\doibase
  {[https://doi.org/10.1016/j.nds.2023.06.001](https://doi.org/10.1016/j.nds.2023.06.001)}}
  {\bibfield  {journal} {\bibinfo  {journal} {Nuclear Data Sheets}\ }\textbf
  {\bibinfo {volume} {190}},\ \bibinfo {pages} {1} (\bibinfo {year}
  {2023})}\BibitemShut {NoStop}%
\bibitem [{\citenamefont {{Jerkstrand}}\ \emph {et~al.}(2011)\citenamefont
  {{Jerkstrand}}, \citenamefont {{Fransson}},\ and\ \citenamefont
  {{Kozma}}}]{Jerkstrand:2011}%
  \BibitemOpen
  \bibfield  {author} {\bibinfo {author} {\bibfnamefont {A.}~\bibnamefont
  {{Jerkstrand}}}, \bibinfo {author} {\bibfnamefont {C.}~\bibnamefont
  {{Fransson}}}, \ and\ \bibinfo {author} {\bibfnamefont {C.}~\bibnamefont
  {{Kozma}}},\ }\href {\doibase 10.1051/0004-6361/201015937} {\bibfield
  {journal} {\bibinfo  {journal} {{Astron.\ and Astrophys.}}\ }\textbf
  {\bibinfo {volume} {530}},\ \bibinfo {eid} {A45} (\bibinfo {year} {2011})},\
  \Eprint {http://arxiv.org/abs/1103.3653} {arXiv:1103.3653 [astro-ph.HE]}
  \BibitemShut {NoStop}%
\bibitem [{\citenamefont {Grebenev}\ \emph {et~al.}(2012)\citenamefont
  {Grebenev}, \citenamefont {Lutovinov}, \citenamefont {Tsygankov},\ and\
  \citenamefont {Winkler}}]{Grebenev:2012}%
  \BibitemOpen
  \bibfield  {author} {\bibinfo {author} {\bibfnamefont {S.~A.}\ \bibnamefont
  {Grebenev}}, \bibinfo {author} {\bibfnamefont {A.~A.}\ \bibnamefont
  {Lutovinov}}, \bibinfo {author} {\bibfnamefont {S.~S.}\ \bibnamefont
  {Tsygankov}}, \ and\ \bibinfo {author} {\bibfnamefont {C.}~\bibnamefont
  {Winkler}},\ }\href {\doibase 10.1038/nature11473} {\bibfield  {journal}
  {\bibinfo  {journal} {Nature}\ }\textbf {\bibinfo {volume} {490}},\ \bibinfo
  {pages} {373} (\bibinfo {year} {2012})},\ \Eprint
  {http://arxiv.org/abs/1211.2656} {arXiv:1211.2656 [astro-ph.HE]} \BibitemShut
  {NoStop}%
\bibitem [{\citenamefont {Huo}\ \emph {et~al.}(2011)\citenamefont {Huo},
  \citenamefont {Huo},\ and\ \citenamefont {Yang}}]{Huo:2011A56}%
  \BibitemOpen
  \bibfield  {author} {\bibinfo {author} {\bibfnamefont {J.}~\bibnamefont
  {Huo}}, \bibinfo {author} {\bibfnamefont {S.}~\bibnamefont {Huo}}, \ and\
  \bibinfo {author} {\bibfnamefont {D.}~\bibnamefont {Yang}},\ }\href {\doibase
  10.1016/j.nds.2011.04.004} {\bibfield  {journal} {\bibinfo  {journal}
  {Nuclear Data Sheets}\ }\textbf {\bibinfo {volume} {112}},\ \bibinfo {pages}
  {1513} (\bibinfo {year} {2011})}\BibitemShut {NoStop}%
\bibitem [{\citenamefont {Watson}\ \emph {et~al.}(2019)\citenamefont {Watson}
  \emph {et~al.}}]{Watson:2019xjv}%
  \BibitemOpen
  \bibfield  {author} {\bibinfo {author} {\bibfnamefont {D.}~\bibnamefont
  {Watson}} \emph {et~al.},\ }\href {\doibase 10.1038/s41586-019-1676-3}
  {\bibfield  {journal} {\bibinfo  {journal} {Nature}\ }\textbf {\bibinfo
  {volume} {574}},\ \bibinfo {pages} {497} (\bibinfo {year} {2019})},\ \Eprint
  {http://arxiv.org/abs/1910.10510} {arXiv:1910.10510 [astro-ph.HE]}
  \BibitemShut {NoStop}%
\bibitem [{\citenamefont {Domoto}\ \emph {et~al.}(2021)\citenamefont {Domoto},
  \citenamefont {Tanaka}, \citenamefont {Wanajo},\ and\ \citenamefont
  {Kawaguchi}}]{Domoto:2021xfq}%
  \BibitemOpen
  \bibfield  {author} {\bibinfo {author} {\bibfnamefont {N.}~\bibnamefont
  {Domoto}}, \bibinfo {author} {\bibfnamefont {M.}~\bibnamefont {Tanaka}},
  \bibinfo {author} {\bibfnamefont {S.}~\bibnamefont {Wanajo}}, \ and\ \bibinfo
  {author} {\bibfnamefont {K.}~\bibnamefont {Kawaguchi}},\ }\href {\doibase
  10.3847/1538-4357/abf358} {\bibfield  {journal} {\bibinfo  {journal}
  {Astrophys. J.}\ }\textbf {\bibinfo {volume} {913}},\ \bibinfo {pages} {26}
  (\bibinfo {year} {2021})},\ \Eprint {http://arxiv.org/abs/2103.15284}
  {arXiv:2103.15284 [astro-ph.HE]} \BibitemShut {NoStop}%
\bibitem [{\citenamefont {Gillanders}\ \emph {et~al.}(2022)\citenamefont
  {Gillanders}, \citenamefont {Smartt}, \citenamefont {Sim}, \citenamefont
  {Bauswein},\ and\ \citenamefont {Goriely}}]{Gillanders:2022opm}%
  \BibitemOpen
  \bibfield  {author} {\bibinfo {author} {\bibfnamefont {J.~H.}\ \bibnamefont
  {Gillanders}}, \bibinfo {author} {\bibfnamefont {S.~J.}\ \bibnamefont
  {Smartt}}, \bibinfo {author} {\bibfnamefont {S.~A.}\ \bibnamefont {Sim}},
  \bibinfo {author} {\bibfnamefont {A.}~\bibnamefont {Bauswein}}, \ and\
  \bibinfo {author} {\bibfnamefont {S.}~\bibnamefont {Goriely}},\ }\href
  {\doibase 10.1093/mnras/stac1258} {\bibfield  {journal} {\bibinfo  {journal}
  {Mon. Not. Roy. Astron. Soc.}\ }\textbf {\bibinfo {volume} {515}},\ \bibinfo
  {pages} {631} (\bibinfo {year} {2022})},\ \Eprint
  {http://arxiv.org/abs/2202.01786} {arXiv:2202.01786 [astro-ph.HE]}
  \BibitemShut {NoStop}%
\bibitem [{\citenamefont {Tarumi}\ \emph {et~al.}(2023)\citenamefont {Tarumi},
  \citenamefont {Hotokezaka}, \citenamefont {Domoto},\ and\ \citenamefont
  {Tanaka}}]{Tarumi:2023apl}%
  \BibitemOpen
  \bibfield  {author} {\bibinfo {author} {\bibfnamefont {Y.}~\bibnamefont
  {Tarumi}}, \bibinfo {author} {\bibfnamefont {K.}~\bibnamefont {Hotokezaka}},
  \bibinfo {author} {\bibfnamefont {N.}~\bibnamefont {Domoto}}, \ and\ \bibinfo
  {author} {\bibfnamefont {M.}~\bibnamefont {Tanaka}},\ }\href@noop {} {\
  (\bibinfo {year} {2023})},\ \Eprint {http://arxiv.org/abs/2302.13061}
  {arXiv:2302.13061 [astro-ph.HE]} \BibitemShut {NoStop}%
\bibitem [{\citenamefont {Sneppen}\ \emph {et~al.}(2024)\citenamefont
  {Sneppen}, \citenamefont {Damgaard}, \citenamefont {Watson}, \citenamefont
  {Collins}, \citenamefont {Shingles},\ and\ \citenamefont
  {Sim}}]{Sneppen:2024ros}%
  \BibitemOpen
  \bibfield  {author} {\bibinfo {author} {\bibfnamefont {A.}~\bibnamefont
  {Sneppen}}, \bibinfo {author} {\bibfnamefont {R.}~\bibnamefont {Damgaard}},
  \bibinfo {author} {\bibfnamefont {D.}~\bibnamefont {Watson}}, \bibinfo
  {author} {\bibfnamefont {C.~E.}\ \bibnamefont {Collins}}, \bibinfo {author}
  {\bibfnamefont {L.}~\bibnamefont {Shingles}}, \ and\ \bibinfo {author}
  {\bibfnamefont {S.~A.}\ \bibnamefont {Sim}},\ }\href {\doibase
  10.1051/0004-6361/202451450} {\bibfield  {journal} {\bibinfo  {journal}
  {Astron. Astrophys.}\ }\textbf {\bibinfo {volume} {692}},\ \bibinfo {pages}
  {A134} (\bibinfo {year} {2024})},\ \Eprint {http://arxiv.org/abs/2407.12907}
  {arXiv:2407.12907 [astro-ph.HE]} \BibitemShut {NoStop}%
\bibitem [{\citenamefont {Arya}\ \emph {et~al.}(2026)\citenamefont {Arya},
  \citenamefont {Damgaard}, \citenamefont {Sneppen}, \citenamefont {Dougan},
  \citenamefont {Sim}, \citenamefont {Ballance},\ and\ \citenamefont
  {Watson}}]{Arya:2026jvs}%
  \BibitemOpen
  \bibfield  {author} {\bibinfo {author} {\bibfnamefont {A.}~\bibnamefont
  {Arya}}, \bibinfo {author} {\bibfnamefont {R.}~\bibnamefont {Damgaard}},
  \bibinfo {author} {\bibfnamefont {A.}~\bibnamefont {Sneppen}}, \bibinfo
  {author} {\bibfnamefont {D.~J.}\ \bibnamefont {Dougan}}, \bibinfo {author}
  {\bibfnamefont {S.~A.}\ \bibnamefont {Sim}}, \bibinfo {author} {\bibfnamefont
  {C.~P.}\ \bibnamefont {Ballance}}, \ and\ \bibinfo {author} {\bibfnamefont
  {D.}~\bibnamefont {Watson}},\ }\href@noop {} {\  (\bibinfo {year} {2026})},\
  \Eprint {http://arxiv.org/abs/2604.05812} {arXiv:2604.05812 [astro-ph.HE]}
  \BibitemShut {NoStop}%
\bibitem [{\citenamefont {Fujibayashi}\ \emph {et~al.}(2023)\citenamefont
  {Fujibayashi}, \citenamefont {Kiuchi}, \citenamefont {Wanajo}, \citenamefont
  {Kyutoku}, \citenamefont {Sekiguchi},\ and\ \citenamefont
  {Shibata}}]{Fujibayashi:2022ftg}%
  \BibitemOpen
  \bibfield  {author} {\bibinfo {author} {\bibfnamefont {S.}~\bibnamefont
  {Fujibayashi}}, \bibinfo {author} {\bibfnamefont {K.}~\bibnamefont {Kiuchi}},
  \bibinfo {author} {\bibfnamefont {S.}~\bibnamefont {Wanajo}}, \bibinfo
  {author} {\bibfnamefont {K.}~\bibnamefont {Kyutoku}}, \bibinfo {author}
  {\bibfnamefont {Y.}~\bibnamefont {Sekiguchi}}, \ and\ \bibinfo {author}
  {\bibfnamefont {M.}~\bibnamefont {Shibata}},\ }\href {\doibase
  10.3847/1538-4357/ac9ce0} {\bibfield  {journal} {\bibinfo  {journal}
  {Astrophys. J.}\ }\textbf {\bibinfo {volume} {942}},\ \bibinfo {pages} {39}
  (\bibinfo {year} {2023})},\ \Eprint {http://arxiv.org/abs/2205.05557}
  {arXiv:2205.05557 [astro-ph.HE]} \BibitemShut {NoStop}%
\bibitem [{\citenamefont {Barnes}\ and\ \citenamefont
  {Kasen}(2013)}]{Barnes:2013wka}%
  \BibitemOpen
  \bibfield  {author} {\bibinfo {author} {\bibfnamefont {J.}~\bibnamefont
  {Barnes}}\ and\ \bibinfo {author} {\bibfnamefont {D.}~\bibnamefont {Kasen}},\
  }\href {\doibase 10.1088/0004-637X/775/1/18} {\bibfield  {journal} {\bibinfo
  {journal} {Astrophys. J.}\ }\textbf {\bibinfo {volume} {775}},\ \bibinfo
  {pages} {18} (\bibinfo {year} {2013})},\ \Eprint
  {http://arxiv.org/abs/1303.5787} {arXiv:1303.5787 [astro-ph.HE]} \BibitemShut
  {NoStop}%
\bibitem [{\citenamefont {Kasen}\ \emph {et~al.}(2013)\citenamefont {Kasen},
  \citenamefont {Badnell},\ and\ \citenamefont {Barnes}}]{Kasen:2013xka}%
  \BibitemOpen
  \bibfield  {author} {\bibinfo {author} {\bibfnamefont {D.}~\bibnamefont
  {Kasen}}, \bibinfo {author} {\bibfnamefont {N.~R.}\ \bibnamefont {Badnell}},
  \ and\ \bibinfo {author} {\bibfnamefont {J.}~\bibnamefont {Barnes}},\ }\href
  {\doibase 10.1088/0004-637X/774/1/25} {\bibfield  {journal} {\bibinfo
  {journal} {Astrophys. J.}\ }\textbf {\bibinfo {volume} {774}},\ \bibinfo
  {pages} {25} (\bibinfo {year} {2013})},\ \Eprint
  {http://arxiv.org/abs/1303.5788} {arXiv:1303.5788 [astro-ph.HE]} \BibitemShut
  {NoStop}%
\bibitem [{\citenamefont {Even}\ \emph {et~al.}(2020)\citenamefont {Even},
  \citenamefont {Korobkin}, \citenamefont {Fryer}, \citenamefont {Fontes},
  \citenamefont {Wollaeger}, \citenamefont {Hungerford}, \citenamefont
  {Lippuner}, \citenamefont {Miller}, \citenamefont {Mumpower},\ and\
  \citenamefont {Misch}}]{Even:2019nbj}%
  \BibitemOpen
  \bibfield  {author} {\bibinfo {author} {\bibfnamefont {W.}~\bibnamefont
  {Even}}, \bibinfo {author} {\bibfnamefont {O.}~\bibnamefont {Korobkin}},
  \bibinfo {author} {\bibfnamefont {C.~L.}\ \bibnamefont {Fryer}}, \bibinfo
  {author} {\bibfnamefont {C.~J.}\ \bibnamefont {Fontes}}, \bibinfo {author}
  {\bibfnamefont {R.}~\bibnamefont {Wollaeger}}, \bibinfo {author}
  {\bibfnamefont {A.}~\bibnamefont {Hungerford}}, \bibinfo {author}
  {\bibfnamefont {J.}~\bibnamefont {Lippuner}}, \bibinfo {author}
  {\bibfnamefont {J.}~\bibnamefont {Miller}}, \bibinfo {author} {\bibfnamefont
  {M.~R.}\ \bibnamefont {Mumpower}}, \ and\ \bibinfo {author} {\bibfnamefont
  {G.~W.}\ \bibnamefont {Misch}},\ }\href {\doibase 10.3847/1538-4357/ab70b9}
  {\bibfield  {journal} {\bibinfo  {journal} {Astrophys. J.}\ }\textbf
  {\bibinfo {volume} {899}},\ \bibinfo {pages} {24} (\bibinfo {year} {2020})},\
  \Eprint {http://arxiv.org/abs/1904.13298} {arXiv:1904.13298 [astro-ph.HE]}
  \BibitemShut {NoStop}%
\bibitem [{\citenamefont {{Tanaka}}\ \emph {et~al.}(2020)\citenamefont
  {{Tanaka}}, \citenamefont {{Kato}}, \citenamefont {{Gaigalas}},\ and\
  \citenamefont {{Kawaguchi}}}]{Tanaka:2019iqp}%
  \BibitemOpen
  \bibfield  {author} {\bibinfo {author} {\bibfnamefont {M.}~\bibnamefont
  {{Tanaka}}}, \bibinfo {author} {\bibfnamefont {D.}~\bibnamefont {{Kato}}},
  \bibinfo {author} {\bibfnamefont {G.}~\bibnamefont {{Gaigalas}}}, \ and\
  \bibinfo {author} {\bibfnamefont {K.}~\bibnamefont {{Kawaguchi}}},\ }\href
  {\doibase 10.1093/mnras/staa1576} {\bibfield  {journal} {\bibinfo  {journal}
  {Mon.\ Not.\ Roy.\ Astron.\ Soc.}\ }\textbf {\bibinfo {volume} {496}},\
  \bibinfo {pages} {1369} (\bibinfo {year} {2020})},\ \Eprint
  {http://arxiv.org/abs/1906.08914} {arXiv:1906.08914 [astro-ph.HE]}
  \BibitemShut {NoStop}%
\bibitem [{\citenamefont {Miller}\ \emph {et~al.}(2019)\citenamefont {Miller},
  \citenamefont {Ryan}, \citenamefont {Dolence}, \citenamefont {Burrows},
  \citenamefont {Fontes}, \citenamefont {Fryer}, \citenamefont {Korobkin},
  \citenamefont {Lippuner}, \citenamefont {Mumpower},\ and\ \citenamefont
  {Wollaeger}}]{Miller:2019dpt}%
  \BibitemOpen
  \bibfield  {author} {\bibinfo {author} {\bibfnamefont {J.~M.}\ \bibnamefont
  {Miller}}, \bibinfo {author} {\bibfnamefont {B.~R.}\ \bibnamefont {Ryan}},
  \bibinfo {author} {\bibfnamefont {J.~C.}\ \bibnamefont {Dolence}}, \bibinfo
  {author} {\bibfnamefont {A.}~\bibnamefont {Burrows}}, \bibinfo {author}
  {\bibfnamefont {C.~J.}\ \bibnamefont {Fontes}}, \bibinfo {author}
  {\bibfnamefont {C.~L.}\ \bibnamefont {Fryer}}, \bibinfo {author}
  {\bibfnamefont {O.}~\bibnamefont {Korobkin}}, \bibinfo {author}
  {\bibfnamefont {J.}~\bibnamefont {Lippuner}}, \bibinfo {author}
  {\bibfnamefont {M.~R.}\ \bibnamefont {Mumpower}}, \ and\ \bibinfo {author}
  {\bibfnamefont {R.~T.}\ \bibnamefont {Wollaeger}},\ }\href {\doibase
  10.1103/PhysRevD.100.023008} {\bibfield  {journal} {\bibinfo  {journal}
  {Phys. Rev.}\ }\textbf {\bibinfo {volume} {D100}},\ \bibinfo {pages} {023008}
  (\bibinfo {year} {2019})},\ \Eprint {http://arxiv.org/abs/1905.07477}
  {arXiv:1905.07477 [astro-ph.HE]} \BibitemShut {NoStop}%
\bibitem [{\citenamefont {Curtis}\ \emph {et~al.}(2023)\citenamefont {Curtis},
  \citenamefont {Miller}, \citenamefont {Fr\"ohlich}, \citenamefont {Sprouse},
  \citenamefont {Lloyd-Ronning},\ and\ \citenamefont
  {Mumpower}}]{Curtis:2023xzm}%
  \BibitemOpen
  \bibfield  {author} {\bibinfo {author} {\bibfnamefont {S.}~\bibnamefont
  {Curtis}}, \bibinfo {author} {\bibfnamefont {J.~M.}\ \bibnamefont {Miller}},
  \bibinfo {author} {\bibfnamefont {C.}~\bibnamefont {Fr\"ohlich}}, \bibinfo
  {author} {\bibfnamefont {T.}~\bibnamefont {Sprouse}}, \bibinfo {author}
  {\bibfnamefont {N.}~\bibnamefont {Lloyd-Ronning}}, \ and\ \bibinfo {author}
  {\bibfnamefont {M.}~\bibnamefont {Mumpower}},\ }\href {\doibase
  10.3847/2041-8213/acba16} {\bibfield  {journal} {\bibinfo  {journal}
  {Astrophys. J. Lett.}\ } (\bibinfo {year} {2023}),\
  10.3847/2041-8213/acba16},\ \Eprint {http://arxiv.org/abs/2212.10691}
  {arXiv:2212.10691 [astro-ph.HE]} \BibitemShut {NoStop}%
\bibitem [{\citenamefont {{Arnett}}(1980)}]{Arnett:1980}%
  \BibitemOpen
  \bibfield  {author} {\bibinfo {author} {\bibfnamefont {W.~D.}\ \bibnamefont
  {{Arnett}}},\ }\href {\doibase 10.1086/157898} {\bibfield  {journal}
  {\bibinfo  {journal} {\apj}\ }\textbf {\bibinfo {volume} {237}},\ \bibinfo
  {pages} {541} (\bibinfo {year} {1980})}\BibitemShut {NoStop}%
\bibitem [{\citenamefont {Cook}\ \emph
  {et~al.}(2025{\natexlab{b}})\citenamefont {Cook}, \citenamefont
  {Guti{\'e}rrez}, \citenamefont {Bernuzzi}, \citenamefont {Radice},
  \citenamefont {Daszuta}, \citenamefont {Fields}, \citenamefont {Hammond},
  \citenamefont {Bandyopadhyay},\ and\ \citenamefont {Jacobi}}]{Cook:2025frw}%
  \BibitemOpen
  \bibfield  {author} {\bibinfo {author} {\bibfnamefont {W.}~\bibnamefont
  {Cook}}, \bibinfo {author} {\bibfnamefont {E.~M.}\ \bibnamefont
  {Guti{\'e}rrez}}, \bibinfo {author} {\bibfnamefont {S.}~\bibnamefont
  {Bernuzzi}}, \bibinfo {author} {\bibfnamefont {D.}~\bibnamefont {Radice}},
  \bibinfo {author} {\bibfnamefont {B.}~\bibnamefont {Daszuta}}, \bibinfo
  {author} {\bibfnamefont {J.}~\bibnamefont {Fields}}, \bibinfo {author}
  {\bibfnamefont {P.}~\bibnamefont {Hammond}}, \bibinfo {author} {\bibfnamefont
  {H.}~\bibnamefont {Bandyopadhyay}}, \ and\ \bibinfo {author} {\bibfnamefont
  {M.}~\bibnamefont {Jacobi}},\ }\href@noop {} {\  (\bibinfo {year}
  {2025}{\natexlab{b}})},\ \Eprint {http://arxiv.org/abs/2508.19342}
  {arXiv:2508.19342 [astro-ph.HE]} \BibitemShut {NoStop}%
\end{thebibliography}

\end{document}